\documentclass[%
 reprint,
nofootinbib,
 amsmath,amssymb,
 aps,
]{revtex4-1}
\usepackage[dvipdfmx]{graphicx}
\usepackage{amsthm}
\usepackage{amscd}
\usepackage{latexsym}
\usepackage{xcolor}
\usepackage{bm}
\usepackage{dcolumn}

\newlength{\figheight}
\newtheorem*{theorem*}{Theorem}
\newtheorem{lemma}{Lemma}

\newtheorem{conjecture}{Conjecture}

\DeclareMathOperator{\ch}{ch}
\DeclareMathOperator{\sh}{sh}
\DeclareMathOperator{\cth}{cth}
\DeclareMathOperator{\tgh}{th}
\DeclareMathOperator{\arch}{arch}

\renewcommand{\tanh}{\tgh}
\renewcommand{\log}{\ln}

\def\re{{\rm e}}
\def\rd{{\rm d}}

\def\la{\lambda}
\def\e{\varepsilon}

\def\te{\text{e}}

\allowdisplaybreaks

\begin{document}


\title{
Equilibrium dynamics of the  XX chain
}
\author{Frank G\"ohmann}
\affiliation{Fakult\"at f\"ur Mathematik und Naturwissenschaften,
Bergische Universit\"at Wuppertal, 42097 Wuppertal, Germany}
\author{Karol K. Kozlowski}
\affiliation{Univ Lyon, ENS de Lyon, Univ Claude Bernard, CNRS,
Laboratoire de Physique, F-69342 Lyon, France}
\author{Jesko Sirker}
\affiliation{Department of Physics \& Astronomy, 
University of Manitoba,
Winnipeg, Manitoba,
Canada R3T 2N2}
\author{Junji Suzuki}
\affiliation{Department of Physics, Faculty of Science, Shizuoka University,Ohya 836, Suruga, Shizuoka, Japan}

\begin{abstract}
The equilibrium dynamics of the spin-$\frac{1}{2}$ XX chain is
re-examined within a recently developed formalism based on the quantum
transfer matrix and a thermal form factor expansion. The transversal
correlation function is evaluated in real time and space. The
high-accuracy calculation reproduces several exact results in limiting
cases as well as the well-known asymptotic formulas obtained by the
matrix Riemann-Hilbert approach. Furthermore, comparisons to
numerical data based on a direct evaluation of the Pfaffian as well as
to asymptotic formulas obtained within non-linear Luttinger liquid
theory are presented.
\end{abstract}
\maketitle


\section{Introduction}\label{sec1}
An increasing knowledge has been acquired on the correlation functions
of the simplest models which are integrable in the sense of the
Yang-Baxter relation.
The vertex operator approach (VOA) has definitely triggered the recent
advance \cite{JimboMiwaBook}. It enables us to evaluate ground state
correlation functions, for instance, of the XXZ chain in the
antiferromagnetic regime in a vanishing magnetic field. A
multiple-integral formula for the reduced density matrix has been
derived naturally within this framework \cite{JMMN92}. The evaluation
of the two- and four-spinon contributions to the dynamical structure
factor of the massive XXZ chain
\cite{BKM98,AbadaBougourziSiLakhal} and the
four spinon ones in the XXX limit \cite{CauxHagemans2006} were
important outcomes of this method.

A complementary approach to the analysis of ground-state correlation
functions of the XXZ chain starts with its algebraic Bethe ansatz
solution.  It utilizes the solution of the quantum inverse problem
\cite{KitanineMailletTerras1999} and a special determinant formula for
the scalar product of on-shell and off-shell Bethe
vectors~\cite{Slavnov89}.
The algebraic Bethe ansatz approach
successfully confirmed \cite{KitanineMailletTerras2000} the multiple-integral
formula for the density matrix obtained within the VOA.
  It properly takes into account the effect of a finite magnetic field in the symmetry
direction of the chain. 
This  approach can be also used to derive and analyze form factor series for dynamical correlation functions in the thermodynamic limit \cite{KKMST09b,KKMST11b,
KKMST11a,KKMST12,DGKS15a,KozlowskiMaillet15,Kozlowski18}. These developments culminated in the extraction, on the basis of exact and first principle calculations, of the long-distance and large-time asymptotic behavior of two point functions in the XXZ chain at zero temperature \cite{Kozlowski19}. Also, it was possible to grasp on exact grounds, the full structure of the edge singularities of the spin structure factors and spectral functions in this model \cite{KozSingularitiesSpectralFctsXXZ}.

In a parallel development, field theoretical approaches have been extended to include non-linearities 
in the spectrum
\cite{ImambekovGlazmanScience,ImambekovSchmidtGlazmanRevModPhys}. It has been argued that using {\it non-linear Luttinger liquid theory} is crucial to obtain proper universal results for threshold singularities in spectral functions and for the long time asymptotics of dynamical correlation functions. For integrable lattice models, in particular, a combination of non-linear Luttinger liquid theory and Bethe ansatz has allowed to derive parameter-free results for edge singularities and high-energy tails of spin structure factors and spectral functions \cite{PereiraSirkerAffleckPRL,PereiraSirkerAffleckJSTAT,PereiraAffleckPRL,PereiraAffleckPRB}. Of particular relevance in the present context are results for the asymptotics of the dynamical longitudinal spin-spin correlation function \cite{KarraschPereiraSirker}
and the transverse spin-spin structure factor \cite{KarimiAffleck} for the XXZ chain at low but finite temperatures. These results play the same role for dynamical correlations as the well-known conformal field theory
formulas do in the static case. In contrast to the latter, however, a verification of the  asymptotic formulas based on a fully microscopic calculation is so far lacking in the dynamical case at low but finite temperature.

It has been observed that the simplest multiple integrals representing
the reduced density matrix of short sub-segments of the XXZ chain
factorize \cite{BoosKorepin2001,BKS03,
SakaiShiroishiNIshiyamaTakahashi2003,BKS04,BST05}.
The algebraic structure of the ground state expectation values of all finite-range operators
was finally understood with the discovery of a `hidden Grassmann
structure' on the space of such operators in \cite{BoosJimboMiwaSmirnovTakeyama2007, 
BoosJimboMiwaSmirnovTakeyama2007II, BoosJimboMiwaSmirnovTakeyama2009}.
While the structure turned out to be relevant for the analysis
of the short-range correlation functions  \cite{SABGKTT11,MiSm19}, 
 the form factor series seem to be more efficient in analyzing the long time,
large distance behavior of the two-point functions.
The microscopic verification \cite{KKMST11b,KozlowskiMaillet15} of the
predictions of conformal field theory \cite{Cardy86} for the XXZ chain
is an important application.

Thanks to the similarity in the structure of the row-to-row transfer matrix
of the six-vertex model and the quantum transfer matrix (QTM) of the XXZ
chain \cite{Kluemper1993}, many results for ground state correlation
functions of the XXZ chain could be generalized to finite temperatures.
This concerns in the first place the multiple-integral representation for
the reduced density matrix of sub-chains \cite{GohmannKlumperSeel2004,
GohmannKlumperSeel2005,GohmannHasencleverSeel2005}. The multiple
integrals factorize even at finite temperatures \cite{BoosGohmannKlumperSuzuki2006,
BoosGohmannKlumperSuzuki2007,BDGKSW08}, which finds its explanation again
in the hidden Grassmann structure \cite{JimboMiwaSmirnov2009}.
Moreover, for the static two-point functions at finite temperature so-called
thermal form-factor expansions, involving form factors of the 
QTM, were introduced and studied in \cite{DGK13a,DGK14a,DugaveGohmannKozlowskiSuzuki2016}.
As an interesting outcome we mention the evaluation of the static
two-point correlation functions in the low-temperature limit in terms
of higher particle-hole excitations \cite{DugaveGohmannKozlowskiSuzuki2016}.
They were conjectured to correspond to the two-, four- and six-spinon contributions in
the VOA. 

As compared to the ground-state case, the inclusion of the time dependence
into the thermal form factors series for the two-point functions requires slightly
more thought. 
It was realized in \cite{Sakai2007} that the solution to
the inverse problem for the QTM has to be adopted for this purpose.
By combining the lattice realization of the two-point functions suggested in \cite{Sakai2007}
with the thermal form factor expansion of \cite{DGK13a}, we recently
proposed \cite{GKKKS2017} a new scheme for the explicit evaluation of
dynamical correlation functions of quantum spin systems in thermal equilibrium. 
As a first test for the validity of the new scheme, we re-derived the well-known formula \cite{Niemeijer1967}
for the longitudinal equilibrium correlation function of the
spin-$\frac{1}{2}$ XX chain.  We also obtained a novel thermal form
factor series for the transversal two-point correlation function
$\langle\sigma_1^-(0) \sigma_{m+1}^+ (t)\rangle$ which will be analyzed
in this work.

It may seem rather amazing that for a simple model as the XX chain,
whose Hamiltonian can be expressed in terms of non-interacting spinless Fermions
\cite{LiebSchultzMattis1961}, the calculation of its transverse
dynamical two-point function at finite temperature still poses
interesting questions. 
It is considerably harder than the longitudinal case, because the
two-point function is non-local in terms of the Jordan-Wigner
Fermions. Consequently, we are facing the problem of evaluating large
determinants. Analytic studies are therefore mainly restricted to the
case where each matrix element has a simple structure so that the
Szeg{\"o} theorem
\cite{McCoyBarouchAbraham1971,BrandtJacoby1976} can be applied.
Results at finite temperatures, that were obtained within approaches
based on the use of Fermions, comprise asymptotic formulae for high
temperatures \cite{BrandtJacoby1976,PeCa77} as well as the numerical
evaluation for finite open systems.
The latter seems most efficient if the correlation function for a
finite length chain is represented in terms of a Pfaffian of two-point
functions of auxiliary Fermions \cite{DK1997,DKS2000}.
Fermion algebra and generalized Wick theorem lead to
an important consequence in the XY chain and its limit,
 a set of nonlinear difference-differential equations. See \cite{Perk1980, MPS1983, PCQN1984, AuYangPerk2009}
and references therein.

A completely different route to the evaluation of the transverse dynamical
two-point functions was taken in \cite{ColomoIzerginKorepinTognetti}.
Following the strategy of \cite{KorepinSlavnov} the authors of
\cite{ColomoIzerginKorepinTognetti} used the coordinate Bethe ansatz
to derive a Fredholm determinant representation of the correlation
function in the thermodynamic limit. 
The integral operator in the Fredholm determinant is of integrable type \cite{KBIBo}.
This formulation was then directly suitable for an asymptotic analysis 
at long times $t$ and large distances $m$ by means of an
associated matrix Riemann-Hilbert problem amenable to a `non-linear
steepest descent method' \cite{DeiftZhouSteepestDescentForOscillatoryRHP}. 
In \cite{ItsIzerginKorepinSlavnov1993}
it was shown that the long time, large distance behavior of the two-point
function for a fixed ratio $m/t$ in a weak magnetic field ($h<4 J$) goes like
$\langle\sigma_1^-(0) \sigma_{m+1}^+ (t)\rangle \sim A t^{\nu} {\rm e}^{- m/\xi}$.
The explicit expressions for $\xi$ and $\nu$ were obtained in the
space-like ($m > 4Jt$) and time-like ($m < 4Jt$) asymptotic regimes.
The result was subsequently extended to strong magnetic field ($h>4 J$)
\cite{Jie}. 

This is one of three companion papers in which we revisit the problem
of the evaluation of the transversal two-point function at finite temperatures.
Our starting point will be the novel thermal form factor expansion for
$\langle\sigma_1^-(0) \sigma_{m+1}^+ (t)\rangle$ that was derived in a
previous communication \cite{GKKKS2017}. In \cite{GKS19I, GKS19II} we
study two different aspects of asymptotic analysis: On the one hand, the high-temperature
analysis for all times and \emph{all} distances and, on the other one, the
evaluation of the amplitude $A$ in the space-like regime at weak
magnetic field ($h < 4J$). In this communication, we shall concentrate
on the numerical evaluation of the series. We will remind the reader in Section
\ref{sec:Fredholm} that the form factor series can be neatly re-summed
in terms of a Fredholm determinant different from the one in
\cite{ColomoIzerginKorepinTognetti}. The kernel of the integral operator
that defines the Fredholm determinant is a strongly oscillating
function for large values of $m$ and $t$. 
We shall demonstrate that, nevertheless, the idea presented in
\cite{Bornemann2010} of a direct evaluation of the determinant can be
applied, if combined with an appropriate choice of the integration
contour in the complex plane.
Then the real time evaluation can be performed in a stable manner
until rather large times $t \gg \frac{m}{4J}$.  Consequently, we are
able to numerically check the asymptotic results of
\cite{ItsIzerginKorepinSlavnov1993,Jie}. We are also able to suggest
a higher-order correction for $h < 4J$ from our numerical analysis.
The advantage of the method of numerical evaluation proposed in this work is that it works
directly in the thermodynamic limit and is free of finite size effects.
 
The paper is organized as follows. In Section \ref{sec:Fredholm}, we
start from a brief review of the results in \cite{GKKKS2017} and then
derive the novel Fredholm determinant representation of the transverse
correlation function. We will discuss the analytic properties of the
functions occurring in the determinant in Section
\ref{sec:analyticity_and_sdp}.  The steepest-descent paths in the
space-like and time-like regimes will be explained. They are crucial for
the numerics, especially in the large-time dynamics and for the
long-distance correlations. Based on these preparations, we present
the result of our numerical analysis in the massive phase in Section
\ref{sec:massive_numerics}. The kinematic poles will become important
at a later stage, and we shall argue how they can be treated
numerically. The situation becomes more complicated in the massless
phase due to the presence of Fermi points on the real axis. This will
be discussed in Section \ref{sec:massless_numerics}. In Section
\ref{sec:evenodd} numerical data are presented showing the difference in the oscillation
amplitudes of correlation functions at odd and even distances. These
differences are then explained based on asymptotic results obtained
within non-linear Luttinger liquid theory. A comparison with the
analytic predictions based on the Fredholm determinant representation
derived in
\cite{ColomoIzerginKorepinTognetti} is given in Section
\ref{sec:asymptotics}. Thanks to the high-precision calculation,
we identify a higher order correction in the massless phase. In Section
\ref{sec:other_method} we compare the new numerical scheme proposed here
and a previously existing method based on the Pfaffian
representation. We summarize our results and point out perspectives in
Section \ref{sec:summary}.  Some of the technical details as well as a
comparison with the exact static values are supplemented in several
appendices.
 
\section{A representation of the transverse correlation function by a Fredholm determinant }
\label{sec:Fredholm}
We consider the spin-$\frac{1}{2}$ XX chain,
\begin{equation}\label{eq:xx_hamiltonian}
     {\cal H} = J \sum_{i=1}^L
                \Bigl( \sigma_i^x \sigma_{i+1}^x +  \sigma_i^y\sigma_{i+1}^y \Bigr)
		-\frac{h}{2} \sum_{i=1}^L \sigma^z_i,
\end{equation}
with periodic boundaries in the thermodynamic limit $L \rightarrow \infty$. We
assume the strength of the spin-spin interaction $J$ and the Zeeman magnetic
field $h$ to be positive. The model has two phases in its ground state
phase diagram, a critical phase (low~$h$) with finite magnetization that depends
of the value of $h$, and a massive phase (high~$h$) in which the magnetization is 
saturated to its maximal possible value $1/2$. The two phases are separated
by the critical field
\begin{equation}
     h_c = 4 J.
\end{equation}

The quantity of interest in this work is the transverse correlation function 
\begin{multline}\label{eq:def_CmtII}
     \langle\sigma_1^-(0) \sigma_{m+1}^+ (t)\rangle_{J,h,T} \\
        = \lim_{L \rightarrow \infty}
	  \frac{{\rm tr}_{1,\cdots,L} \bigl({\re}^{-(\frac{1}{T}+it) {\cal H}}
	        \sigma^-_1 {\re}^{it {\cal H}} \sigma^+_{m+1}\bigr)}
	       {{\rm tr}_{1,\cdots,L}  \bigl({\re}^{-\frac{1}{T} {\cal H}} \bigr)}.
\end{multline}
We shall consider it as a function of distance~$m$ and time~$t$ that depends
parametrically on $J$ and $h$ and on the temperature $T$. The subscripts $J,
h, T$ will often be suppressed below.

We divide the $m$-$t$ space-time plane into two regimes: the space-like regime with
$t < t_c$ and the time-like regime with $t>t_c$, where
\begin{equation}
     t_c = \frac{m}{4J}.
\end{equation}
We shall see that the qualitative behavior of the transverse correlation
function falls into 4 categories: massive time-like, massive space-like,
massless time-like and massless space-like. Our aim is to understand
the different categories quantitatively based on the QTM method.

 
The application of the QTM method and the derivation of a thermal
form factor expansion for the transverse correlation function
(\ref{eq:def_CmtII}) of the XX model was described in detail in
Section 3.5 of \cite{GKKKS2017}. We give a brief summary of the
result obtained there in Appendix~\ref{app:summary_thermal_ffseries}.
%
%
\subsection{A representation by a Fredholm determinant}
\begin{figure}[hbtp]
\centering
\includegraphics[width=3.5cm]{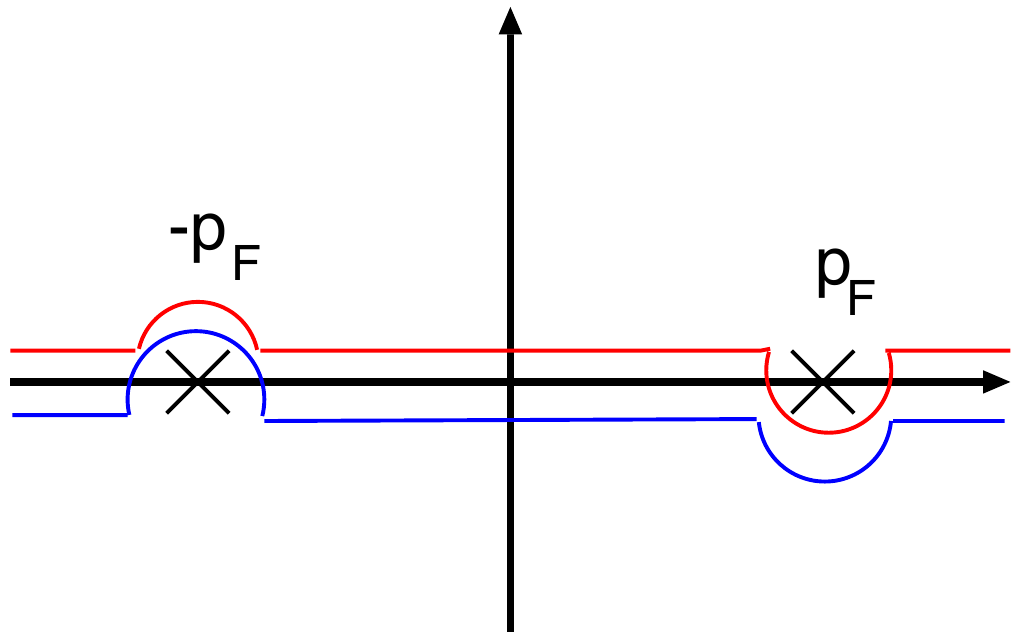}
\caption{The contours ${\cal E}$(red) and   $\bar{\cal E}$(blue) for  $h<h_c$.}
\label{fig:contourE}
\end{figure}

In our previous work \cite{GKKKS2017} we expressed the integrals in the
thermal form factor series in terms of rapidity variables $\la$ (see
(\ref{eq:ffseriesxxtransd})). This is natural for integrable systems in
general, but for the numerical study presented here, it turns out
to be more convenient to switch to momentum variables $p$. Then the
one-particle dispersion takes its familiar functional form
\[
     \epsilon(p) = h-4 J \cos(p) .
\]
By the common abuse of notation we use the same symbol $\epsilon$
to denote the energy as a function of the momentum or rapidity
variables. The substitution $\lambda \mapsto p(\lambda)$ in
(\ref{eq:ffseriesxxtransd}) with $p(\la)$ according to (\ref{eq:pepslambda})
then leads to
\begin{align} \label{eq:ffseriesxxtransp}
     & \langle\sigma_1^- (0)\sigma_{m+1}^+ (t) \rangle  
       = (-1)^{m+1}
	 {\cal A} (m)
         \\ & \times
         \sum_{n=1}^\infty \frac{(-1)^{n-1}}{n! (n-1)!}
         \int_{\cal E} \prod_{r=1}^n \rd p_r
	 \mu(p_r)  \re^{i (m p_r - t \epsilon (p_r))}
	 \notag \\ & \times
         \int_{\bar{\cal E}} \prod_{s=1}^{n-1} \rd q_s
         \bar{ \mu}(q_s)  \re^{-i (m q_s - t \epsilon (q_s))}
	 {\cal D}_p^2 \bigl( \{p_r\}_{r=1}^n, \{q_s\}_{s=1}^{n-1} \bigr), \notag
\end{align}
where $ {\cal A} (m)$  is defined in  (\ref{defamt}), 
\begin{align}
\label{eq:measures}
  &\mu(p)= \frac{{\re}^{\sigma_+(p)} }{2\pi (1-\re^{\epsilon(p)/T} )},
  \quad
  \bar{ \mu}(q)= \frac{{\re}^{-\sigma_-(q)}}{2\pi  (1-\re^{-\epsilon(q)/T})},
  \notag \\[1ex]
  &\sigma(p)=  \int_{\cal E} \frac{dq}{2\pi i} \frac{1}{\tan\bigl(\frac{p-q}{2}\bigr)}
   \ln \Bigl( \frac{1+\re^{-\epsilon(q)/T}}{1-\re^{-\epsilon(q)/T}}\Bigr),
   \notag \\[1ex]
  & {\cal D}_p \bigl( \{p_r\}_{r=1}^n, \{q_s\}_{s=1}^{n-1} \bigr)=  \nonumber \\
   &\phantom{ab} \frac{    \underset{1\le j < k \le n}{\prod}\sin \bigl( \frac{p_j-p_k}{2}\bigr)   \underset{1\le j < k \le n-1} {\prod}\sin \bigl( \frac{q_j-q_k}{2} \bigr)}
   {\prod_{j=1}^n \prod_{k=1}^{n-1} \sin\bigl( \frac{p_j-q_k}{2} \bigr) },
\end{align}
and  $\sigma_+(p)$ (respectively  $\sigma_-(p)$) is the boundary value as $p$ approaches
${\cal E}$ from above (respectively \ below). We call $\{p_r\}_{r=1}^n$ the hole momenta
and $\{q_s\}_{s=1}^{n-1}$ the particle momenta.
The contour ${\cal E}$ is a straight line $[-\pi+i\delta,
\pi+i\delta]$ in the massive phase, $h>h_c$.  Similarly, $\bar{\cal
E}$ is a contour just below the real axis. When $h<h_c$, the contours
will be slightly deformed in order to avoid the Fermi points%
\footnote{The Fermi momentum $k_F$ with the proper dimension
will be introduced later in Section \ref{sec:evenodd}.} $\pm p_F$
(defined by $\epsilon(\pm p_F)=0$) as shown in Figure~\ref{fig:contourE}.  
The contours ${\cal E}$ and $\bar{\cal E}$ are the images of ${\cal C}$
and $\bar{\cal C}$ (see Appendix~\ref{app:summary_thermal_ffseries}) under the transformation
$\lambda \mapsto p(\lambda)$. 

Using a technique developed in \cite{KorepinSlavnov} we can rewrite
(\ref{eq:ffseriesxxtransp}) as an explicit factor times a Fredholm
determinant of a special form. For this purpose, let us prepare
some notations in advance,
\begin{align}
     \tilde{V}(q_i,q_j) & = \int_{\cal E} \rd p \, \mu(p)
        \varphi(p,q_i)\varphi(p,q_j)\re^{2 t u_{m,t}(p)}, \notag \\
        V(q_i,q_j) & = \re^{-t u_{m,t} (q_i)}
	\tilde{V}(q_i,q_j) \re^{-t u_{m,t} (q_j)}, \notag \\
        \tilde{v}(q_i) & =  \int_{\cal E} \rd p\,
	                    \mu(p) \varphi(p,q_i) \re^{2 t u_{m,t}(p)}, \notag \\
        v(q_i) & = \re^{-t u_{m,t} (q_i)} \tilde{v}(q_i), \notag \\
        \Omega(m,t)&=  \int_{\cal E} \rd p \,\mu(p) \re^{2 t u_{m,t}(p)}, \notag \\
        P(q_i,q_j)&=\frac{v(q_i) v(q_j)}{\Omega(m,t)},  \label{eq:defP}
\end{align}
where
\begin{equation}
     \varphi(p,q) = \frac{\re^{i\frac{q-p}{2}}}{\sin \bigl(\frac{q-p}{2}\bigr)}, \quad
     u_{m,t}(p)=\frac{i}{2}  (\frac{m}{t}p- \epsilon(p)).
\end{equation}
We further define integral operators,
\begin{align}
(\hat{V}f) (q) &= \int_{\bar{\cal E}} dq' V(q,q') f(q') \bar{\mu}(q'),  \notag \\
(\hat{P}f) (q) &=  \frac{v(q)}{\Omega(m,t)}\int_{\bar{\cal E}} \rd q' \, v(q') f(q') \bar{\mu}(q').
\end{align}
The operator $\hat{P}$ is obviously a one-dimensional projector.
After these preparations, by a calculation very similar to that
in \cite{GKS19I}, we obtain the following formula.
\begin{theorem*}\label{prop:Fredholm_representation}
The transverse correlation function of the spin-$\frac{1}{2}$ XX model
can be represented using a Fredholm determinant that involves the integral
operator $\hat{P} - \hat{V}$,
\begin{multline} \label{eq:Fredholm_representation}
     \langle\sigma_1^-(0) \sigma_{m+1}^+ (t)\rangle
        = (-1)^{m+1}
	  {\cal A} (m) \Omega(m,t) \\
          \times {\rm det}_{\bar{{\cal E}}} (1+ \hat{P}-\hat{V}).
\end{multline} 
\end{theorem*}
The symbol ${\rm det}_{\cal S}$ means the  Fredholm determinant
where the integral operator is acting on functions supported on ${\cal S}$.


Equation ~(\ref{eq:Fredholm_representation}) is a novel analytic expression
representing the transverse dynamical correlation function
(\ref{eq:def_CmtII}) of the infinite system at finite temperatures.
In recent years it has become increasingly clear \cite{Bornemann2010}
that Fredholm determinants can be very efficiently evaluated numerically
by simply approximating them by determinants of finite matrices,
very much like integrals are approximated by finite sums. The numerical
error decreases with the number of discretization points. It has been
estimated for kernels with various analytic properties in
\cite{Bornemann2010}.

When we evaluate (\ref{eq:Fredholm_representation}) numerically,
we are considering the kernel as a function of the momentum
variables that depends parametrically on distance $m$ and time $t$.
When $m$ and $t$ become large, the kernel function becomes
a rapidly oscillating function of the momentum variables on the
contour $\bar{\cal E}$. This affects the numerical accuracy of
the calculation in a similar way as in the case of integrals
over rapidly oscillating functions. As we shall see, the cure to
this problem as well is similar as in the case of integrals
over rapidly oscillating functions. We have to deform the
integration contour in the complex plane into a saddle point
contour. Since we are dealing with meromorphic functions with
infinitely many poles the deformation is an intricate issue.
This will be, on a technical level, the main subject of the
discussions in the following sections.

We would like to remark that $\hat{V}$ belongs to the class of
integrable integral operators \cite{KBIBo}. This is easily seen
if we rewrite its kernel in the following form,
\begin{equation}\label{eq:Vandv} 
\widetilde{V}(q_i,q_j) = \frac{{\rm e}^{i\frac{(q_i-q_j)}{2}} \widetilde{v}(q_j) -
{\rm e}^{-i\frac{(q_i-q_j)}{2}} \widetilde{v}(q_i) }{\sin\bigl( \frac{q_i-q_j}{2}\bigr)}.
\end{equation}
For integrable integral operators there exist powerful tools in
order to analyze the asymptotic behavior analytically
\cite{DeiftZhouSteepestDescentForOscillatoryRHP}. A representation of
the correlation function (\ref{eq:def_CmtII}) involving a Fredholm
determinant with a similar but manifestly different kernel, which
belongs to the family of integrable integral operators as well, was
analyzed in \cite{ItsIzerginKorepinSlavnov1993}. The leading terms
of the large space and time asymptotic behavior of the transverse
correlation function of the XX model have been successfully derived
from that representation.  Here, however, we are interested in a
quantitative study on an {\it arbitrary} space scale and time scale.
%
%
\section{The analytic properties of integrands and the steepest
descent paths}\label{sec:analyticity_and_sdp}
The contours ${\cal E}, \bar{\cal E}$ are  optimal choices
in the static limit.  As time evolves, the phase factors $\re^{\pm i(mp- \epsilon(p)t)}$
bring instability and we eventually need to deform the contours for a reliable calculation.
One may encounter singularities of the integrands during the deformation.
We thus have to find a balance between the advantage of reducing the instability from
$\re^{\pm i(mp- \epsilon(p)t)}$ and the cost of passing through many singularities
of the integrands. Below we shall discuss the analytic properties of the integrands 
and the optimal choice of integration paths.

\subsection{The saddle points and the steepest descent paths}\label{sec:steepest_descendent_path}

We need to take account of the  steepest descent paths for ${\rm e}^{\pm u_{m,t}(p) t}$
in the asymptotic region $m, t \gg 1$. The locations of saddle points for the
system  in the space-like regime are qualitatively different  from those for
the system in the time-like regime. On the other hand, they do not depend on
whether the system is in the massive or in the massless phase.
\subsubsection{ The space-like regime $t<t_c$}
\begin{figure}[hbtp]
\centering
\setlength{\tabcolsep}{20pt}
\includegraphics[height=\figheight]{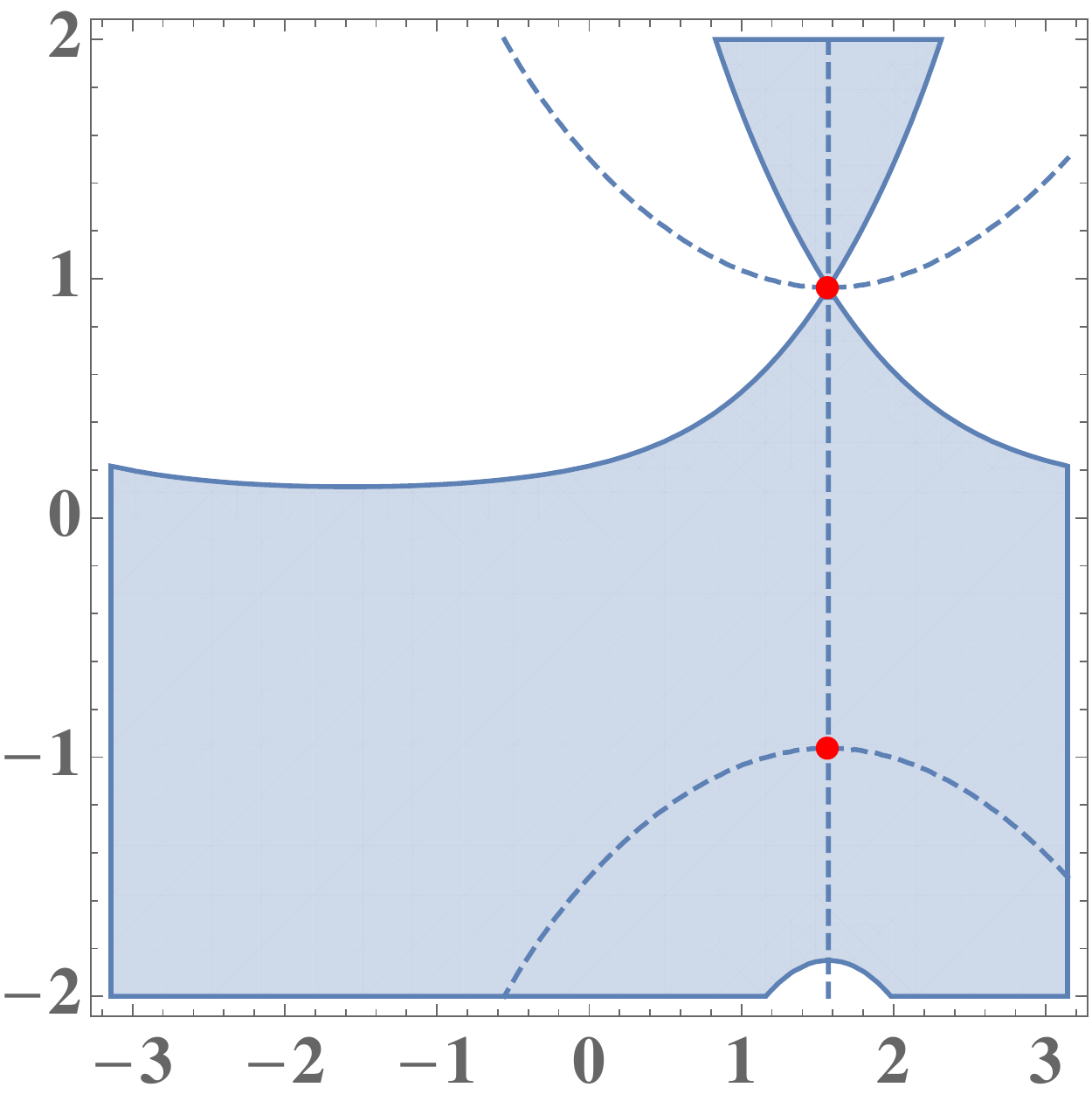} \hspace{0.5cm}
\includegraphics[height=\figheight]{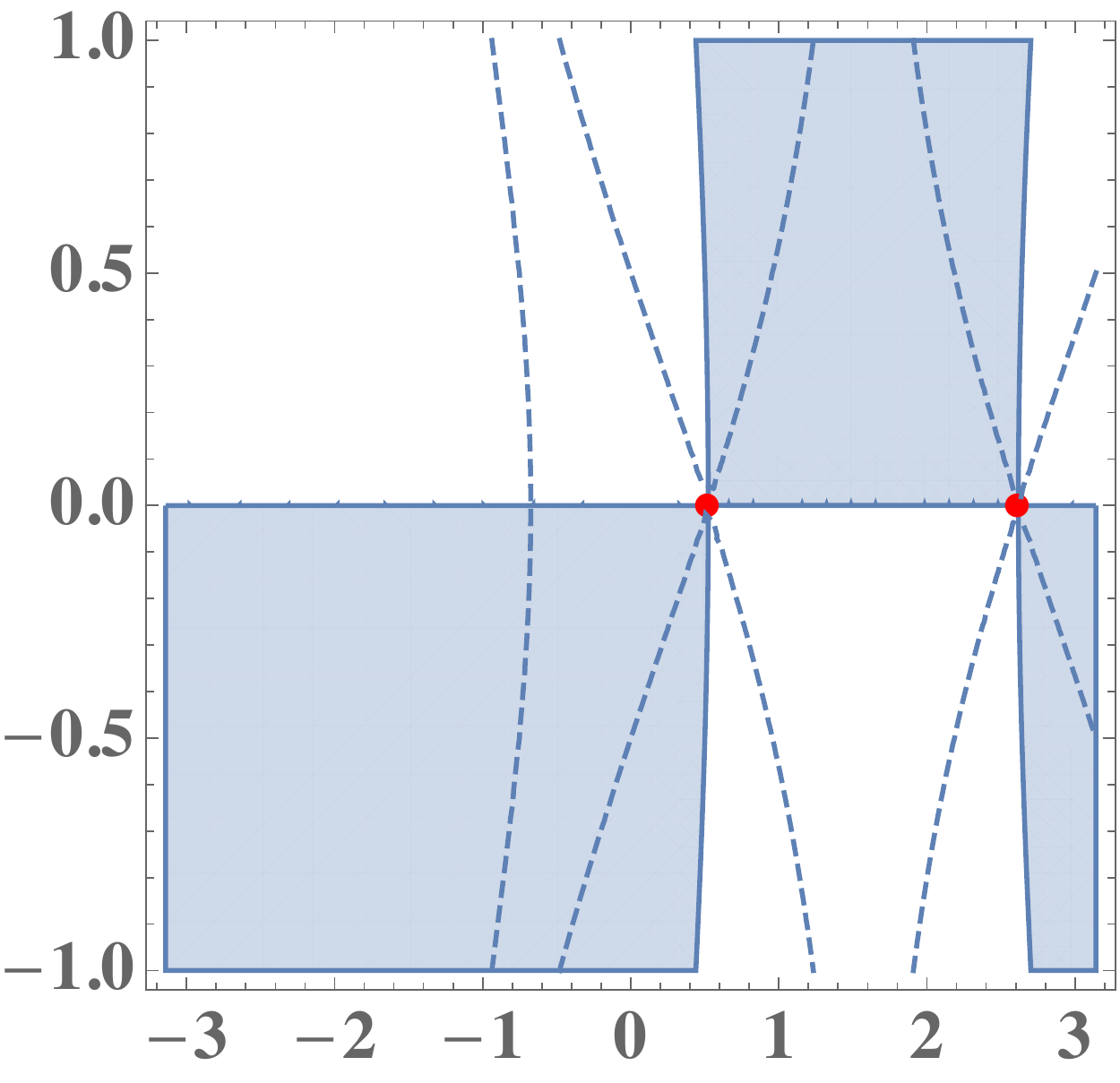}
\caption{In the space-like regime (left panel) the steepest descent
path for the hole momenta passes $p_+$ ``horizontally". In the time-like
regime (right panel) the steepest descent path passes through both,
$p_+$ and $p_-$.}
\label{fig:steepest_space}
\end{figure}

  The saddle points $p_{\pm}$  in the space-like regime  lie at
\begin{equation}\label{eq:saddlepointInspace}
p_{\pm} = \frac{\pi}{2} \pm i \arch \bigl(\frac{m}{4 J t }\bigr).
\end{equation}
The red points in Figure~\ref{fig:steepest_space} denote  them.
The dashed curve corresponds to the \textit{loci} of points $p$ 
such that $\operatorname{Im}\bigl(  u_{m,t}(p) \big)=\operatorname{Im}  \big( u_{m,t}(p_+)\big)$, and 
the shaded region satisfies $|{\rm e}^{ u_{m,t}(p)}|> |{\rm e}^{ u_{m,t}(p_+)}|$.
Thus, the steepest descent path for the hole momentum 
passes through $p_+$  horizontally and never enters into the shaded region.
We have an upside-down figure for the path of the particle momentum
which passes through $p_-$, as it  has the  conjugate phase ${\rm e}^{- u_{m,t}(p) t}$. 
As a result,  the optimal path for the hole (particle) momentum in the
space-like regime is similar to ${\cal E}$ ($\bar{\cal E}$): it stays
above (below) the real axis.

%
\subsubsection{The time-like regime $t>t_c$}

In this regime the saddle points $p_{\pm}$ are on the real axis,
\begin{equation}\label{eq:def_p_pm_time}
p_{\pm} = \frac{\pi}{2} \pm  \arccos \bigl( \frac{m}{4 J t} \bigr).
\end{equation}
They are depicted by red dots in the right panel of Figure~\ref{fig:steepest_space}.
The shaded region  indicates $|{\rm e}^{ u_{m,t}(p)}|>1$.
Thus, for the hole momentum, the steepest descent path runs through $p_+$ and $p_-$,
while staying inside of the un-shaded region.
Accordingly, it cannot stay above the real axis, but passes below the real axis
for  $p_-< \operatorname{Re} (p)< p_+$. The figure for the particle momentum is obtained by 
turning the figure for the hole momentum upside down.

\subsection{\boldmath The analyticity of $\mu$ and $\bar{\mu}$} \label{sec:analyticitymu}

We only have to know $\mu(p)$ ($\bar{\mu}(p)$) in (\ref{eq:measures}) 
as functions of $p$ in the upper (lower) half plane in the space-like regime.
As was already discussed, one needs to continue them analytically into
the whole complex plane in the time-like regime. The analytic properties
turn out to be quite different for the massless ($h<h_c$) phase and for the
massive ($h>h_c$) phase.

\subsubsection{The massive regime $h > h_c$}
We set
\begin{equation}\label{eq:defqj}
q_j =\arccos( \frac{h+i j \pi T}{4 J}) \qquad (j\in \mathbb{Z}),
\end{equation}
so that 
\begin{equation}\label{eq:def_qj}
1+{\rm e}^{-\epsilon( q_{2j-1}) /T}=0,
\qquad
1-{\rm e}^{- \epsilon( q_{2j})/T} =0.
\end{equation}
We further define ``upper roots" $q^u_{j}$ and ``lower" roots $q^d_{j}$ by
\[
q^u_{j} = 
\begin{cases}
 q_j&  j \le 0, \\
-q_j&  j \ge 1,
\end{cases}
\qquad
q^d_{j} = 
\begin{cases}
 -q_j&  j \le 0, \\
 q_j&  j \ge 1,
\end{cases}
\]
so that $\operatorname{Im} (q_j^u) >0$ and  $\operatorname{Im}(q_j^d) <0$.

The roots closest to the real axis are
$q^u_0= i \arch \bigl(\frac{h}{4J}\bigr) $ and $ q^d_0= -q^u_0$.

\begin{figure}[hbtp]
\centering
\setlength{\tabcolsep}{20pt}
\includegraphics[height=2.7cm]{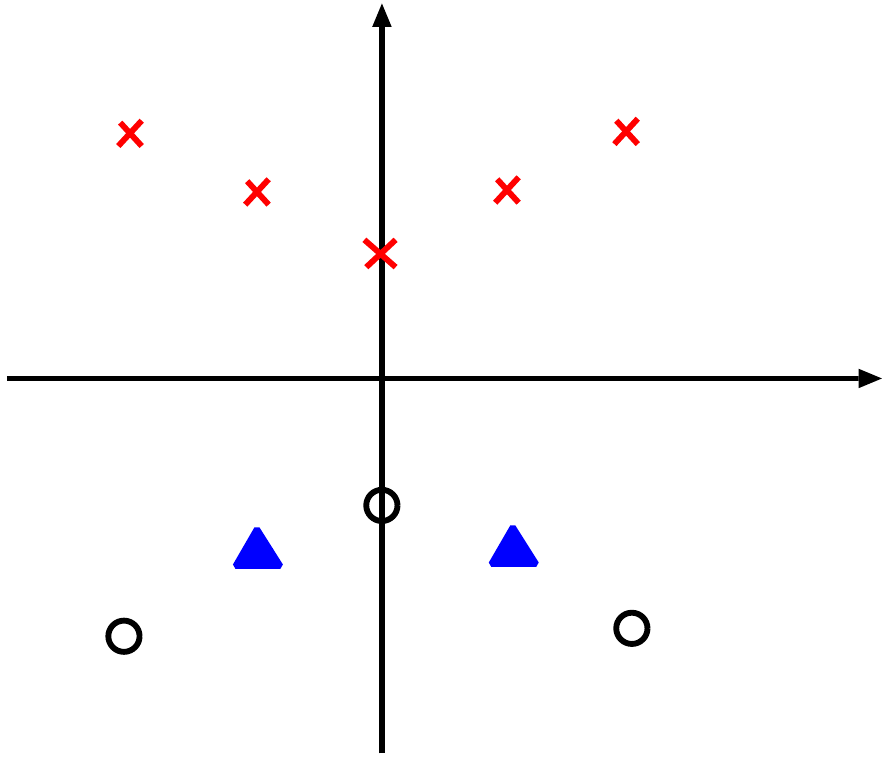}   \hspace{0.5cm}
\includegraphics[height=2.7cm]{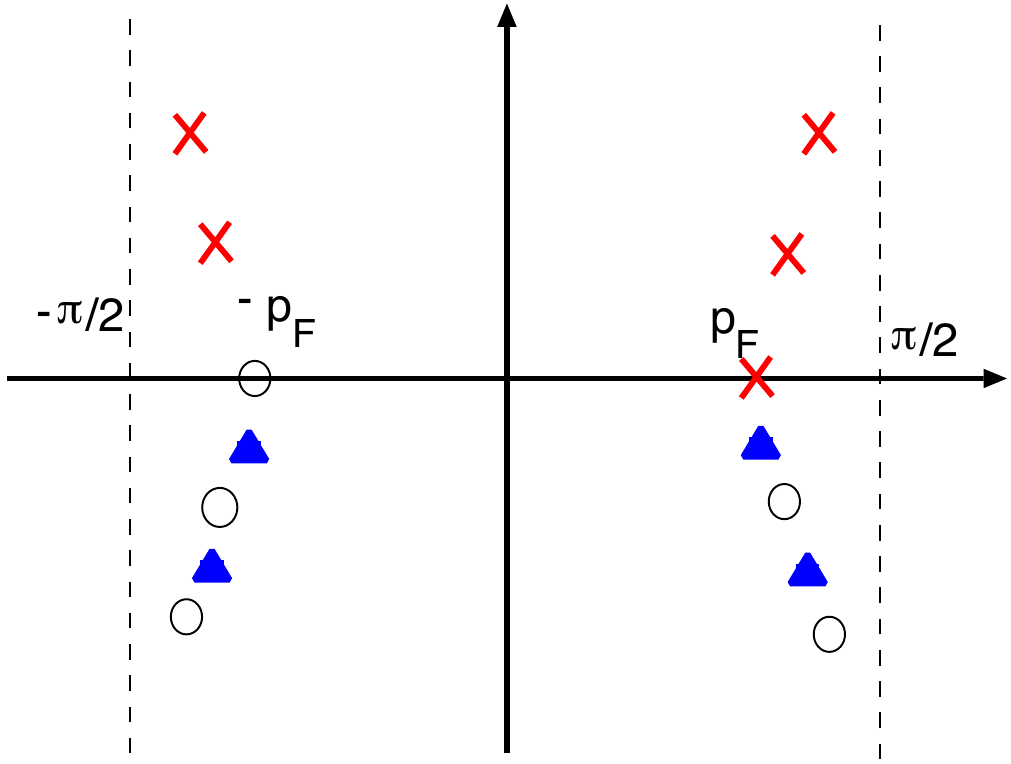}
\caption{A schematic picture of the analyticity of $\mu$ in the massive (left) and in the massless phase (right).
}
\label{fig:mu_zerospoles}
\end{figure}

The following analytic properties of  $\mu$  are derived in  \ref{app:derivation_mu}.
\begin{lemma}
The function $\mu(p)$  has \\
i) simple poles at   $q^u_j$ in the upper half plane, \\
ii) double poles at   $q^d_{2j+1}$ in the lower half plane, \\
iii)  simple zeros at  $q^d_{2j}$ in the lower half plane.\\
 There is a strip including the real axis which is free of zeros and poles
(of width  $2 \arch \bigl(\frac{h}{4J}\bigr) $).
\end{lemma} 
Figure~\ref{fig:mu_zerospoles} (left panel) illustrates the situation. 
The red crosses represent single poles, while blue triangles are double
poles. The circles denote the single zeros.\\
The zeros and poles for $\bar{\mu}(p)$ are obtained by taking the mirror
image with respect to the real axis.

Note that the width of the analytic strip is independent of the temperature in the massive phase.
Once a finite magnetic field is imposed, the width stays finite. Thanks to
this fact, the evaluation of the correlation function is simpler in the massive
regime.

\subsubsection{The massless regime $h<h_c$}

We denote by $\{q^r_j\}_{ j\in \mathbb{Z}}$, the (right) ``Bethe roots",
which are identical to the $q_j$ in eq.~(\ref{eq:defqj}).
Note that $p_F (= q^r_0 )\le \Re {\rm e} (q^r_j )<\frac{\pi}{2}$.  
The (left) ``Bethe roots" are defined by
$q^{\ell}_j  = -q^{r}_{-j}$. \\

In~\ref{app:derivation_mu}\hspace{5pt} we show that $\mu$ ($\bar{\mu}$)
has the following analytic properties.

\begin{lemma}
The function $\mu(p)$  has \\
i)  simple poles at   $q^r_{-2j}$ and at $ q^{\ell}_{-2j}$ ($j \ge 1$) in the upper half plane, \\
ii) a simple pole at $p=p_F$ and a simple zero at $p=-p_F$ on the real axis, \\
iii)   double poles at   $q^r_{2j+1}$ and at $ q^{\ell}_{2j+1}$ ($j \ge 0$) in the lower half plane, \\
iv) simple zeros at  $q^r_{2j}$ and at $ q^{\ell}_{2j}$ ($j \ge 1$). \\
The zeros and poles of $\bar{\mu}(p)$ are obtained by taking the mirror
image w.r.t.\ the origin.
\end{lemma} 
The massless case is illustrated in the right panel of Figure~\ref{fig:mu_zerospoles}.
The differences $ q^{\ell}_{-2j}-q^{\ell}_{-2j-2}$ and $ q^{\ell}_{2j+1}-q^{\ell}_{2j-1}$ become
${\cal O}(T)$ as $T \rightarrow 0$. Thus, singularities will approach towards the
real axis as $T \rightarrow 0$ and the choice of the contours becomes difficult.
This poses a technical problem in the massless regime. 
%
%
\section{\boldmath Numerical study of the massive regime $h>h_c$}\label{sec:massive_numerics}
In this and in the next sections we present the results of our study
of the transverse correlation function based on the numerical evaluation
of (\ref{eq:Fredholm_representation}). We follow the proposal in
\cite{Bornemann2010} and evaluate the Fredholm determinant by a
Gauss-Legendre integration with $n$ discrete points.

We shall start from the simpler case $h>h_c$. In this case the Fermi points
are away from the real axis. 
Thus, we can separately treat two technical problems, how to deal with  
the Fermi points and how to deal with the steepest descent path.

\subsection{\boldmath The space-like regime $t<t_c$}\label{sec:massive_space}
In spite of having discussed the steepest descent path above, we can use 
simple straight integration contours in order to produce accurate results,
as long as  $m \lesssim 10$ in this regime.
In order to demonstrate this, we first consider the static limit where
many results are available. 
The exact static short-range correlation functions at arbitrary $T$ and $h$
were obtained, e.g., in \cite{BDGKSW08} (for the XXZ model in general).
We compare them with the static results obtained from 
(\ref{eq:Fredholm_representation}) in \ref{appendix:static_finiteT}. 
The results match with reasonable precision.
We have to choose, of course, the appropriate number of discretized 
points $n$ to achieve agreement. For example, fixing $h=4.1 J,
T=0.1 J$ and $m=0$ we find
 \[
\langle \sigma_1^-(0) \sigma_1^+(0) \rangle =
 \begin{cases} 
 0.0186123688\cdots&    \text{exact},\\ 
 0.0186094\cdots &   n=64,  \\
  0.0186123689\cdots &   n=512.
  \end{cases}
  \]
Thus, we can say that with $n=512$, eq.~(\ref{eq:Fredholm_representation})
practically reproduces the exact value.
For larger segments, some indirect evidence is presented in \ref{appendix:static_zeroT}.
Encouraged by this success, we adopt the straight contours in the space-like
regime as they are conveniently simple for the calculation. Indeed, they produce
accurate results even beyond $t_c$.
 
\begin{figure}[!h]
\centering
\setlength{\tabcolsep}{20pt}
\includegraphics[height=\figheight]{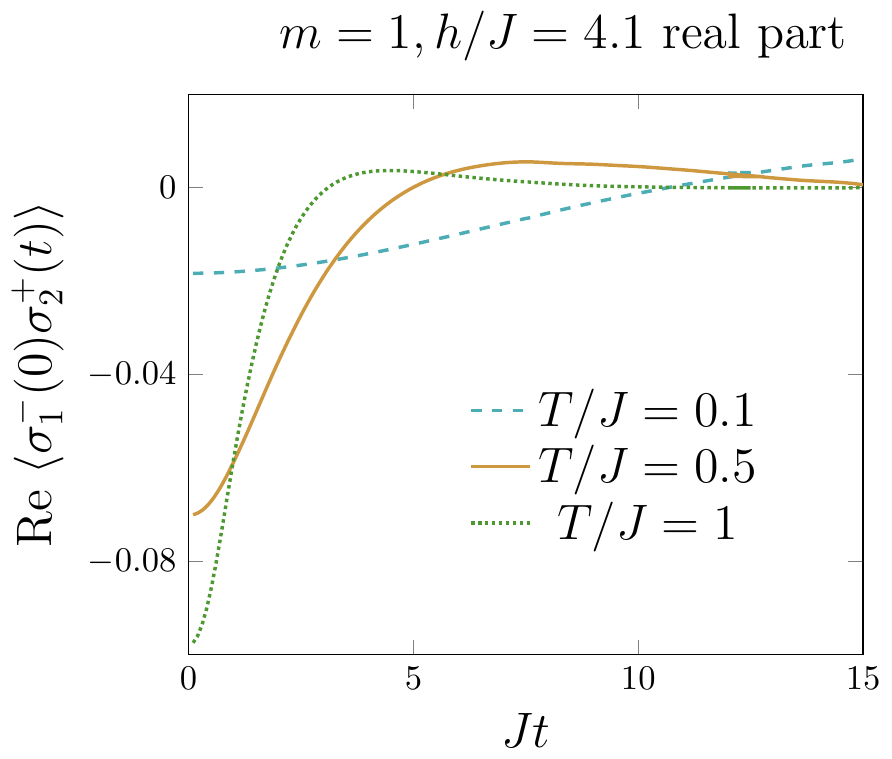}\hspace{-0.2cm} 
\includegraphics[height=\figheight]{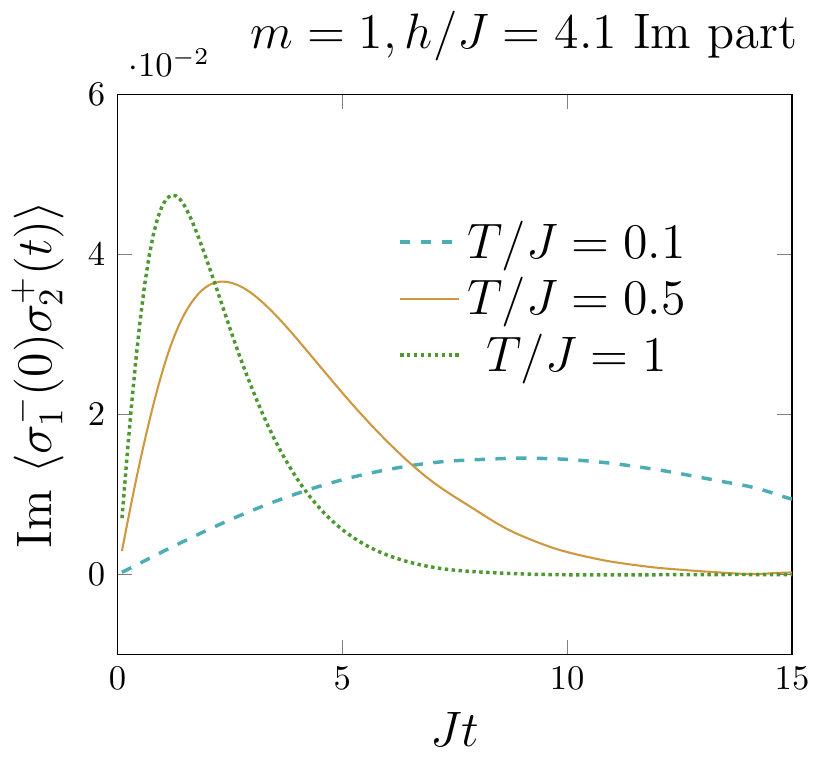}
\caption{
The real part (left) and the imaginary part (right) of 
$\langle \sigma^-_1(0) \sigma^+_2(t) \rangle  $
for $h=4.1 J$.  
}
\label{fig:m1h41nonsaddle}
\end{figure}

\begin{figure}[!h]
\centering
\setlength{\tabcolsep}{20pt}
\includegraphics[height=\figheight]{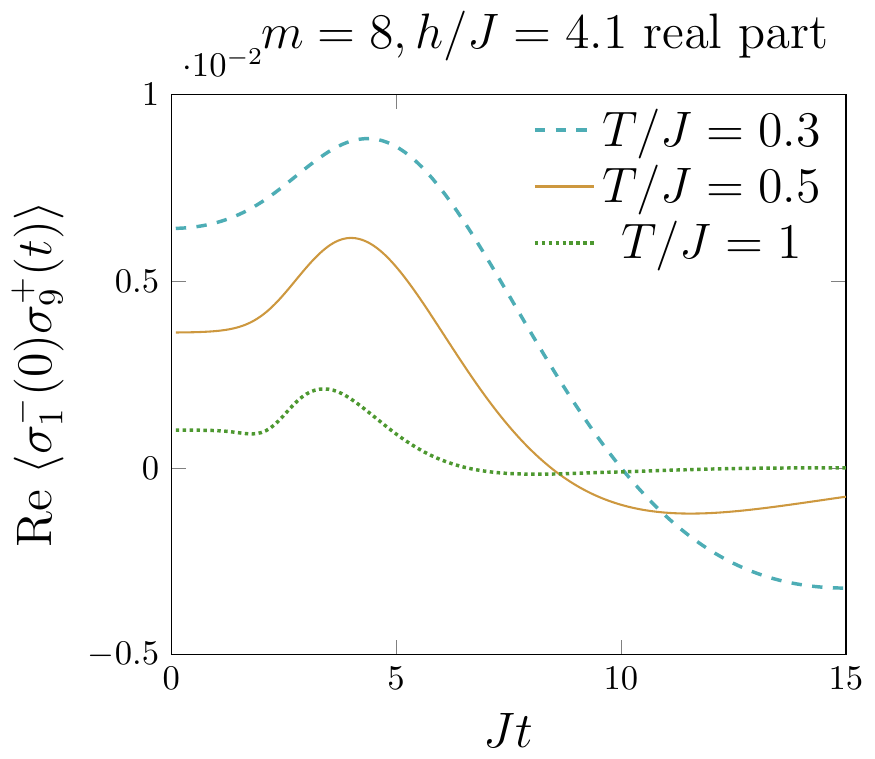} \hspace{-0.2cm} 
\includegraphics[height=\figheight]{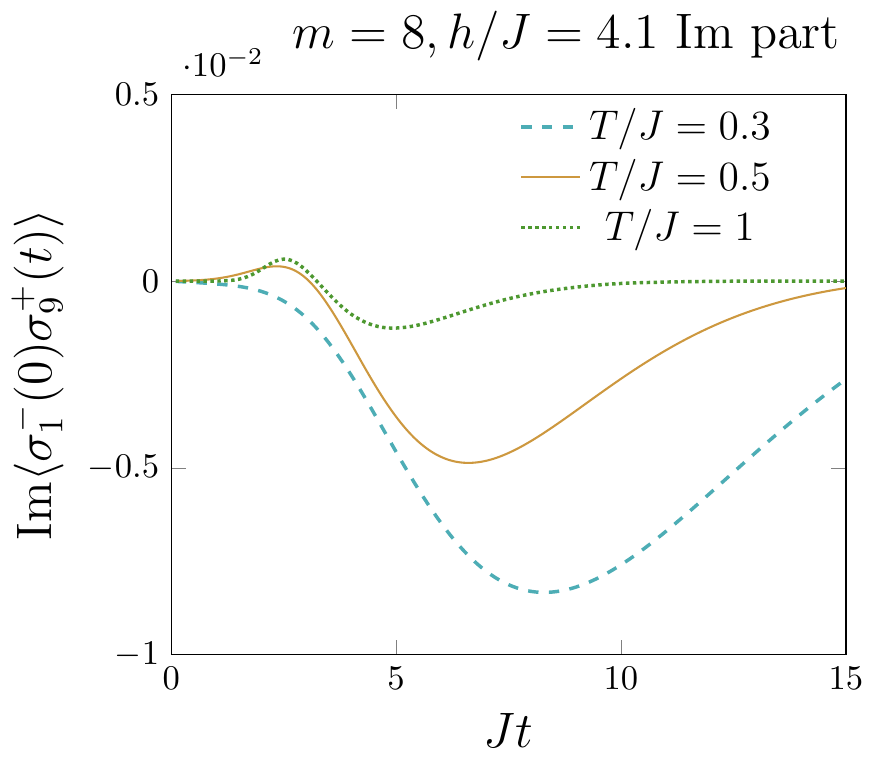}
\caption{
The real part (left) and the imaginary part (right) of
$\langle \sigma^-_1(0) \sigma^+_9(t) \rangle $
for $h=4.1 J$.
}
\label{fig:m8h41nonsaddle}
\end{figure}

Figures~\ref{fig:m1h41nonsaddle} and \ref{fig:m8h41nonsaddle} show examples
for $h=4.1 J$, $m=1$ and $m=8$ for various $T$.
The horizontal axes indicate $J t$.
Although the space-like regime is limited to $J t<0.25$ when $m=1$ and  $J t<2$ when $m=8$,
the straight contours produce stable values even after $t_c$.

The real part stays almost flat in the space-like regime. This can be better
seen in the case of $m=8$ in Figure \ref{fig:m8h41nonsaddle}. The amplitude
is enhanced around $t=t_c$.

For larger values of $m$ (typically $m \sim 100$), we need to adopt
contours which take into account the steepest descent paths. The
saddle points in eq.~(\ref{eq:saddlepointInspace}) are located away
from the real axis, and we shift the contour ${\cal E}$ ( $\bar{{\cal
E}}$ ) to a straight line passing though $p_+$ ($p_-$).
The integration contours cross poles of $\mu(p)$ and $\bar{\mu}(q)$ if
$\frac{h}{4} < \frac{m}{4t }$, which is clearly seen in 
Figure~\ref{fig:mu_zerospoles} (left).
This modifies our formula  (\ref{eq:Fredholm_representation}) slightly,
following the argument in \cite{DugaveGohmannKozlowskiSuzuki2016}, 
which will be summarized in \ref{appendix:det_poles_formula} for
the reader's convenience.

\subsection{\boldmath The time-like regime $t>t_c$}
\begin{figure}[hbtp]
\centering
\includegraphics[width=5.3cm]{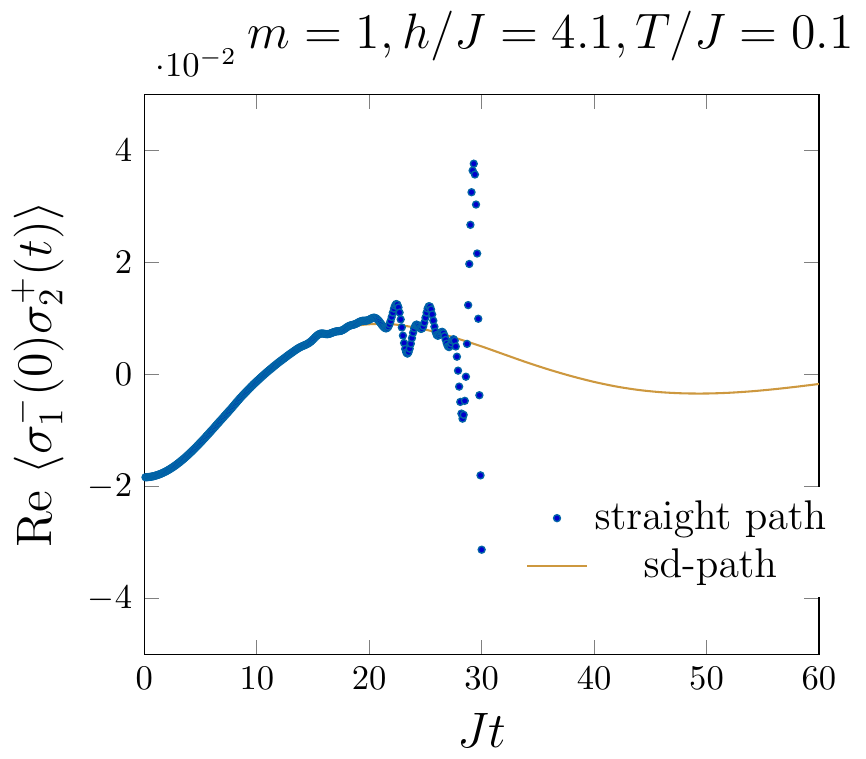}
\caption{
The real part of   $\langle \sigma^-_1(0) \sigma^+_2(t) \rangle $
for $T=0.1 J, h=4.1 J$. The dots are obtained using the straight
integration contours, and the line is obtained by using the steepest descent (sd) path.
}
\label{fig:massivenonsaddle}
\end{figure}

The calculation with the straight integration contours gradually becomes
unreliable  for larger $J t$. An example is given in Figure~\ref{fig:massivenonsaddle}.

\begin{figure}[hbtp]
\centering
\includegraphics[width=3.5cm]{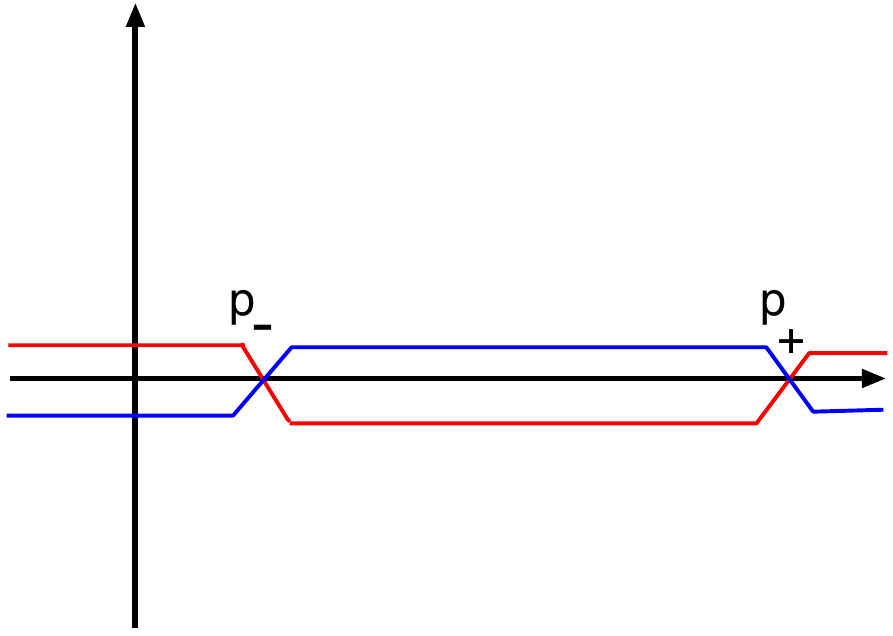}
\caption{
Plausible paths for the hole momentum (red) and for the particle momentum (blue).
}
\label{fig:steepestdescendent}
\end{figure}

We thus deform the contours in the time-like regime.  
The most naive choices for the
hole momentum ($p$) and for the  particle momentum $(q) $ may be the ones depicted in 
Figure  \ref{fig:steepestdescendent}, 
which respect the steepest descent paths discussed in
Section~\ref{sec:steepest_descendent_path}.

\begin{figure}[!h]
\centering
%
\setlength{\tabcolsep}{20pt}
\includegraphics[height=\figheight]{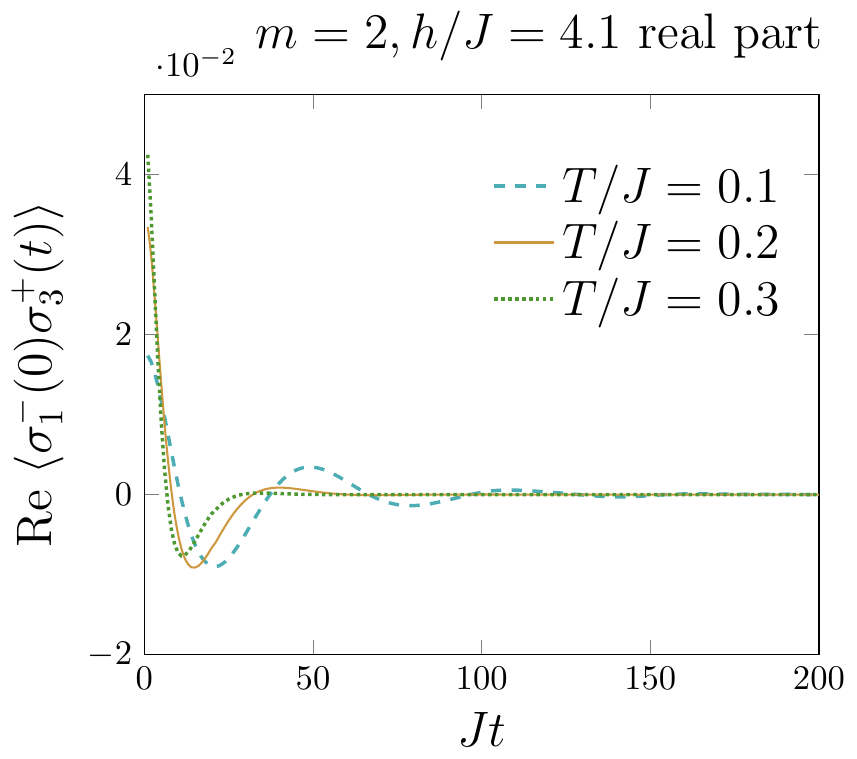} \hspace{-0.1cm} 
\includegraphics[height=\figheight]{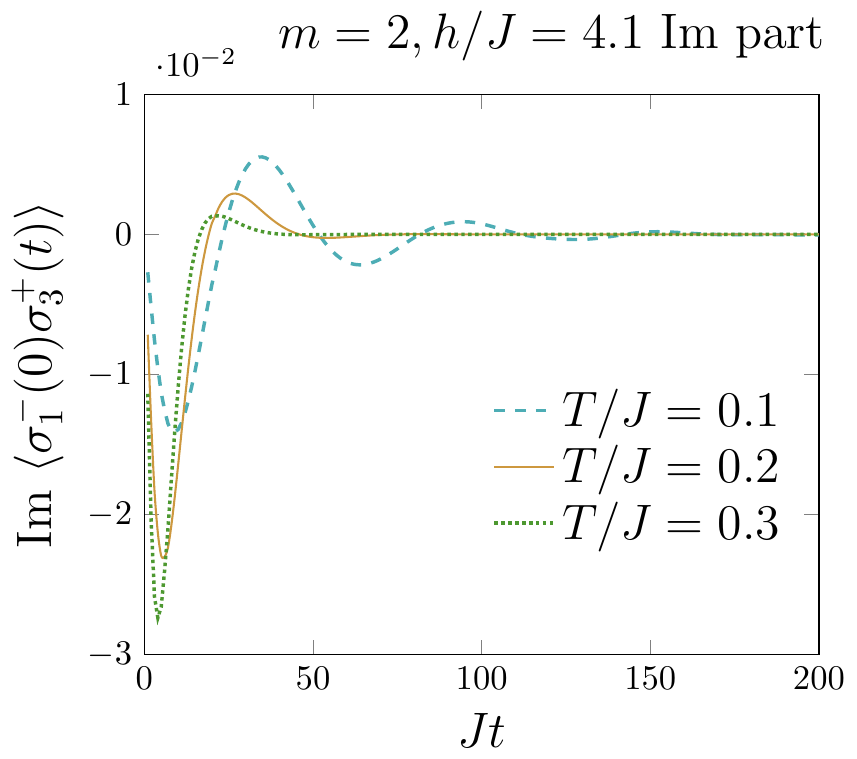}
%
\caption{
The real part (left) and the imaginary part (right) of 
$\langle \sigma^-_1(0) \sigma^+_3(t) \rangle$ for $h=4.1J$
in the time-like regime ($J t>0.5$).
}
\label{fig:m2h41}
\end{figure}

The paths are determined only by the phases $\re^{t u_{m,t}(p)}$ and
$\re^{-t u_{m,t}(q)}$. One, however, needs to be careful: Since $q$
and $p$ are swapped, there appears an additional term in $\tilde{v}$,
\[
\tilde{v}(q)= \int_{{\cal C}_p}  dp\, \mu(p) {\rm e}^{2t u_{m,t}(p)} \varphi(p,q)  +
4\pi i    {\rm e}^{2t u_{m,t}(q)} \mu(q),
\]
due to a pole of $\varphi(p,q)$ at $p=q$. Here ${\cal C}_p$ denotes the red path in 
Figure~\ref{fig:steepestdescendent}.
As we interpret $p$ ( respectively  $q$) as the momentum of a hole ( respectively a particle),
we refer to the singularity as the kinematic pole in analogy with scattering theory. 
Consequently, the kernel $K(q_i,q_j)$ of the integral operator
$\hat K = \hat P - \hat V$ contains a term 
\[
-\frac{16 \pi^2}{\Omega(m,t)}  {\rm e}^{t u_{m,t}(q_i)+t u_{m,t}(q_j) } \mu(q_i)  \mu(q_j).
\]
This becomes very large for $J t \gg 1$ if $\operatorname{Im} (q) \sim {\cal O}(1)$
and makes the calculation again unstable. We thus choose the red contour in
Figure~\ref{fig:steepestdescendent} for the $p$ variables while we adopt a
straight contour for the $q$ variable such that 
$\operatorname{Im} (q) \sim 0^-$ and $ \operatorname{Im} (p)< \operatorname{Im} (q)$ for $p_-<\operatorname{Re} (q) <p_+$. \\
Still, there is a problem in dealing with the intersection points $p_+, p_-$.
We simply exclude them from the set of discretized sampling points
in the integrals over the $q$ variable.\\
After the above modification and by increasing the number of
sampling points for the Fredholm determinant ($n=100 - 512$, typically),
a large scale calculation is possible. As an illustration, the plots for
$m=2, h=4.1 J$ are shown in Figure \ref{fig:m2h41}.

\begin{figure}[!h]
\centering
\setlength{\tabcolsep}{20pt}
\includegraphics[height=\figheight]{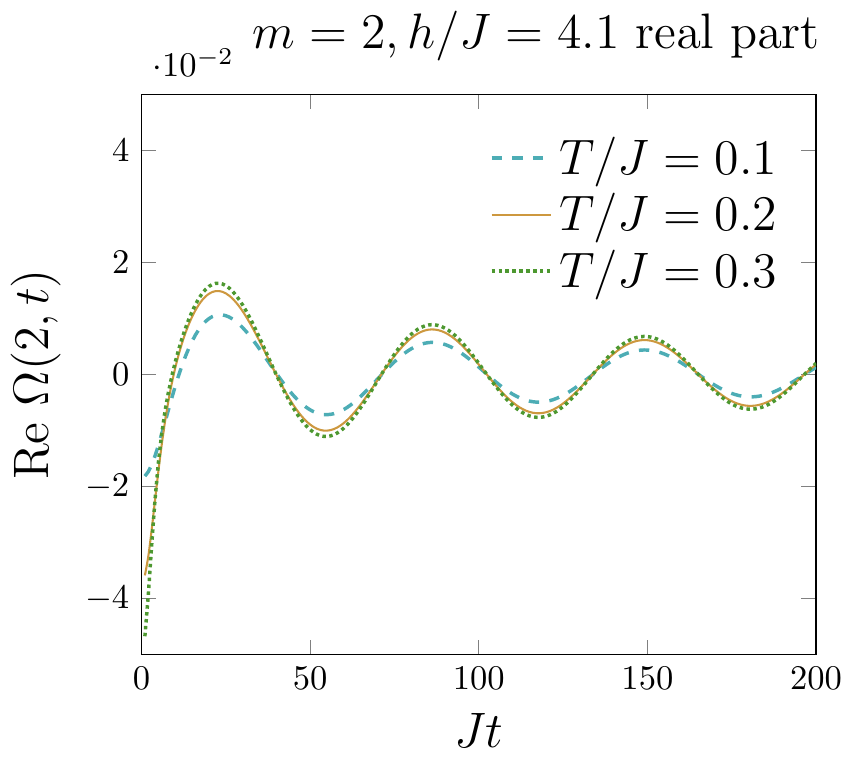} \hspace{-0.1cm} 
\includegraphics[height=\figheight]{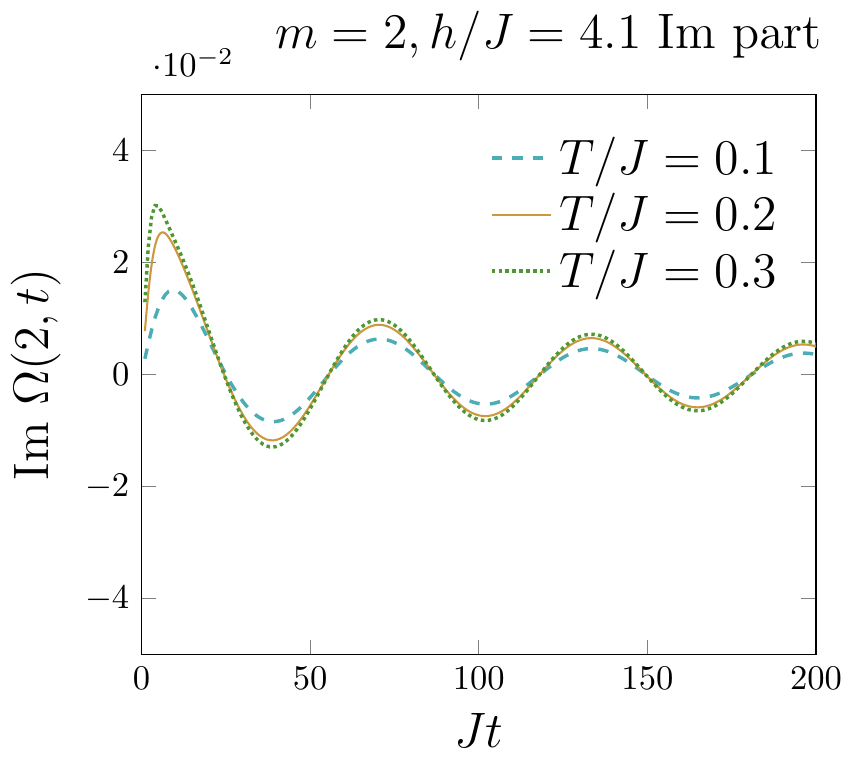}
\caption{
The real part (left) and the imaginary part (right) of $\Omega(m,t)$ for
$m=2, h=4.1 J$. The oscillation periods for $T=0.1,\,0.2,\, 0.3 J$ seem almost identical.
}
\label{fig:m2h41omega}
\end{figure}

\begin{figure}[!h]
\centering
\setlength{\tabcolsep}{20pt}
\includegraphics[height=\figheight]{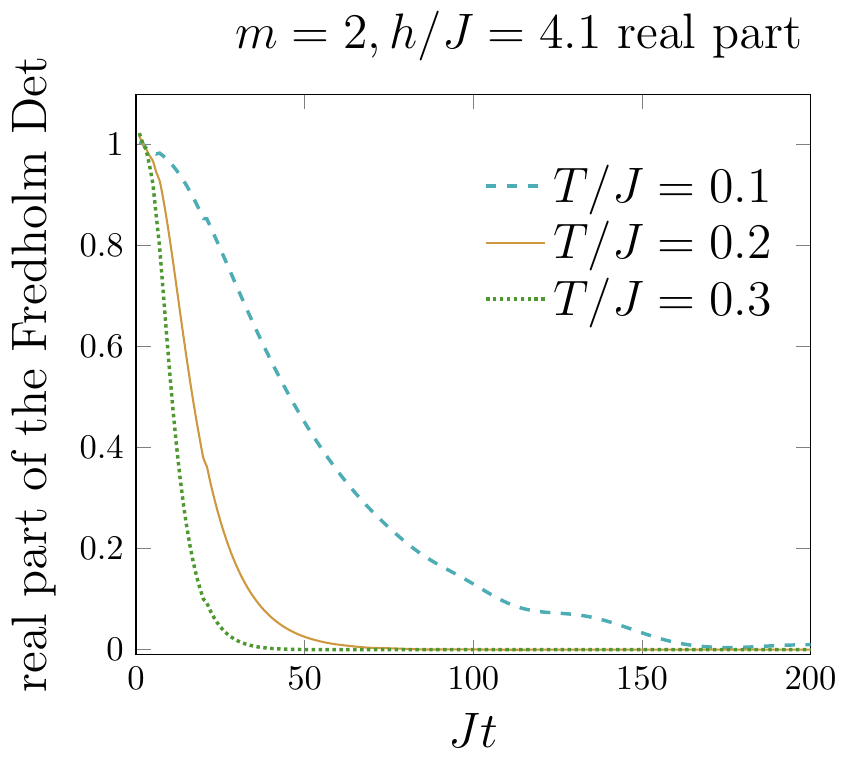}\hspace{-0.1cm} 
\includegraphics[height=\figheight]{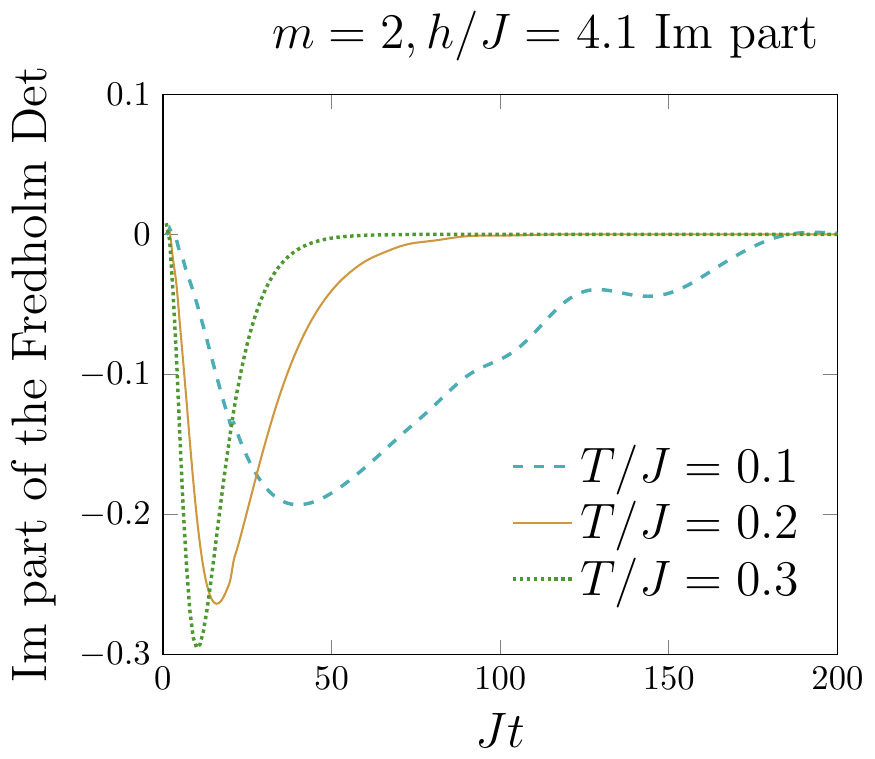}
\caption{
Time evolution of the Fredholm determinants corresponding to Figure \ref{fig:m2h41},
real part left and imaginary part right.
}
\label{fig:m2h41Fredholm}
\end{figure}

One immediately notices the {\it very long period} oscillation.
This may be attributed to the oscillatory behavior of $\Omega(m,t)$
of which  the period seems to depend only weakly on temperature
(Figure \ref{fig:m2h41omega}).
The amplitude of $\Omega(m,t)$ decreases slowly in time and the decay of
the amplitude of the correlation functions mainly comes from that of the
Fredholm determinant, \textit{c.f.} Figure \ref{fig:m2h41Fredholm}, especially for higher $T$.

\begin{figure}[!h]
\centering
\setlength{\tabcolsep}{20pt}
\includegraphics[height=\figheight]{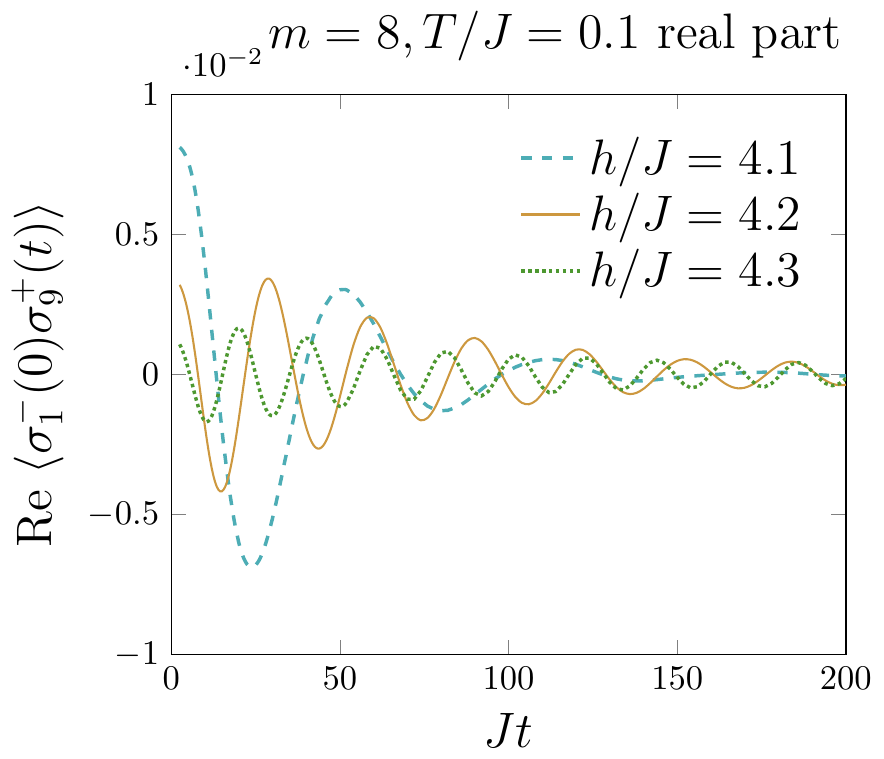} \hspace{-0.2cm} 
\includegraphics[height=\figheight]{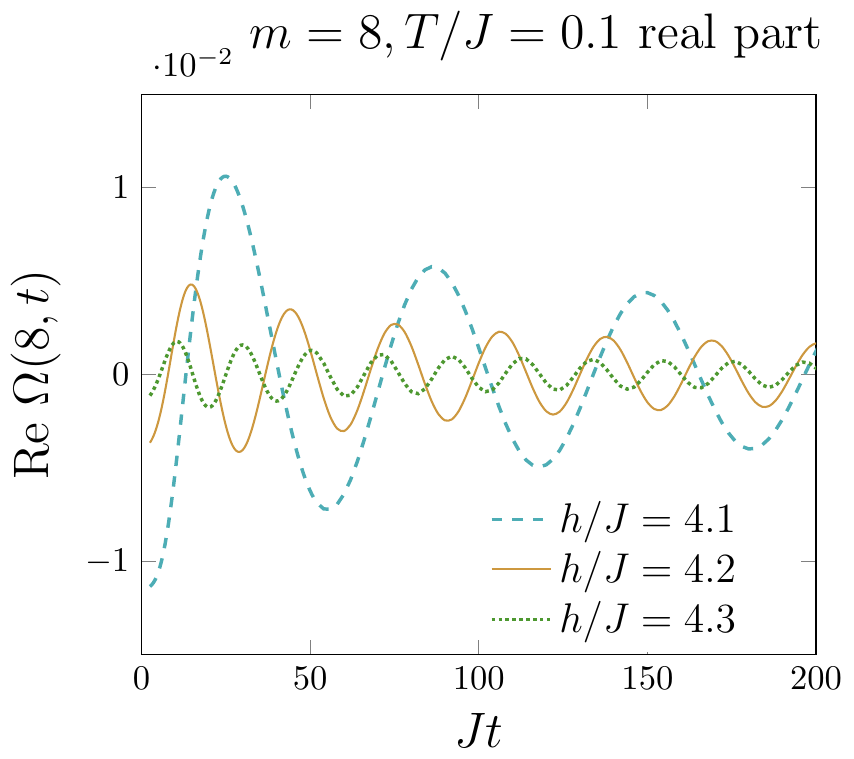}
\caption{
Different periods of oscillation for $h/J=4.1,4.2,4.3$, $T/J=0.1$ and $m=8$.
Left panel shows the real part of the correlation functions, right panel
shows the real part of the corresponding function $\Omega(m,t)$.
}
\label{fig:m8T01massiveoscillation}
\end{figure}

Unlike on temperature, the period of oscillation obviously does depend on the
magnetic field, and it grows as $h$ approaches $h_c$ (Figure~\ref{fig:m8T01massiveoscillation}).
It may be easier to check this by looking at the imaginary part of $\Omega(m,t)$
(cf.\ right panel).

The oscillatory period of $\Omega(m,t)$ is easily understood in the asymptotic
region $t\gg t_c$ and  $T \searrow 0$. A saddle point analysis  yields
\begin{align}\label{eq:estimate_Omega}
\Omega(m,t) &\sim   \sqrt{ \frac{\pi}{2 \left |u''_{m,t}(p_+) t \right| } } \mu(p_+) \re^{i(mp_+ -\epsilon(p_+) t) } \nonumber \\
&+ \sqrt{ \frac{\pi}{2 \left |u''_{m,t}(p_-) t  \right| } } \mu(p_-) \re^{i(mp_- -\epsilon(p_-) t)}.
\end{align}
Note that  $t\gg t_c$,  $p_- \sim 0$ and $p_+ \sim \pi$. Due to the asymmetry 
of the integrand $q \rightarrow -q$  in (\ref{eq:measures}) for
$p=0$ or $p=\pi$, one can show that $\sigma_+(p_{\pm}) \sim 0$ so that
$\re^{\sigma_+(p_{\pm})} \sim 1$. On the other hand $\epsilon(p_-)/T 
\sim( h-4 J )/T \ll \epsilon(p_+)/T$. Thus,
\[
\left|\mu(p_-) \right|=\left|\frac{\re^{\sigma_+(p_-)}}{2\pi(1- \re^{\epsilon(p_-)/T})} \right| \gg   \left|\mu(p_+) \right|
\]
and we can safely drop the first term in (\ref {eq:estimate_Omega}).
Thus, we are left with a single oscillating term and conclude that 
the period of the oscillation diverges as $ C (h_c -h)^{-1}$, for some constant $C$, when  $h \rightarrow h_c$.
Note that $h = h_c$ is an exceptional point as
the Fermi points pinch the real axis and the above argument is not valid. 

%
%
\section{\boldmath Numerical study in the massless regime $h<h_c$}\label{sec:massless_numerics}

The existence of  Fermi points on the real axis makes the evaluation technically 
more involved than in the massive case. This is due to the fact that
the Fermi-point singularity problem and the steepest-descent path problem are
coupled. The situation is slightly simpler in the  space-like regime from which
we start our consideration.

\subsection{\boldmath The space-like regime $t<t_c$}

We again transform ${\cal E} (\bar{\cal E})$ to a straight contour in
the upper (lower) half plane for the hole (particle) momentum if $m$
is not too big.  By this deformation, in contrast to the massive case,
we inevitably pick up the contribution from the Fermi points. This
modification can be treated in a similar manner as the contributions
of the other poles, shortly commented on at the end of
Section~\ref{sec:massive_space} on the massive phase.  The details are
explained in \ref{appendix:det_poles_formula}.  We choose a
straight contour $[-\pi+i\delta, \pi+i\delta]$ for $p$.
Although the choice of $\delta>0$ is in theory arbitrary, in practice
one should not take it too small. Table~\ref{tb:m1staticdelta}
illustrates this with the examples of $m=1$, $t=0$ and $h=0.5 J$ for
various choice of $\delta$. The rightmost column gives the exactly
known static values.

\begin{table}[htb]\label{tb:m1staticdelta}
  \begin{center}
    \begin{tabular}{|c|c|c|c||c|} \hline
       T&      $\delta=0.01$&    $\delta=0.05$&  $\delta=0.1$&  exact  \\ \hline 
      0.1 &     -0.34556050  &       -0.31549848&   -0.31547691&  -0.31547744 \\
      0.5 &     -0.32068971 &        -0.30676642&    -0.30675704&  -0.30675704  \\
     1 &         -0.28351570 &        -0.27640356 &   -0.27639925 & -0.27639925  \\
      \hline
    \end{tabular}
    \caption{The static correlation  function
    $\langle \sigma^-_1(0) \sigma^+_2(0) \rangle $ 
    for $h/J=0.5$ and various temperatures.
    The Fredholm determinant is evaluated by means of an approximation by a 256$\times$ 256 matrix.}
  \end{center}
\end{table}

This  can be easily understood as the contour passes  near the singularities of 
$\mu$ (at $p_F$) or $\bar{\mu}$ (at $-p_F$). Hence, too small values of $\delta$ spoil
the numerical accuracy.

\begin{figure}[!h]
\centering
\setlength{\tabcolsep}{20pt}
\includegraphics[height=3.4cm]{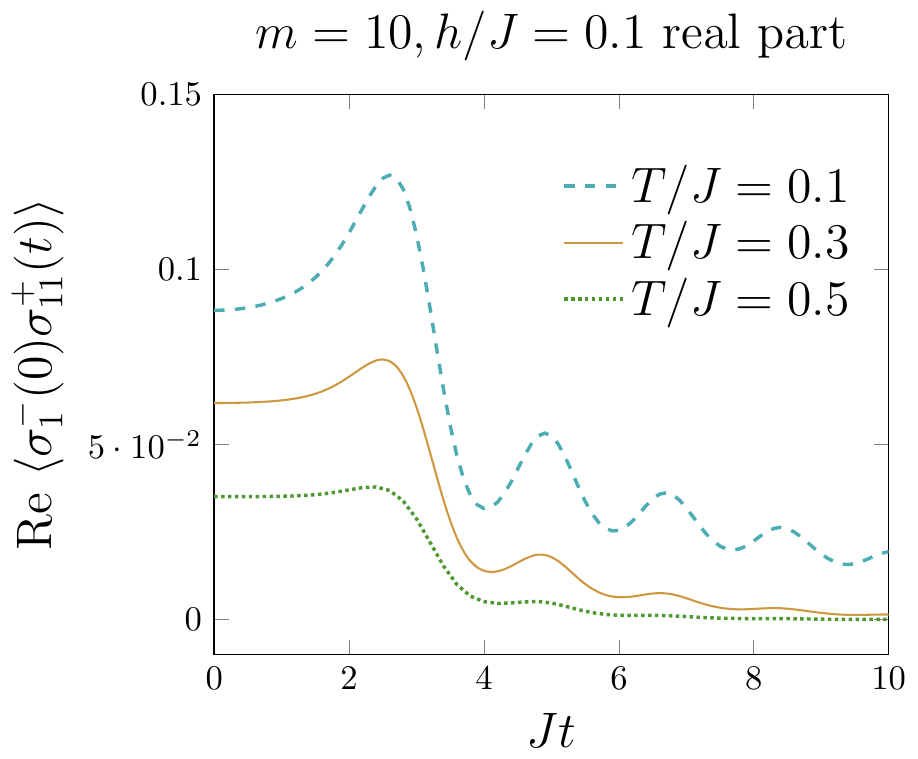} \hspace{-0.1cm} 
\includegraphics[height=3.4cm]{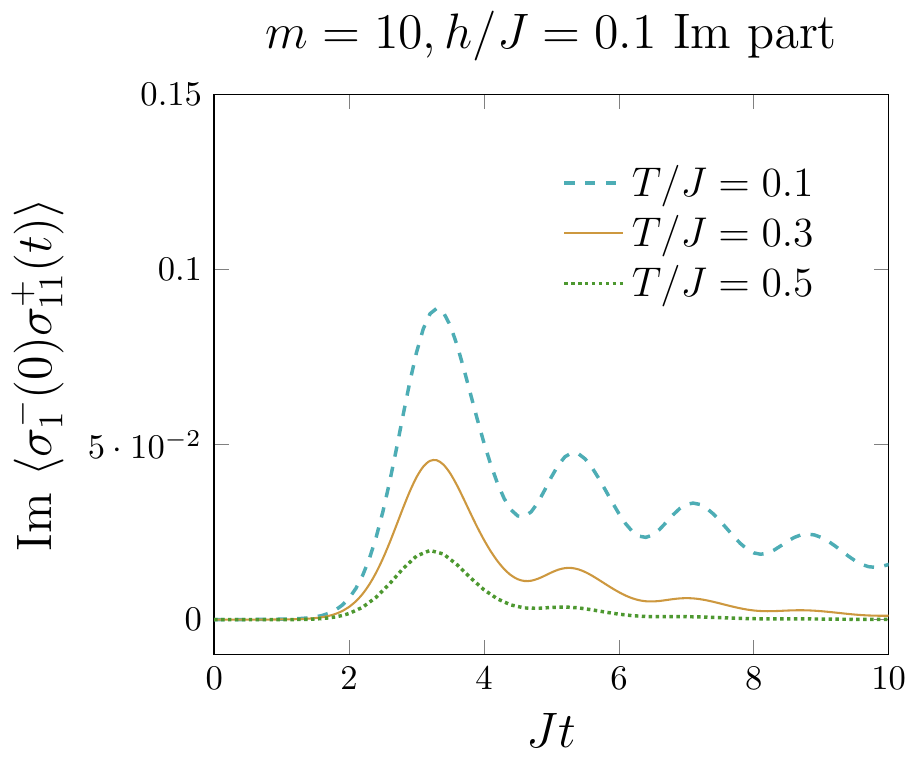}
\caption{
The  right (left) figure depicts the real (imaginary) part of 
$\langle \sigma^-_1(0) \sigma^+_{11}(t) \rangle$ for $h=0.1 J$ and $T=0.1,0.2,0.3 J$, 
evaluated with straight line contours.
}
\label{fig:m10h01masslessnonsaddle}
\end{figure}

\begin{figure}[!h]
\centering
  \includegraphics[height=\figheight]{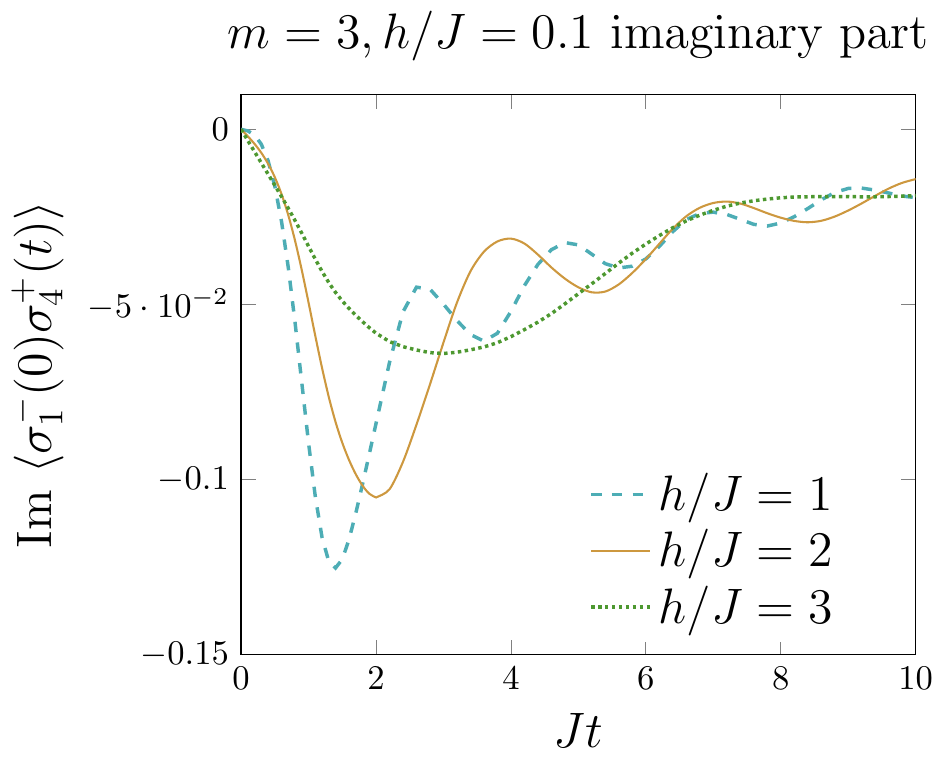}
\caption{
Imaginary part of  
$\langle \sigma^-_1(0) \sigma^+_4(t) \rangle$ 
in the massless regime for $T/J=0.1$ and several values of $h/J$,
evaluated using the straight line contours.
}
\label{fig:m3T01_massless_oscillation}
\end{figure}

With suitable choices of $\delta$ we can again adopt the straight line contours 
well beyond the space-like regime, see Figure~\ref{fig:m10h01masslessnonsaddle}.
The oscillation frequency in the time-like regime $t>t_c$ for small temperatures is given by the effective bandwidth 
in the Fermionic model, $4J-h$, and goes to zero as $h$ approaches $h_c=4J$ as demonstrated in Figure~\ref{fig:m3T01_massless_oscillation}.

For larger values of $m$, $\delta$ is determined by the location of the
saddle point (\ref{eq:saddlepointInspace}) and one has to the take into account
the pole contributions as described in \ref{appendix:det_poles_formula}.
Some numerical results will be supplemented later (cf.\ Section~\ref{sec:other_method}).

%
%
\subsection{\boldmath The time-like regime $t>t_c$}
\begin{figure}[!h]
\centering
\includegraphics[height=\figheight]{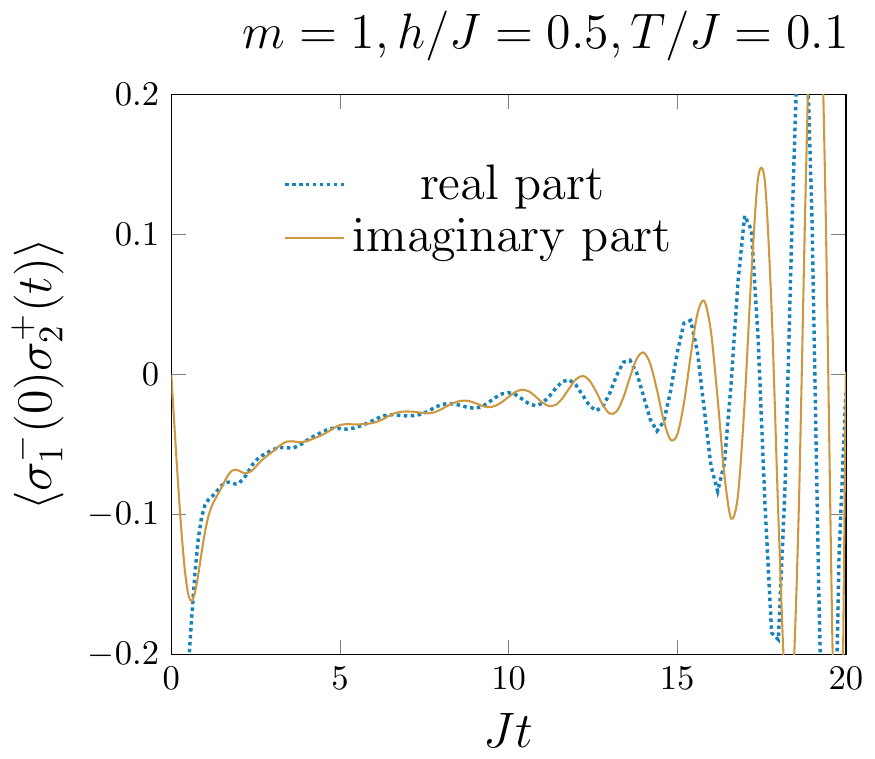}
\caption{
The real and the imaginary part of $\langle \sigma^-_1(0) \sigma^+_2(t) \rangle$
(using the straight line contours) for $h/J=0.5 $ and $T/J=0.1$. The numerical results become unstable for $Jt\gtrsim 10$.
}
\label{fig:m1h05masslessnonsaddle}
\end{figure}

As in the massive phase, the straight integration contours fail to
give reliable results as time
evolves. Figure~\ref{fig:m1h05masslessnonsaddle} shows an example.

\begin{figure}[!h]
\centering
\setlength{\tabcolsep}{20pt}
\includegraphics[height=3.5cm]{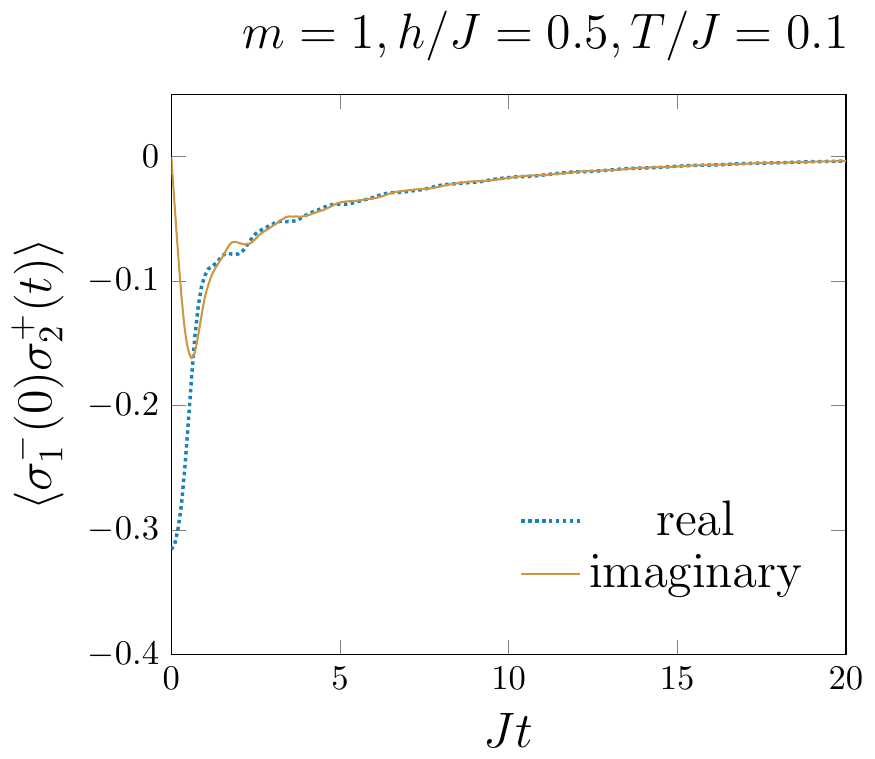} \hspace{0.05cm}
\includegraphics[height=3.5cm]{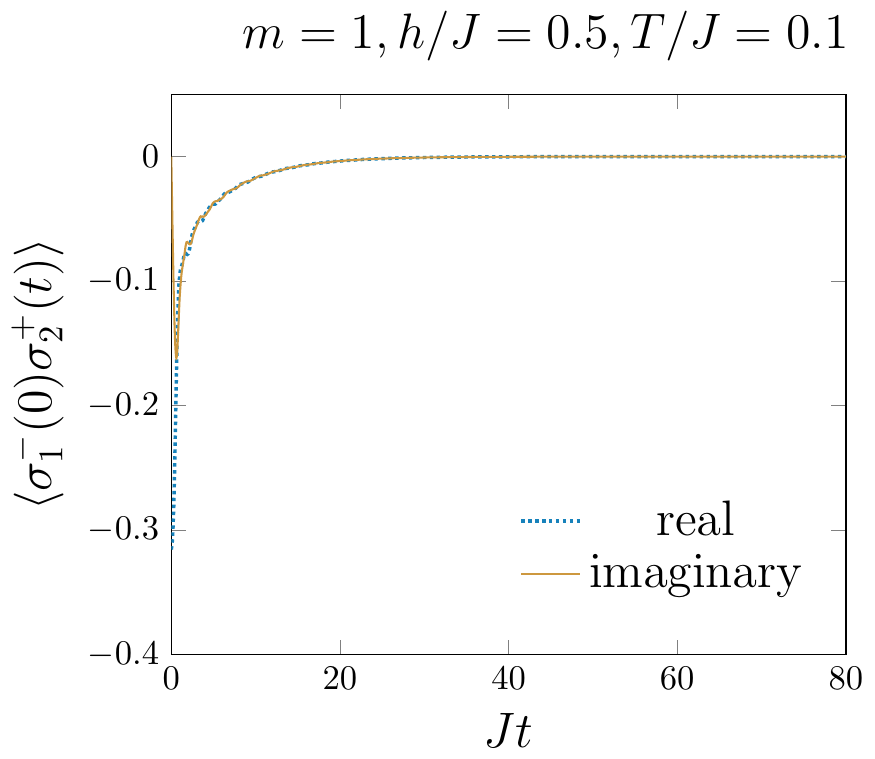}

\caption{
The real and the imaginary part of $\langle \sigma^-_1(0) \sigma^+_2(t) \rangle$ 
(using the steepest descent paths) for $h/J=0.5 $ and $T/J=0.1$.
The right panel shows the same correlation function on a longer time scale.
The Fredholm determinant is evaluated from a 1024 $\times$ 1024 matrix.
}
\label{fig:m1h05massless_saddle}
\end{figure}

We thus utilize the steepest descent paths. 
The existence of Fermi points on the real axis makes the situation
more involved than in the massive case. At the same time, one must pay
attention to the singularities of $\mu(p), \bar{\mu}(q)$. The choice
of the paths thus becomes a complicated technical issue, and we
decided to summarize it in \ref{app:choice_contour}. \\
By applying the prescription presented there, the result are now
stable as shown in Figure~\ref{fig:m1h05massless_saddle}, where we
have chosen the same parameters as in
Figure~\ref{fig:m1h05masslessnonsaddle}. The right panel represents
the result up to $Jt=80$, and the real part of $\langle \sigma^-_1(0)
\sigma^+_2(t) \rangle$
decays to the order of $10^{-7}$.

\begin{figure}[!h]
\centering
\setlength{\tabcolsep}{20pt}
\includegraphics[height=3.4cm]{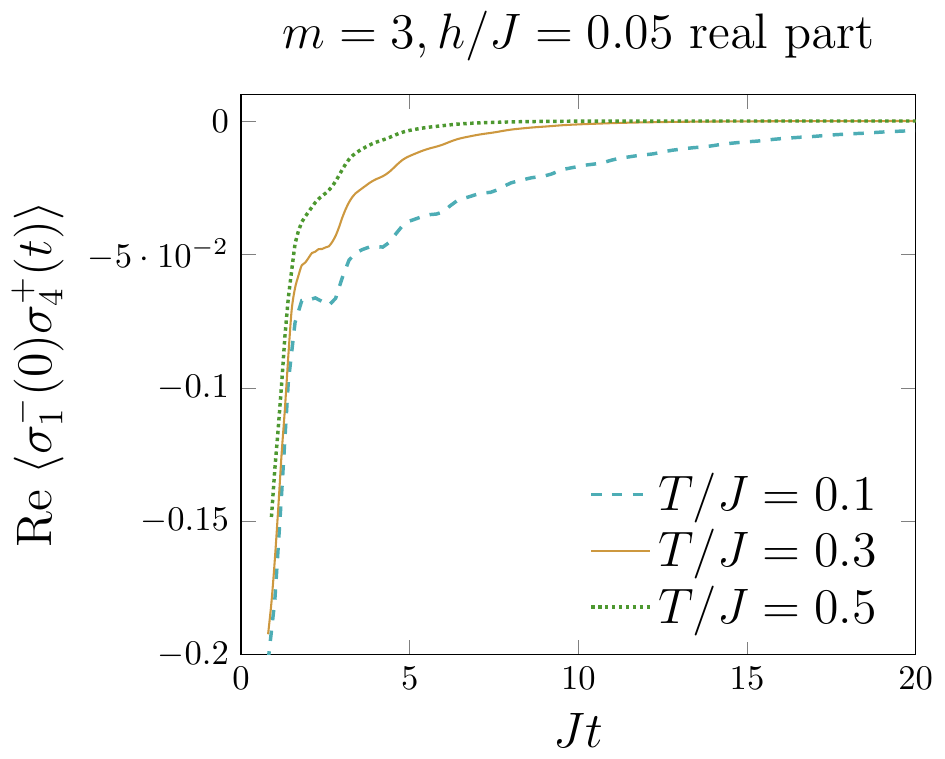} \hspace{-0.1cm}
\includegraphics[height=3.4cm]{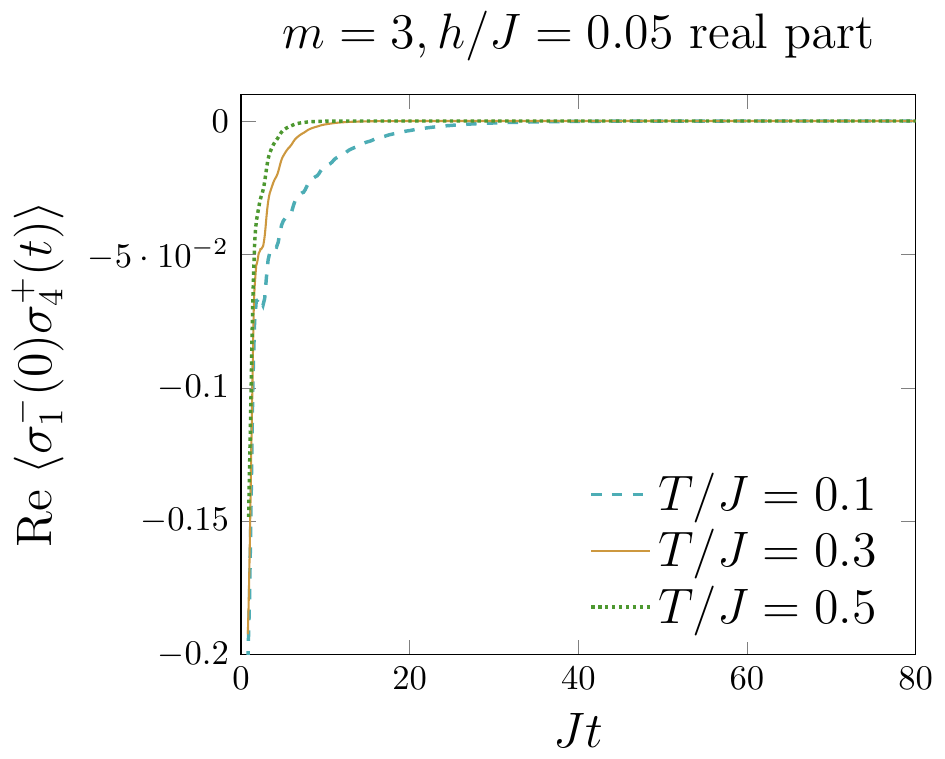}
%
\caption{
The real part of  $\langle \sigma^-_1(0) \sigma^+_4(t) \rangle$, evaluated from the
``steepest descent" paths, for $h/J=0.05$ and various $T$, up to $J t=20$
(left) and up to $J t=80$ (right).
}
\label{fig:m3h05_steepest}
\end{figure}

Examples with slightly different parameters are given in
Figure~\ref{fig:m3h05_steepest}, showing again smooth behavior until they
exhibit sufficient decay.

\begin{figure}[!h]
\centering
{\includegraphics[height=\figheight]{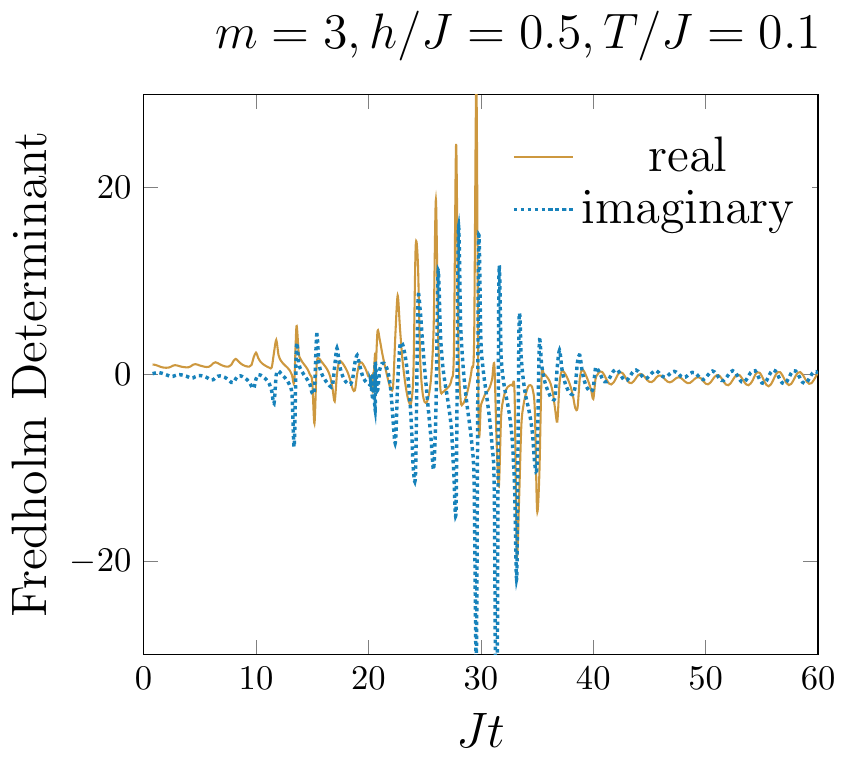}
}
\caption{
The real and  the imaginary part of the Fredholm determinant corresponding
to  $\langle \sigma^-_1(0) \sigma^+_4(t) \rangle$
for $h=0.5 J, T=0.1 J$ with $n=256$ discretization points.
There is a clear enhancement of fluctuations around $20<J t<40$.
}
\label{fig:m3h05T01Fredholm}
\end{figure}

The fluctuation is enhanced at lower temperatures, where one has to be
particularly careful. Independent calculations also suggest this.  For
example, the authors of \cite{DKS2000} calculated $\langle
\sigma^x_{i}(0) \sigma^x_{i+50}(t) \rangle$ in the finite XX-chain
(400 sites) using the Pfaffian representation. Their result does not
converge numerically for various choices of $i$ around $Jt=30 \sim 40$
for $T/J=0.04$ (in the present normalization). The authors claim that
this is due to a ``boundary effect".  Our results using the
Pfaffian representation of the correlation function, which we present
in Section \ref{sec:other_method}, do not show such
instabilities. This suggests that the observed instabilities are
numerical in nature (a critical step is a numerically stable direct
evaluation of the Pfaffian) and not related to boundary effects.
In the framework of our QTM approach we are dealing with an infinite
size system from the beginning so that boundary effects are certainly absent. 
Nevertheless, a similar instability is observed at lower temperatures.
We examined each factor in (\ref{eq:Fredholm_representation}) and
found that the instability mainly comes from the Fredholm determinant
(Figure~\ref{fig:m3h05T01Fredholm}).

\begin{figure}[!h]
\centering
\setlength{\tabcolsep}{20pt}
\includegraphics[height=\figheight]{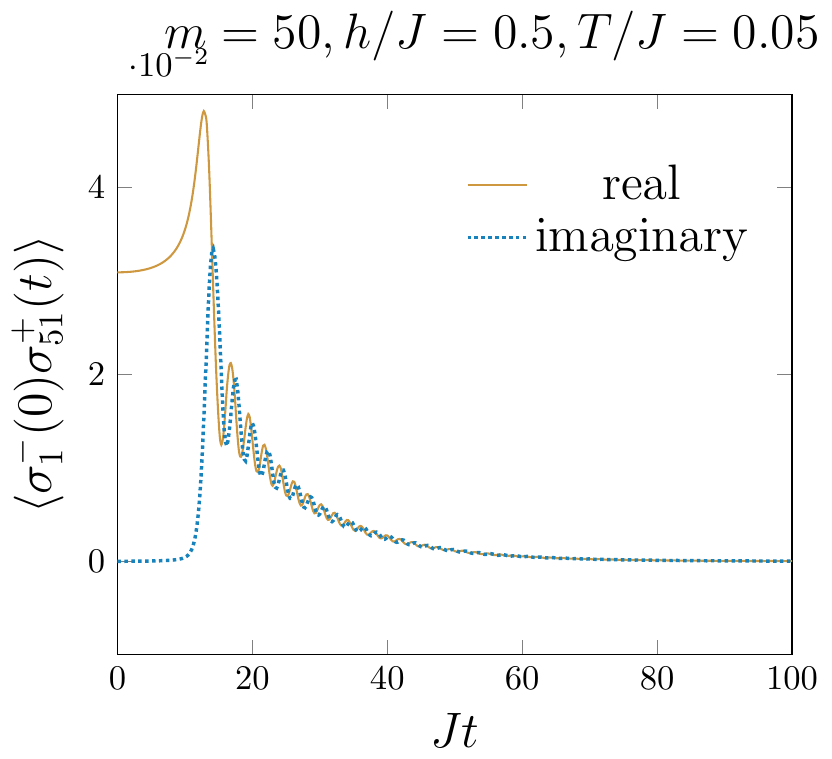} \hspace{0.1cm}
\includegraphics[height=\figheight]{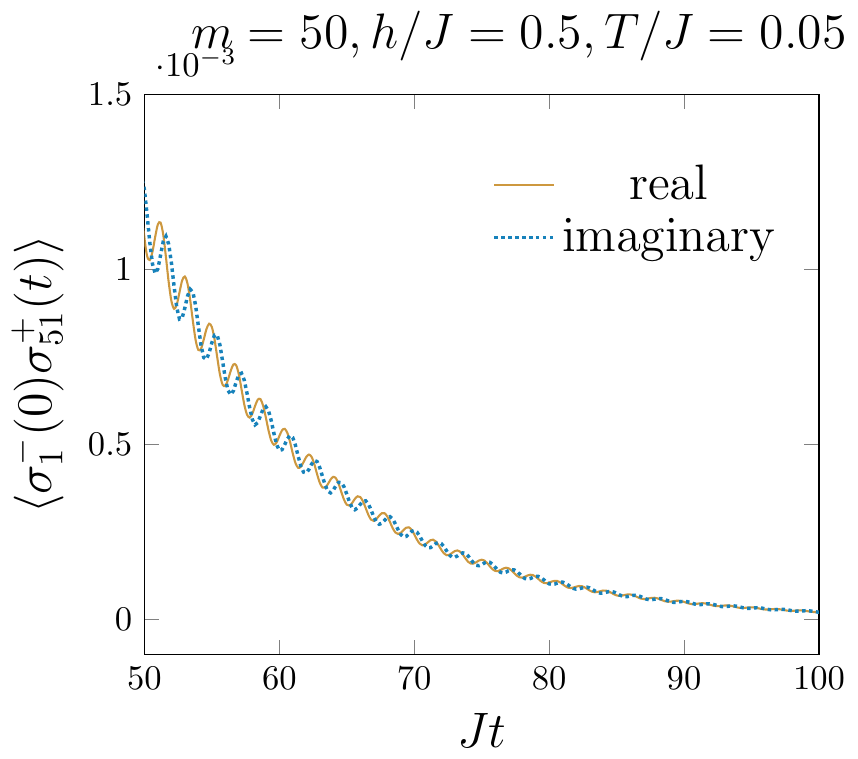}
\caption{
The real part and the imaginary part of $\langle \sigma^-_1(0) \sigma^+_{51}(t) \rangle$
at $T=0.05 J,h=0.5 J$ with $n=1536$ (left).
}
\label{fig:m50T005h05n}
\end{figure}

In order to achieve higher accuracy in the evaluation of the Fredholm determinant,
we increase the number $n$ of discretization points. We defer the discussion 
of the right choice of $n$ to \ref{app:choice_on_n}. 
With suitable values of $n$, we are able to perform a precise evaluation of
$\langle \sigma^-_1(0) \sigma^+_{m+1}(t) \rangle$ 
on a longer time scale.  An example is given in
Figure~\ref{fig:m50T005h05n}. The right panel presents a zoom of the
curves for large $Jt$. This exhibits a slowly decaying
oscillating pattern.

%
%
\section{Even-odd effect}\label{sec:evenodd}
\begin{figure}[!h]
\centering
{\includegraphics[width=3.7cm]{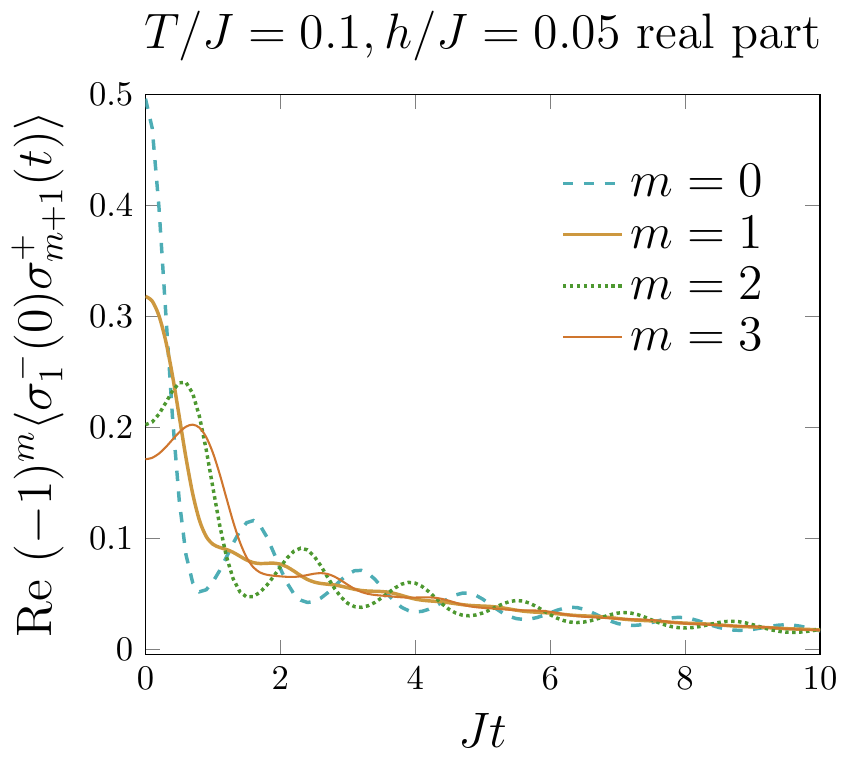}
\hspace{0.01cm}
\includegraphics[width=3.7cm]{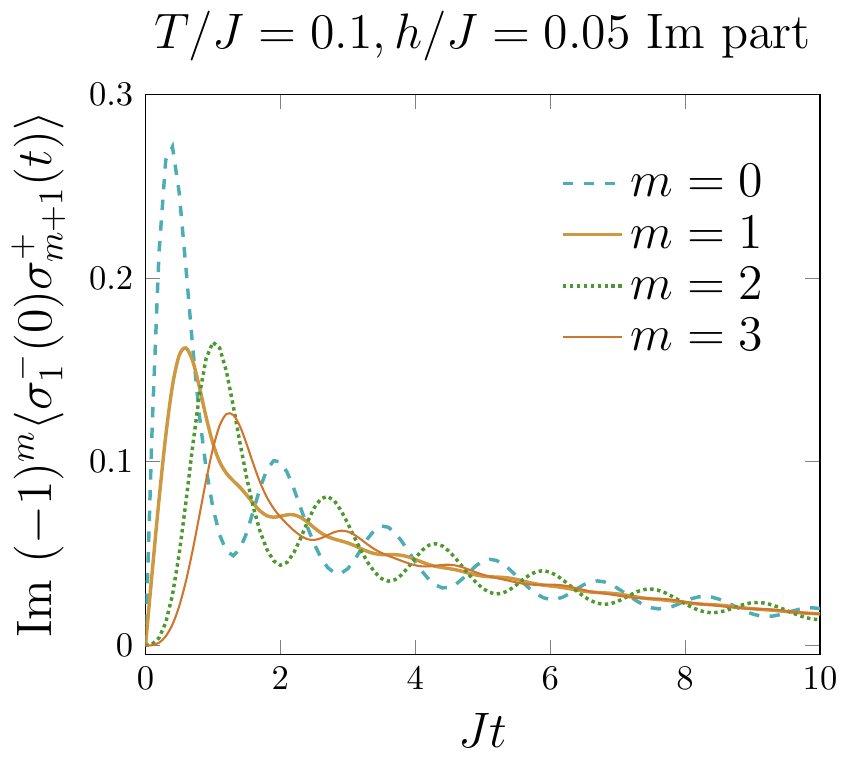}
}
\caption{
The real (imaginary)  part of 
$(-1)^m \langle \sigma^-_1(0) \sigma^+_{m+1}(t) \rangle$
at $T=0.1 J,h=0.05 J$ for $m=0 \sim 3$ in the left(right) plot .
}
\label{fig:mevenodd_massless}
\end{figure}

There are similarities and differences between the transverse correlator
in the massive and in the massless regime, as we observed above.
We comment on one more difference which seems to have been overlooked 
in past publications: the even-odd effect.
By this we mean a drastic suppression of the oscillation amplitude
of $\langle \sigma^-_1(0) \sigma^+_{m+1}(t) \rangle$  
for $m$ being odd and for small  $h/J$.  The plots in Figure~\ref{fig:mevenodd_massless}
present examples in the massless phase. 

\begin{figure}[!h]
\centering
{\includegraphics[width=\figheight]{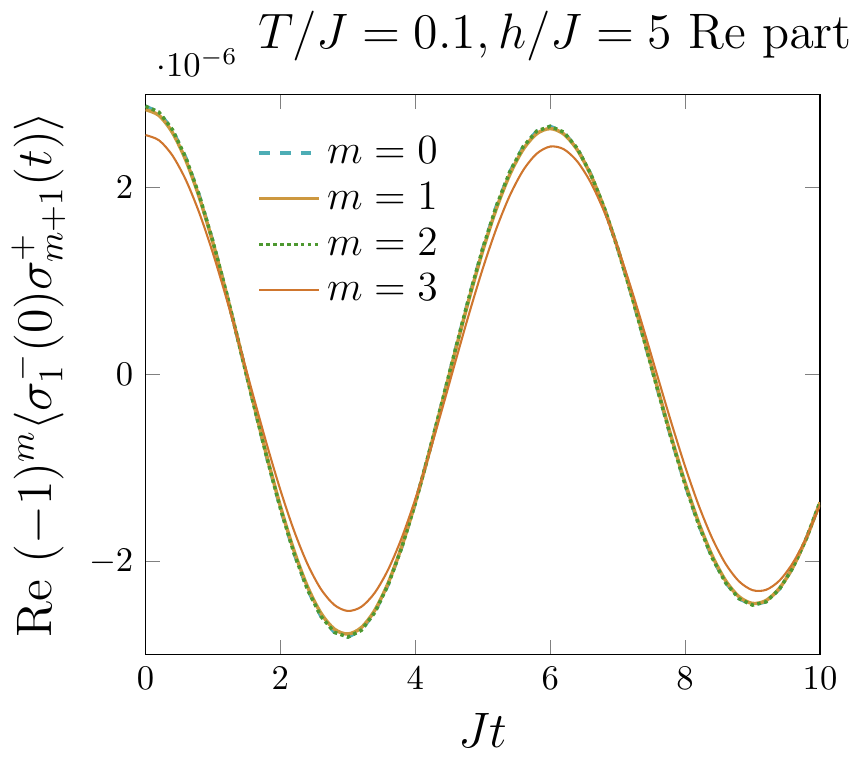}
\hspace{0.1cm}
\includegraphics[width=\figheight]{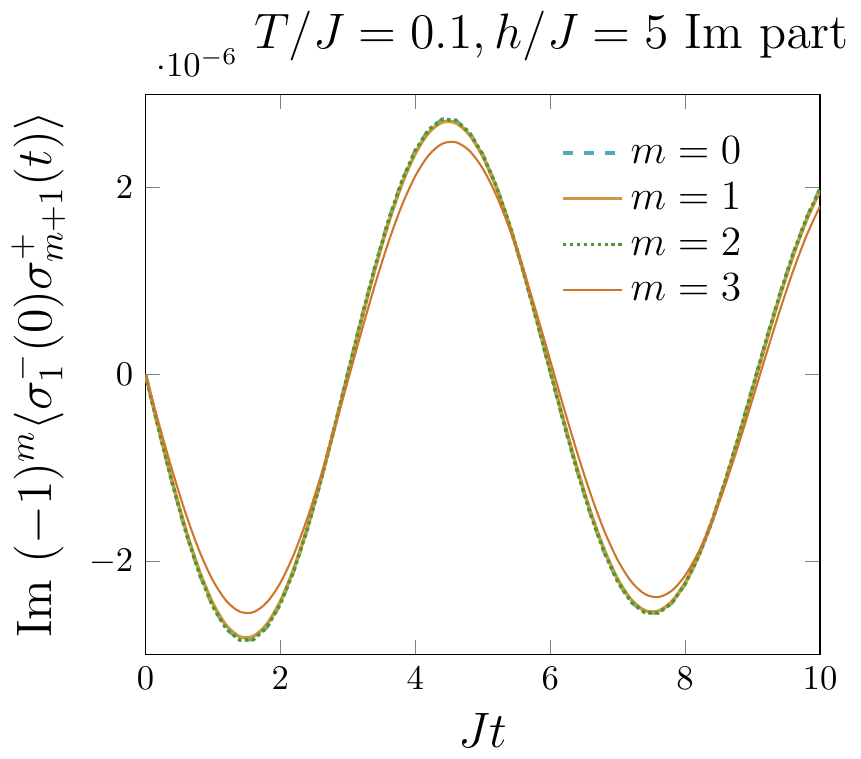}
}
\caption{
The real (imaginary)  part of  $(-1)^m \langle \sigma^-_1(0) \sigma^+_{m+1}(t) \rangle$
at $T=0.1 J,h=5 J$ for $m=0 \sim 3$ in the left(right) plot .
}
\label{fig:mevenodd_massive}
\end{figure}

On the other hand, this even-odd difference is not observed in the massive case.  See Figure \ref{fig:mevenodd_massive}.
Our formula  consists of several factors (\ref{eq:Fredholm_representation}).
The behavior of $\Omega$ is the same for $m$ odd and $m$ even. We numerically
find that in the massless case the oscillation phase of $\Omega(m,t)$ and
that of the Fredholm determinant part almost cancel each other
and that this results in the monotonous time evolution for odd $m$.
By way of contrast the two contributions do not cancel each other for $m$
even or in the massive case. 

The even-odd effect in the massless case can be explained
straightforwardly within non-linear Luttinger liquid theory. A first
step in the derivation of the long time asymptotics of the transverse
spin-spin correlation function of the XXZ chain is a Jordan-Wigner
transformation $\sigma^-_m=(-1)^m\exp(i\pi\sum_{l<m}c_l^\dagger c_l)
c_m$ onto spinless Fermions with annihilation operator $c_m$. Next,
the dispersion is linearized around the Fermi points $c_m\sim
\te^{ik_F x} \Psi_R(x) + \te^{-ik_Fx}\Psi_L(x)$ with $x=ma$ where $a$
is the lattice constant. After using the standard Bosonization
identities, this approach then allows to calculate the low-energy
contributions to the transverse spin-structure factor which are
located at momentum $k a\sim\pi$ and $k a\sim 2\pi M$ where $M$ is the
magnetization per site. Standard Bosonization thus predicts that there
are two low-energy contributions to the transverse spin-spin
correlation function. At the XX point, the contribution from
$ka\sim\pi$ is dominant and is given for small temperatures $T\ll 4J$
by \cite{KarimiAffleck}
%
\begin{multline}
\label{NLL1}
\langle \sigma^-_0(0)\sigma^+_m(t)\rangle^{(0)}\propto \left(\frac{\pi T}{v}\right)^{1/2}
\\ \times \frac{\te^{i\pi x/a}}{\sh^\frac{1}{4}\left(\frac{\pi T}{v}(x-vt-i\eta^+)\right)
                \sh^\frac{1}{4}\left(\frac{\pi T}{v}(x+vt+i\eta^+)\right)} \,.
\end{multline}
%
Here $v=4J\sin(k_F a)$ is the sound velocity and
$\eta^+$ denotes a small regularization parameter. 
Note, in particular, that for $T=0$ and $vt\gg x$ standard Bosonization predicts a decay $\sim 1/\sqrt{vt}$ at long times. Furthermore, there is no part which oscillates in time at this level. To obtain the oscillating contribution one needs to keep---in addition to the Fermi point contributions covered by standard Bosonization---also the saddle point contributions which appear at real momenta $\bar{p}_{\pm}$ for $t>t_c$, see eq.~(\ref{eq:def_p_pm_time})
\footnote{ Here the momentum has a proper dimension  $\bar{p}_{\pm} = {p}_{\pm}/a$}.
Doing so leads to non-linear Luttinger liquid theory. In the XX case, the theory is particularly simple because the saddle point and the Fermi point contributions do not interact with each other. We now use the extended ansatz $c_m\sim
\te^{ik_F x} \Psi_R(x) + \te^{-ik_Fx}\Psi_L(x) +\te^{i\bar{p}_+ x}d(x)$ where $d$ denotes a particle near the saddle point. It is then a fairly straightforward calculation to show that the additional saddle point contribution is  \cite{KarimiAffleck}
%
\begin{multline}
\label{NLL2}
\langle \sigma^-_0(0)\sigma^+_m(t)\rangle^{(1)} \\
\propto \te^{i \bar{p}_+x}\cos(k_F x)
\langle \sigma^-_0(0)\sigma^+_m(t)\rangle^{(0)}\langle d(0,0)d^\dagger(x,t)\rangle \,.
\end{multline}
%
We are interested here in understanding the asymptotic behavior in the
time-like regime for $vt\gg x$ and $T\ll 4J$. In this limit we have
$\bar{p}_+\, a\to\pi$ and we can approximate the propagator for the high-energy
particle by $\langle d(0,0)d^\dagger(x,t)\rangle\propto
\te^{i(4J-h)t}/\sqrt{vt}$. Putting together the two contributions (\ref{NLL1}) 
and (\ref{NLL2}) for this case we arrive at
%
\begin{multline}
\label{NLL3}
\langle \sigma^-_0(0)\sigma^+_m(t)\rangle \\ \stackrel{vt\gg m}{\propto} 
(-1)^m\frac{(\pi T/v)^{1/2}}{\sh^{1/2}(\pi Tt)}\left[ 1 +A \cos(k_F x)\frac{\te^{i(4J-h)t}}{\sqrt{vt}}\right]
\end{multline}
%
with some unknown amplitude $A$. In the zero temperature limit, in
particular, eq.~(\ref{NLL3}) predicts that the correlation function
asymptotically consists of a uniform part which decays as
$1/\sqrt{vt}$ and part which decays as $1/vt$ and oscillates with
frequency $4J-h$.

Crucially, the oscillating part contains a factor $\cos(k_F x)$. This
means that for $h\to 0$ ($k_F \,a \to \pi/2$) the oscillating contribution
is absent if the spatial distance $m$ in the two-point correlation
function is odd!

\begin{figure}[!h]
\centering
{\includegraphics*[width=8cm]{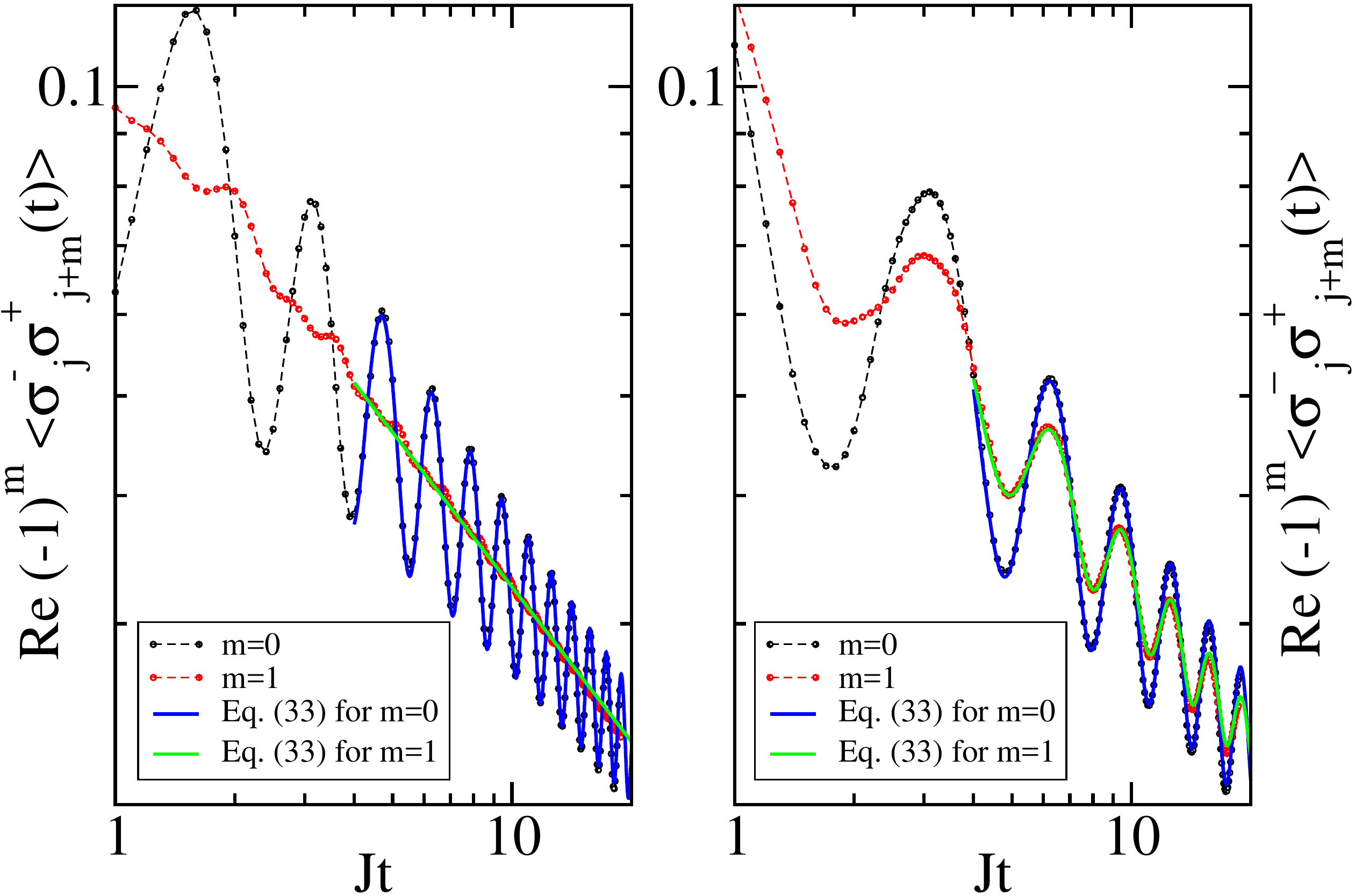}
}
\caption{The real parts of the auto- and nearest-neighbor correlation functions for $T=0.01$ with magnetic field $h/J=0$ (left) and $h/J=2$ (right).}
\label{Fig_NLL}
\end{figure}

To check the predictions of Eq.~(\ref{NLL3}) we present in Figure \ref{Fig_NLL} low-temperature numerical data for the auto- and nearest-neighbor
 correlation function at $h/J=0$ (left panel) and $h/J=2$ (right panel).
>From the log-log plot it is clear that for $tT\ll 1$ the correlation functions decay following a
power law in time for all cases.
 For $h/J=0$ ($k_F \,a=\pi/2$)
the autocorrelation shows oscillations while the nearest-neighbor correlation function does not whereas both are oscillating for
$h/J=2$. A comparison with eq.~(\ref{NLL3}) using the amplitudes of the
uniform and the oscillating part as fitting parameters confirms that the long time behavior is described by the non-linear Luttinger liquid result and that the factor $\cos(k_Fx)$ is indeed the reason for the observed even-odd effect at zero magnetic field. Finally, we note that higher order corrections to eq.~(\ref{NLL3}) do exist, i.e.~terms which decay faster than $1/vt$ at $T=0$. Such terms are responsible for the remaining small oscillations observed in the nearest-neighbor
 correlation function at zero field.

%
%
\section{Comparison with asymptotic formulae}\label{sec:asymptotics}

Starting from a slightly different Fredholm determinant Its {\it et al}.\
\cite{ItsIzerginKorepinSlavnov1993} reformulated the problem of the asymptotic analysis 
of the transverse correlation function in terms of a matrix Riemann Hilbert problem.
They obtained a closed expression for the leading terms of
the asymptotic expansion of the transverse correlation function in 
$t \rightarrow \infty$ and $m \rightarrow \infty$ with a fixed direction $\phi$ s.t. 
$\cot \phi =m/4J t$  at arbitrary $T$ and  $h<h_c$.
 The space-like  regime (\ref{eq:Its_space})  corresponds to 
 $0\le \phi <\pi/4$ while the  time-like one  (\ref{eq:Its_time})  does to 
$\pi/4< \phi \le \pi/2$. 
Subsequently their method was
applied to the massive case $h > h_c$ in a PhD thesis by Jie \cite{Jie}.
An interesting feature of our novel thermal form factor series
(\ref{eq:ffseriesxxtransd}) is that it is more suitable for
the long time, large distance analysis in the space-like regime than
the Fredholm determinant representation of Its {\it et al}.\ 
\cite{ItsIzerginKorepinSlavnov1993} in that the asymptotics can be easily
extracted from the first term in the series. For this reason we can obtain
the constant term in the asymptotic expansion which was heretofore unknown
in the massless phase. In this section, we provide a quantitative comparison of the
available asymptotic formulae with our numerical results.  
\subsection{\boldmath Space-like regime in the massless phase $h < h_c$}

\begin{figure}[!h]
\centering
\includegraphics[height=1.2\figheight]{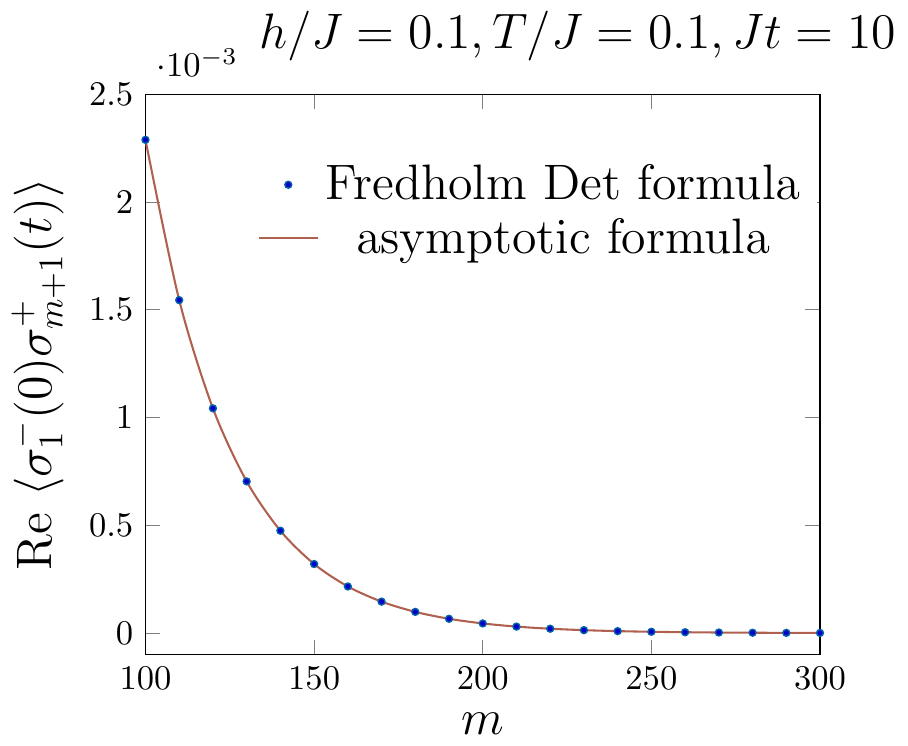} 
\caption{
The real  part of  $\langle \sigma^-_{1}(0)  \sigma^+_{1+m}(t)\rangle $
with $T/J=0.1, h/J=0.1, Jt=10$ and $m=100\sim 300$ evaluated from the Fredholm
determinant (dots) and from (\ref{eq:Its_space}) (after appropriate renormalization).
}
\label{fig:spacelike_asymptotic}
\end{figure}

Its {\it et al}.\ \cite{ItsIzerginKorepinSlavnov1993} use a 
Hamiltonian $J=-1, h \rightarrow 2h$ in (\ref{eq:xx_hamiltonian})
and  provide formulae for the conjugate transversal correlation
function
\begin{equation}\label{eq:def_Its_g}
g(m,t, h)= \langle \sigma_{m+1}^+ (t) \sigma_1^- (0) \rangle_{J=-1, 2h,T}.
\end{equation}
By a unitary transformation and an
adaption of para\-meters, it is related to ours by
\begin{equation}\label{eq:gauge_transformation}
\langle \sigma_1^- (0) \sigma_{m+1}^+ (t) \rangle_{J=1,- h, T}
=(-1)^m ( g(m,t,h/2))^\ast.
\end{equation}
In  the space-like regime of the massless phase their asymptotic formula translates
into
\begin{align}\label{eq:Its_space}
\langle \sigma^-_{j}(0) & \sigma^+_{j+m}(t)\rangle 
 =(-1)^mC (T,h) \nonumber \\
 &\times  \exp \Bigl( \left |\frac{m}{2\pi} \right | \int^{\pi}_{-\pi} {\rd}p\, 
  \log \left | \tanh \frac{h-4J \cos p}{2T} \right | \Bigr)
\end{align}
after properly recovering the exchange coupling $J$.
The constant $C (T,h)$ was not explicitly calculated in
\cite{ItsIzerginKorepinSlavnov1993}.
Within our framework it is not hard to reproduce this formula and to obtain
an explicit expression for $C (T,h)$ from 
eq.~(\ref{eq:ffseriesxxtransd}),
\begin{align*}
C (T,h) = \frac {2 T \tilde{\Phi}(\lambda_F^-)}{ \epsilon'(\lambda_F^- )  }
      &\exp \biggl\{  -
	             \int_{{\cal C}' \subset {\cal C}} \rd \la  \\
		   &\int_{\cal C} \rd \mu \, \cth' (\la - \mu) z(\la) z(\mu) \biggr\} \, ,
\end{align*}
where $\lambda_F^-$ is an inverse image of $-p_F$, $p(\lambda_F^-) =-p_F$ in
(\ref{eq:pepslambda}). See \cite{GKS19II} for  details. We supplement convincing
numerical support in  Figure \ref{fig:spacelike_asymptotic}.
We have to adopt the steepest descent paths even in the space like regime for
larger values of $m$ in order to achieve sufficient accuracy. 

We assumed that the other poles (of $\mu$) are sufficiently away from $\lambda_F^-$
in the above derivation. When $T \searrow 0$, however, they accumulate towards the Fermi
points, and we inevitably have contributions from them which are neglected in the above
as exponentially small corrections. Their estimation is an interesting problem which we
hope to discuss in the future. 

\subsection{\boldmath Time-like regime in the massless phase $h<h_c$}

\begin{figure}[!h]
\centering
{\includegraphics[height=\figheight]{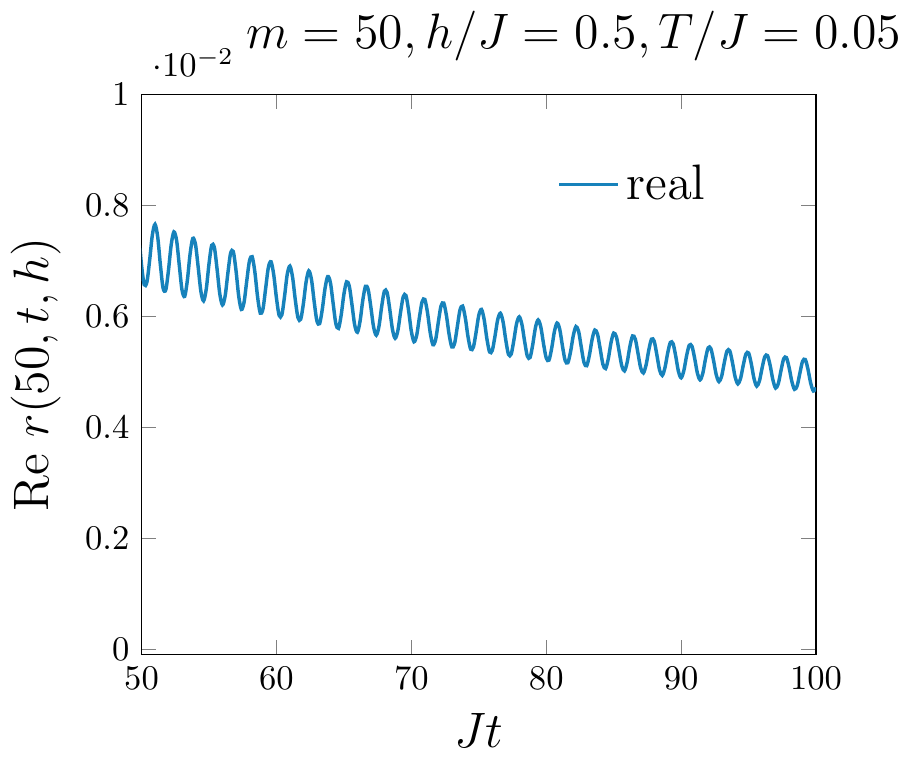}}
\caption{
The real part of $r(50,t,h)$ at $T=0.05 J, h=0.5 J$ for $n=1536$. 
}
\label{fig:m50T005h05:Its}
\end{figure}

The  asymptotic formula  of \cite{ItsIzerginKorepinSlavnov1993} becomes
 more involved in the time-like regime,  
 \begin{equation}\label{eq:Its_time}
g(m,t, h/2)  =  A_0 (T,h) I_m(t)
  \end{equation}
 with
\begin{align}\label{eq:def_Im}
 &I_m(t):=  (t J) ^{2 \nu_+^2+ 2 \nu_-^2} \nonumber \\
&\phantom{ab}\times \exp 
\Bigl(
\frac{1}{2\pi} \int_{-\pi}^{\pi} \rd k\,  \left |m-4t  J \sin k \right| \log \left |\tanh \frac{\varepsilon(k)}{2T} \right |
\Bigr),
\end{align} and
\[
\nu_{\pm} =\frac{1}{2\pi} \log \left | \tanh \frac{ h\mp 4  J\sqrt{1-  (\frac{m}{4Jt})^2} }{2T} \right |.
\]
The authors of \cite{ItsIzerginKorepinSlavnov1993} 
did not present the explicit form of  $A_0 (T,h)$ ($C$ in their notation) but 
commented that the higher-order correction modifies
$A_0 (T,h) \rightarrow A_0 (T,h) (1+c(t,m))$, where  $c(t,m)\sim t^{-\frac{1}{2}}$.
We  again use (\ref{eq:gauge_transformation}) and re-introduce $J$.
As $I_m$ is real-valued for  positive $t$, it is natural
to consider the ratio 
\[
r(m,t,h):=(-1)^m \langle \sigma^-_{1}(0)  \sigma^+_{m+1}(t)\rangle_{-h,T}/I_m(t). 
\]
In view of asymptotic analysis, $r(m,t,h)$ depends on $m$ only through $m/t$. \\
%
Based on our numerics (with $n=1536$) we  propose
\begin{conjecture}\label{conj:timelike_asymptotic}
For  fixed $T, h, m$ and $t \gg 1$, the ratio  $r(m,t,h)$ consists of both non-oscillating 
and oscillating parts, 
\[
r(m,t,h)\sim  r_{\rm non-osc}(m,t,h)  + r_{\rm osc}(m,t,h).
\]
The non-oscillating  part $r_{\rm non-osc}(m,t,h)$ includes $A_0 (T,h)$.
The  period $2\pi/\omega$ of the oscillating part $r_{\rm osc}(m,t,h)$ behaves as $\omega \sim h_c+h$ as $h \rightarrow -h_c$, and its amplitude
is also expanded by  powers of $ t^{-\frac{1}{2}}$.
\end{conjecture}
We, however, do not have any estimate of the time scale when the
above asymptotic form becomes valid.
For example, let us look at $r(50,t,h)$ for $T=0.05 J, h=0.5 J$ with $n=1536$ in 
Figure~\ref{fig:m50T005h05:Its}.
The average value does not seem to saturate, which suggests 
that  
$\langle \sigma^-_{1}(0)  \sigma^+_{51}(t)\rangle_{-h,T} $
is yet to reach the region described by the asymptotic formula if $t<100$.
%
There is, however, a limitation on the range of $t$, for numerical accuracy.
We thus need to find other ways than dealing with $t \gg 100$.
Intuitively, the smaller $m$ is, the shorter we have to wait until 
$r(m,t,h)$ reaches  ``equilibrium".
Thus, we focus on the extreme case $m=0$ (the auto-correlation).  
To be precise, this corresponds to the direction $\phi=0$ and the 
direct application of the formula (\ref{eq:Its_time}) may need justification.
We however assume its validity in the following.

\begin{figure}[!h]
\centering
\includegraphics[height=\figheight]{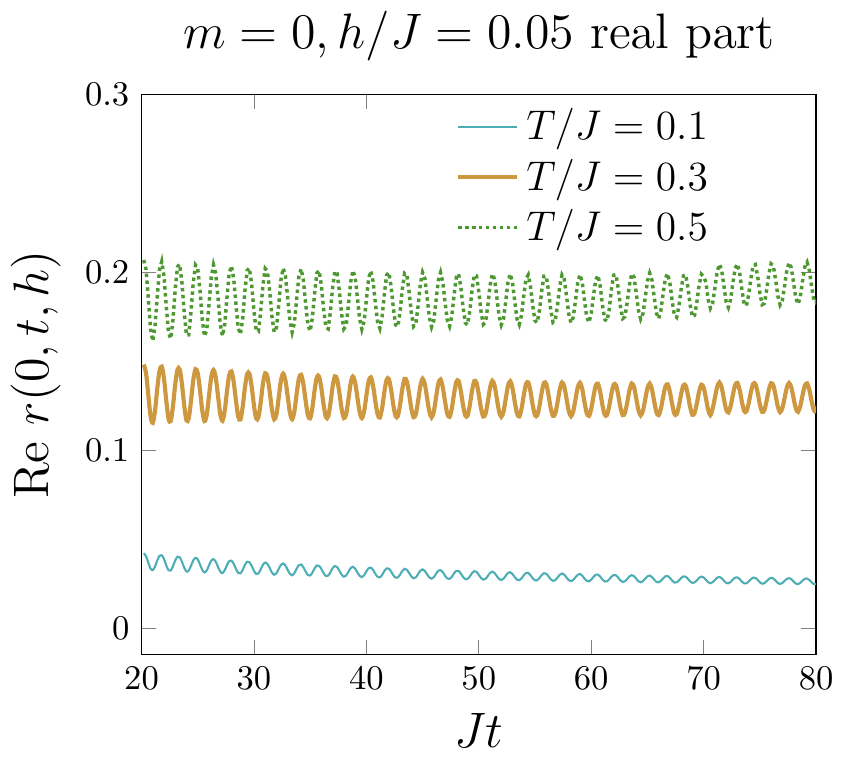}
\caption{
The real part of $r(0,t,h)$ for $h/J=0.05$. 
The temperature takes values $T/J=0.1, 0.3, 0.5$. Numerical inaccuracies develop for $Jt\gtrsim 70$ at $T/J=0.5$.
}
\label{fig:m0h005:Its}
\end{figure}

The real parts of $r(0,t,h)$ are plotted in Figure~\ref{fig:m0h005:Its} for
$T/J=0.1, 0.3, 0.5$ with $h/J=0.05$.

\begin{figure}[!h]
\centering
\setlength{\tabcolsep}{20pt}
\includegraphics[height=3.5cm]{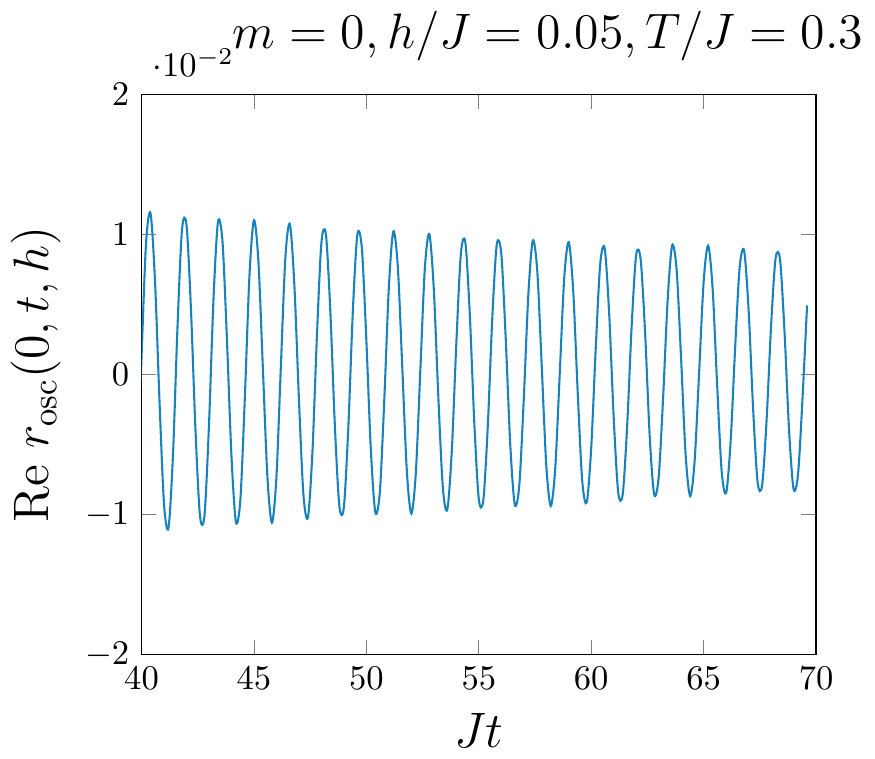} \hspace{-0.1cm}
\includegraphics[height=3.5cm]{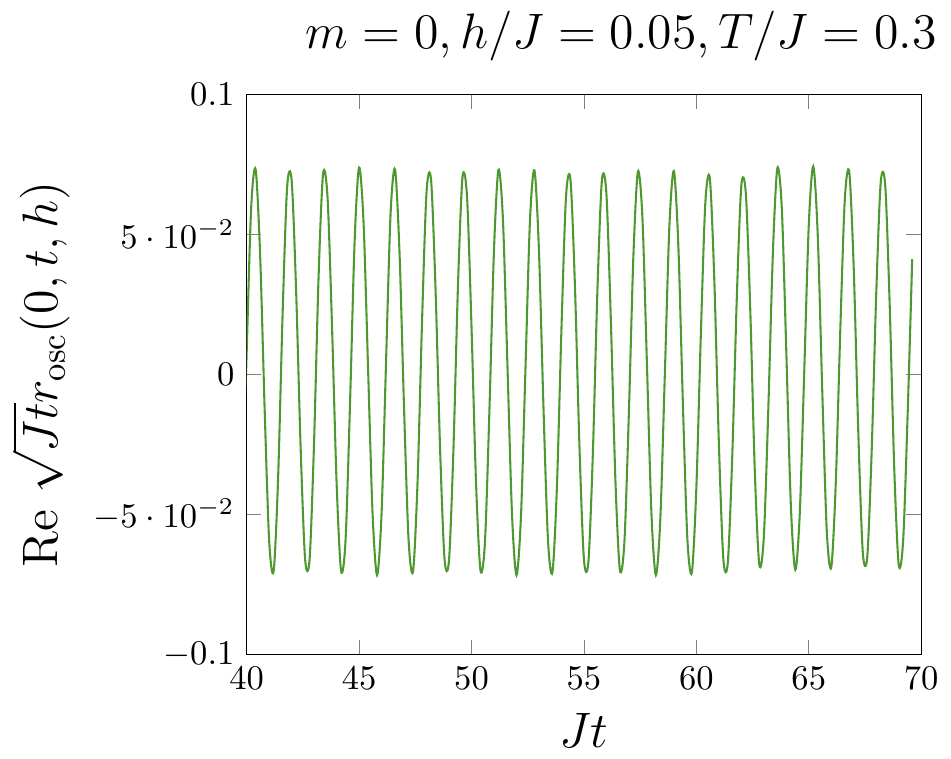}
\caption{
Estimate of the real part of $r_{\rm osc}(0,t,h)$ and $ \sqrt{Jt} r_{\rm osc}(0,t,h)$. 
}
\label{fig:m0T03h005:amplitude}.
\end{figure}

While $\operatorname{Re}\bigl( r_{\rm non-osc}(0,t,h)\bigr)$ is still slowly decreasing for $T/J=0.1$,
it already seems to approach ``equilibrium" for $T/J=0.3,0.5$ around $J t \sim 40$.
The amplitude  of the oscillating part  $\operatorname{Re}\bigl(r_{\rm osc}(0,t,h)\bigr)$  seems to be {\it very slowly} decreasing.
In order to examine this quantitatively, we fit $r_{\rm non-osc}(0,t,h)$ for $m=0, h=0.05 J, T=0.3 J$
using  the data for $40<t J<70$,
\begin{align}
r_{\rm non-osc}(0,t,h) &\sim 0.12390+0.12372 i   \, +  \nonumber \\
&\frac{0.034657+0.033381i}{\sqrt{Jt}}+O(\frac{1}{Jt}).  \label{eq:r_nonosc_estimate}
\end{align}
The sum of the first two terms in the right-hand side is identified with $A_0(0.3J,0.05 J)$. The sub-leading terms agree with the 
remark by Its {\it et al}. 
Taking this for granted, 
the real part of  $r_{\rm osc}(0,t,h)$ is estimated as in Figure \ref{fig:m0T03h005:amplitude}. 
The plot suggests that the amplitude of  $ \sqrt{Jt} r_{\rm osc}(0,t,h) $ stays constant, that is, 
$r_{\rm osc}(0,t,h) \sim t^{-\frac{1}{2}}$. This is again consistent with the estimate by Its {\it et al}..

\begin{figure}[!h]
\centering
\includegraphics[height=\figheight]{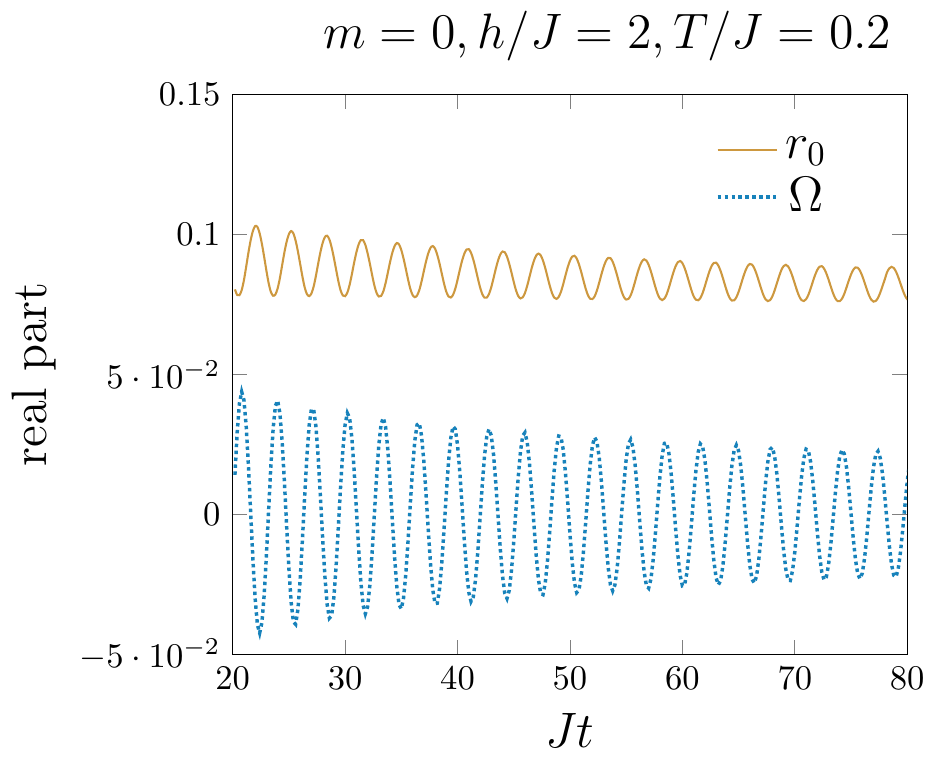}
\caption{
Comparison of the real parts of $r(0,t,-h)$  
and $\Omega(0,t)$ for $T/J=0.2, h/J=2$.
}
\label{fig:comparison_Omega}
\end{figure}

There are remarks. One notices that $\operatorname{Re}\bigl(r(0,t,h)\bigr)$ for $T/J=0.5 $ shows a discontinuity
at $T/J \sim 70$. 
This is a numerical artifact due to inaccuracy:  
$|\langle \sigma^-_{1}(0)  \sigma^+_{1}(t)\rangle |$ already reaches $O(10^{-25})$. Thus,
it is too difficult to have precise control over the numerics.
The oscillatory part remains observable after sufficiently long time
(except for $h=h_c$).
The origin of the oscillation can be attributed to that of $\Omega(m,t)$. 
The comparison of  the real parts of $r(0,t,-h)$ 
and $\Omega(0,t)$ is given in  Figure~\ref{fig:comparison_Omega}
for $T/J=0.2, h/J=2$.
We can easily check that their frequencies coincide. 
One expects from eq.~(\ref{eq:Fredholm_representation}) that
$\langle \sigma^-_{1}(0) \sigma^+_{1}(t)\rangle$  
is proportional to $\Omega(0,t)$, but it is not so simple. The
inverse power of $\Omega$ in the kernel of the Fredholm determinant
partially cancels the oscillation, and this may contribute to the
non-oscillating part $A_0$. The center of oscillation of
$\langle \sigma^-_{1}(0)  \sigma^+_{1}(t)\rangle$ 
is thus different from zero, while it is zero for $\Omega(0,t)$.

\subsection{Consistency with the result from nonlinear Luttinger Liquid theory}

Finally, we comment on the relation between the above analysis and  the results from
the continuum theory, (\ref{NLL1}) and (\ref{NLL3}).  
We shall neglect logarithmic corrections in $t$ and consider only the leading $T\rightarrow 0$
behavior.
Let us start from the space-like case, $m, t \rightarrow \infty$
and $t<t_c$. 
One applies a Sommerfeld type argument to the integral in (\ref{eq:Its_space}) and finds
\[
\langle \sigma^-_{j}(0) \sigma^+_{j+m}(t)\rangle 
\sim (-1)^m {\rm e}^{-\frac{m\pi T}{2v}}.
\]
This is consistent with (\ref{NLL1})  for $T$ small but finite and $\pi T x/v \gg 1$.
Next consider the time-like case $m, t \rightarrow \infty$
and $t>t_c$. 
The integral  $I_m$ in (\ref{eq:def_Im}) has different forms 
for $p_-<p_F$ and for $p_F<p_-$,%
\footnote{The symbol $p_-$ is defined in (\ref{eq:def_p_pm_time}).}
\[
I_m(t) \sim
\begin{cases}
{\rm e}^{ - \frac{\pi t T}{2}}&  p_-<p_F, \\
{\rm e}^{ - \frac{m\pi T}{2v}}&  p_->p_F. \\
\end{cases}
\]

Thus if $p_-<p_F$,
together with Conjecture \ref{conj:timelike_asymptotic}
and relation (\ref{eq:gauge_transformation}), we conclude that
\begin{multline*}
\langle \sigma_1^- (0) \sigma_{m+1}^+ (t) \rangle \\
=(-1)^m {\rm e}^{-\pi t T/2} A_0 \Bigl(1+ \frac{\tilde{A}(m)}{\sqrt{Jt}} {\rm e}^{i(-h+4J) t}  \Bigr).
\end{multline*}
One identifies the terms on the right hand side  with the dominant and sub-dominant
terms in (\ref{NLL3}) if $ tT \gg 1$.
This condition is consistent with the asymptotic analysis as $T$ is small but finite.
 In view of (\ref{NLL3}), the third and fourth terms in (\ref{eq:r_nonosc_estimate}) look irrelevant and
 may be attributed to the weak  transient behavior which still exists.  We however do not have a conclusive evidence.\\
The above result suggests that the meaning of the space-like and time-like regimes
 undergoes a change in the low-temperature limit, or, in other words, that the ‘light cone’ needs to be redefined.
  At elevated temperatures, it is defined by $m/t = 4J$, 
  while in the low-$T$ limit,  it is defined by the Fermi velocity, $m/t = v  (\le 4J)$.
This means that the space-like regime becomes enlarged in the low-T limit
and that it becomes more enlarged if the magnetic field is higher.


\subsection{\boldmath Time-like regime in the massive case $h>h_c$}

\begin{figure}[!h]
\centering
\setlength{\tabcolsep}{20pt}
\includegraphics[height=3.4cm]{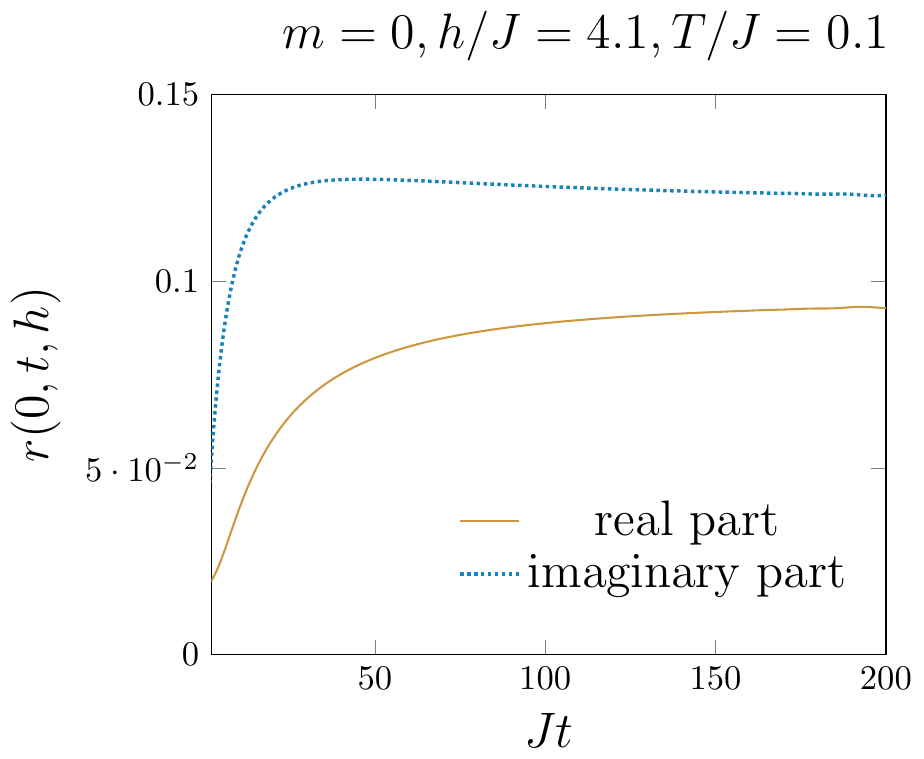} \hspace{0.1cm}
\includegraphics[height=3.4cm]{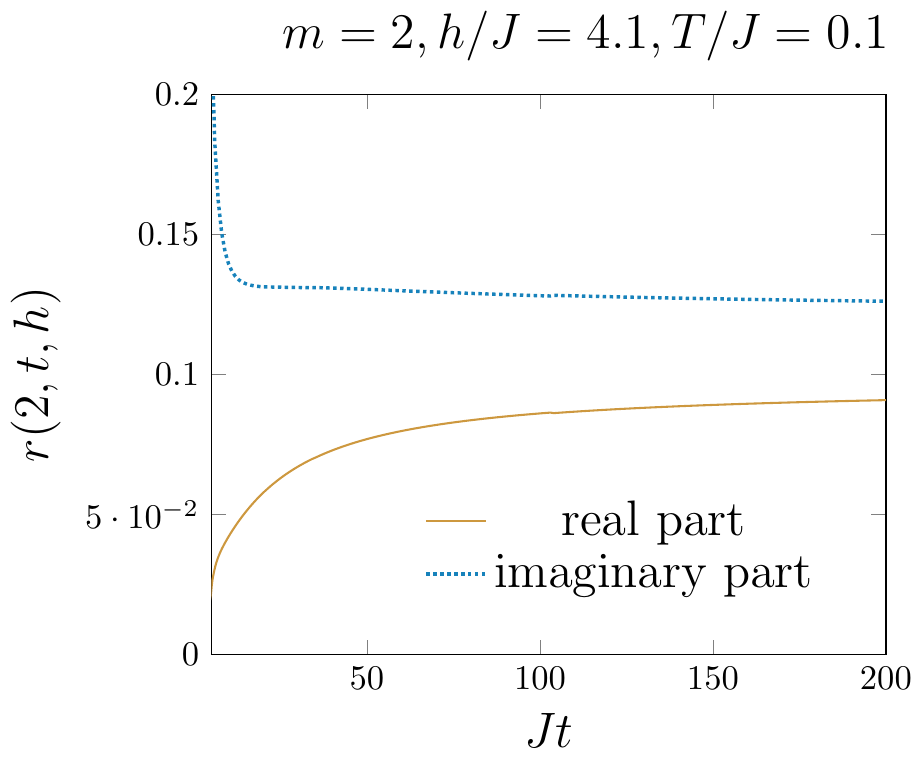}
\caption{
The real and the imaginary parts of $r(0,t,h)$ (left) and $r(2,t,h)$ (right)
for $h=4.1 J$ and $T=0.1J$.
}
\label{fig:comparison_with_Jie}
\end{figure}

A similar long time, large distance asymptotic formula for the massive
case $h>h_c$ was proposed by Jie \cite{Jie}. Set
\begin{align*}
%
\nu_{1} &=-\frac{1}{\pi} \log  \tanh \bigg( \frac{ h- 4 J \sqrt{1-  (\frac{m}{4t J})^2} }{2T} \bigg) \\
\nu_{2} &=\frac{1}{\pi} \log  \tanh \bigg(\frac{ h + 4 J \sqrt{1-  (\frac{m}{4t J})^2} }{2T} \bigg)\, .
\end{align*}
Then Jie argued that\footnote{An obvious typo in the second term ($(-J t)^{i\nu_1}$ instead of $(-J t)^{i\nu_2}$)
is corrected.}
%
\begin{multline*}
g(m,t, h/2) =
  {\mathfrak C} I_m(t) (-Jt)^{-\frac{1}{2}} {\rm e}^{-ih t} \\ \times
\Bigl( u_- (-J t)^{-i\nu_1} {\rm e}^{4it J \cos \alpha +i m \alpha} \\
+(-1)^m v_- (-J t)^{i\nu_2} {\rm e}^{-4it J \cos \alpha -i m \alpha}  +o(1) \Bigr),
\end{multline*}
where $\alpha = \pi-\arcsin \bigl(\frac{m}{4t J}\bigr)$ and $I_m(t)$ is
defined in (\ref{eq:def_Im}). 
The coefficient ${\mathfrak C} $ is a smooth function of $\alpha, T, m$ and $h$.
The explicit forms of $u_-, v_-$ and ${\mathfrak C}$ are too complicated to be reproduced here.
We only remark that  $|v_-| \gg |u_-|$ as $T \ll 1$. 
Thus, only the second term survives in this limit, and
it results the long-period oscillation $\omega \sim  h-4 J $ observed numerically.
%
The ratio 
\begin{multline*}
r(m,t,h)= \\ (-1)^m \langle \sigma^-_{1}(0)  \sigma^+_{m+1}(t)\rangle_{h,T}/({\mathfrak C}^{-1}g(m,t, -h/2))^*
\end{multline*}
is plotted in Figure~\ref{fig:comparison_with_Jie}. It seems to reach a constant value as
time evolves. 

\section{Comparison with the Pfaffian representation}\label{sec:other_method}
In this section we compare our results with results obtained by more
conventional methods.  Among them, the Pfaffian representation of the
correlation function is the most well-established for the dynamics of
the XX model. It deals with an open chain of $L$ sites. As the
translational invariance is broken, we rather consider $\langle
\sigma_j^-(0) \sigma_{j+m}^+(t) \rangle$ instead of (\ref{eq:def_CmtII})
and typically center the two-point correlation function at
$L/2$. After a simple calculation using the Fermion algebra, one
represents $\langle \sigma_j^-(0)
\sigma_{j+m}^+(t) \rangle$ by the Pfaffian of an anti-symmetric
$(2j + 2m +2)\times (2j + 2m +2)$ dimensional matrix, see e.g.\
\cite{McCoyBarouchAbraham1971,DKS2000} for details.

\begin{figure}[!h]
\centering
\setlength{\tabcolsep}{20pt}
\includegraphics[height=\figheight]{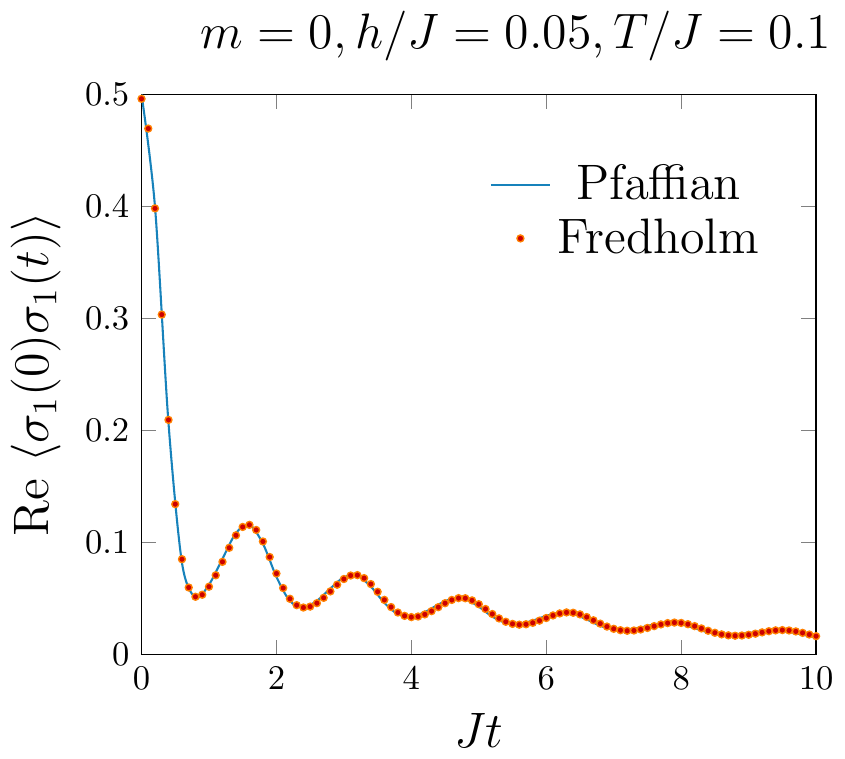} \hspace{-0.1cm}
\includegraphics[height=\figheight]{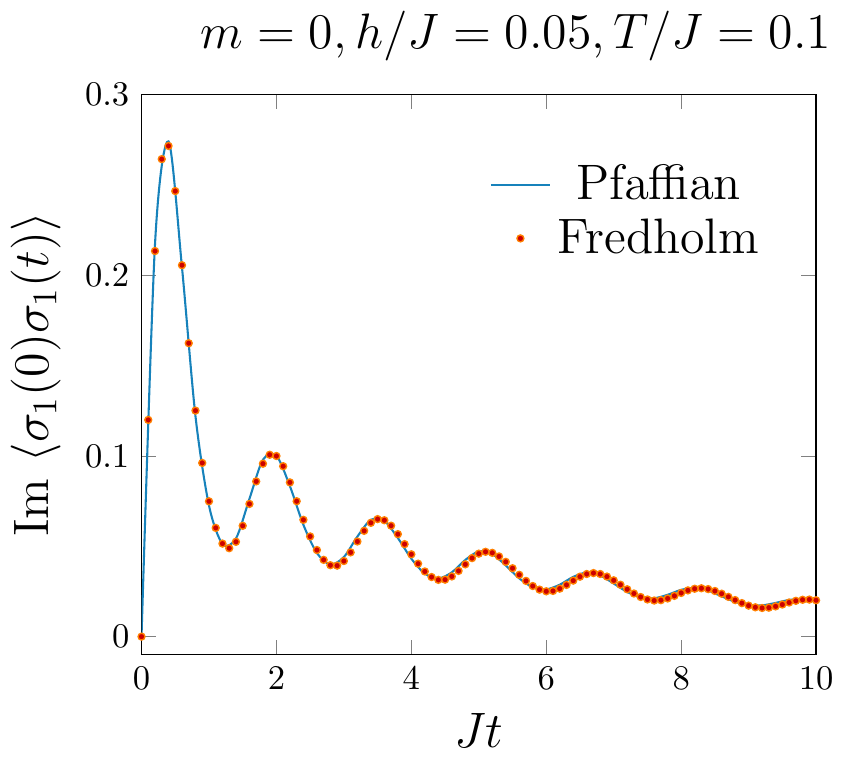}
\caption{
The real (the imaginary) part of $\langle \sigma_j^-(0) \sigma_{j}^+(t) \rangle$ on a small time scale 
in the left (right) plot.
}
\label{fig:m0T01h005:Pfaff}
\end{figure}

For small $m$ and $t$ the two results, one obtained by our formalism and the other one by
the Pfaffian method,\footnote{We refer the numerical calculation based on the  Pfaffian 
representation of the correlation function as ``the Pfaffian method" for short.} match perfectly.  
See Figure~\ref{fig:m0T01h005:Pfaff} for the autocorrelation function for $m=0$, $h/J=0.05$, $T/J=0.1$
and $L=128$.

\begin{figure}[!h]
\centering
{
\includegraphics[height=\figheight]{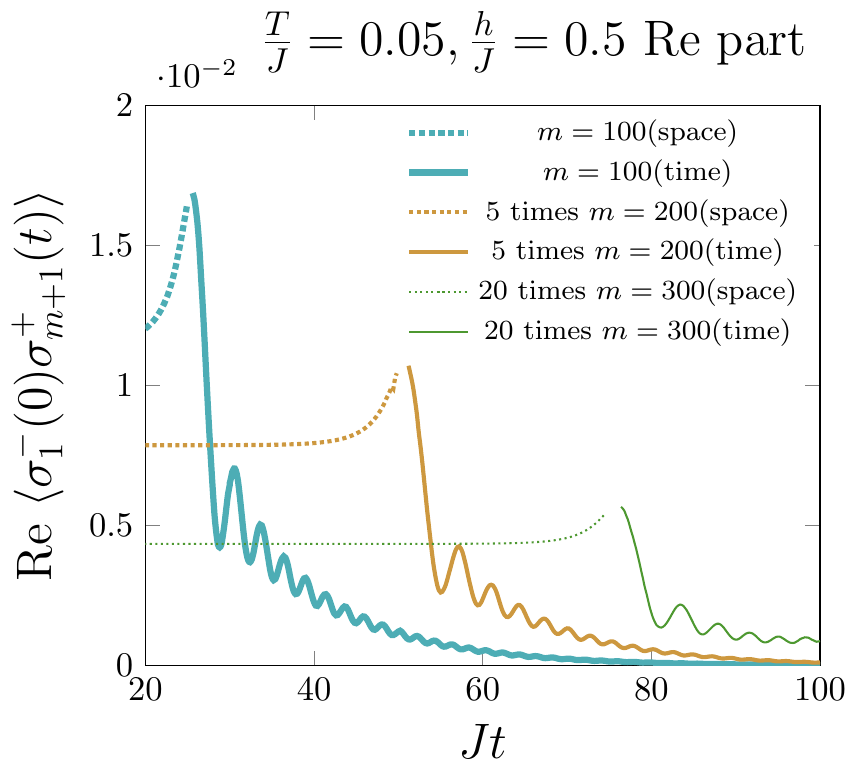}
}
\caption{
The real part of $\langle \sigma_1^-(0) \sigma_{101}^+(t) \rangle,  
5 \times \langle \sigma_1^-(0) \sigma_{201}^+(t) \rangle 
$  and  $ 20 \times \langle \sigma_1^-(0) \sigma_{301}^+(t) \rangle$
.}\label{fig:T005h05_m_large}
\end{figure}

One of the advantages of the present approach in comparison with the Pfaffian method
lies in the fact that the distance $m$ appears as a mere parameter. This allows us to
deal with large $m$ immediately. For illustration, the data for $m=100, 200$ and $300$,
obtained from  eq.~(\ref{eq:Fredholm_representation}) are plotted in
Figure~\ref{fig:T005h05_m_large} for $T/J=0.05, h/J=0.5$.\footnote{
As $m$ is large enough,  we use the steepest descent method also in the space-like regime.
It becomes inaccurate around $t_c=m/4J$ and some points are omitted near $t_c$ in the plots.}

\begin{figure}[!h]
\centering
\setlength{\tabcolsep}{20pt}
\includegraphics[height=\figheight]{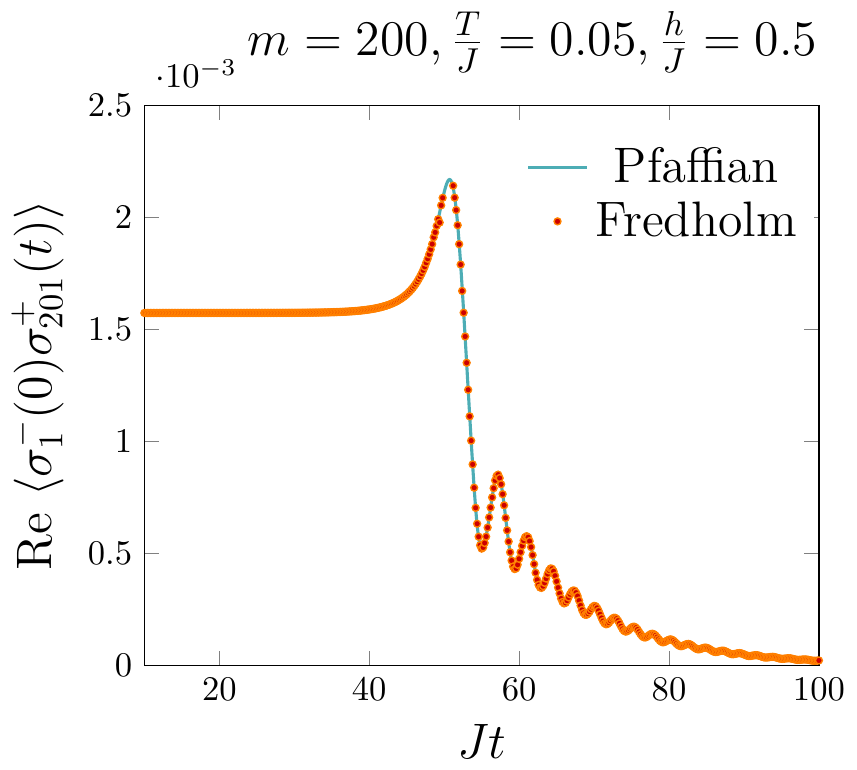} \hspace{-0.1cm}
\includegraphics[height=\figheight]{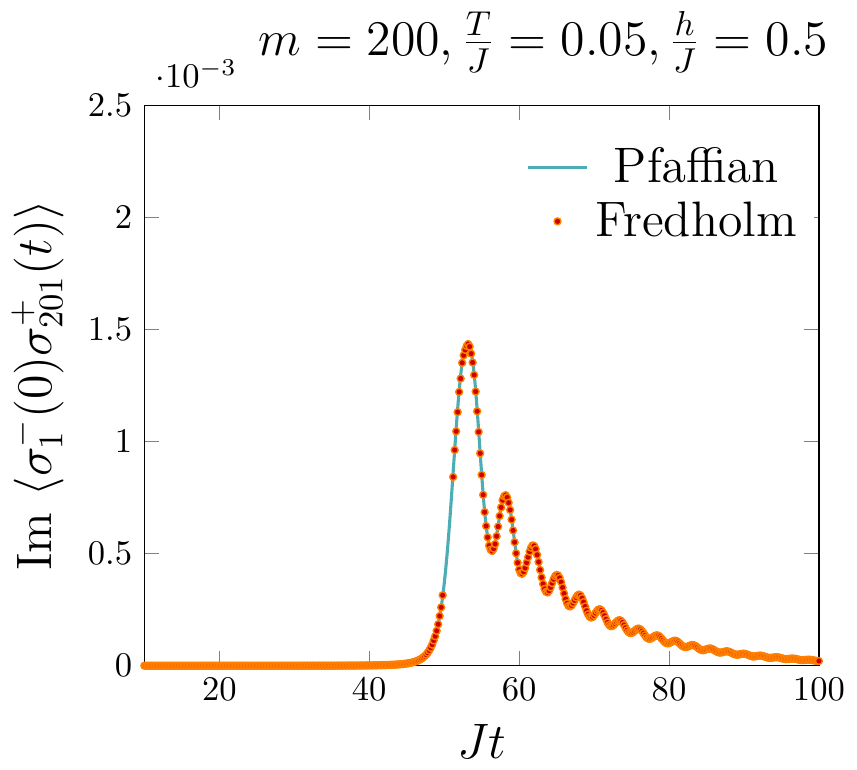}
\caption{
The real part (left)  and the imaginary part (right) of
$\langle \sigma_1^-(0) \sigma_{201}^+(t) \rangle$. 
The curves are results using the Pfaffian method and dots are obtained by the Fredholm determinant.
}\label{fig:m200_cf_Pfaffian_Fredholm}
\end{figure}

As indicated above, the Pfaffian formulation deals with a finite-size
system.  In order to eliminate the boundary effects, one has to choose
the system size $L$, the spin distance $m$ and the time $t$ so that
$L/2-m/2> v t$ is satisfied.  Here $v$ stands for the sound
velocity. Thus, for $m$ and $t$ becoming larger, one has to deal with
a larger-size matrix. It is fair to note that one also needs to
increase the number of discretization points $n$ in the evaluation of
the Fredholm determinant in order to keep the numerical precision for
larger $t$. An example is discussed in \ref{app:choice_on_n}.
 With these tunings and using a stable direct evaluation of the
Pfaffian (instead of calculating it as the square root of the
determinant), the two results coincide also for large $m$ and $t$.
An example with $m=200, T/J=0.05$ and $h/J=0.5$ is shown in
Figure~\ref{fig:m200_cf_Pfaffian_Fredholm}.
  
We stress, however, that the numerical error in the present scheme
solely originates from the discretization and should be distinguished
form the error due to the finite size effect explained above. In the
former case the error was estimated in \cite{Bornemann2010}. For the
type of kernel occurring in our Fredholm determinant the error decreases
algebraically with the number of discretization points.

\section{Summary and conclusion}\label{sec:summary}
In this communication we revisited the equilibrium dynamics of the XX chain.
We employed a new scheme for the exact evaluation of correlation functions,
based on the QTM and on a thermal form factor expansion. We rewrote the thermal
form factor series for the transversal correlation function as an explicit
factor times a Fredholm determinant. This representation was the starting
point of our numerical analysis, for which we utilized a direct discretization
of the determinant as suggested in \cite{Bornemann2010}.
We demonstrated that the method yields high-precision results for the
transverse correlation function for a wide range of time, distance, temperature
and magnetic field, if we properly use the freedom in the choice of the
integration contour involved in the definition of the Fredholm determinant. 
The algorithm works for long times and large distances, far in the
asymptotic regime, where we could confirm all existing analytic results
\cite{ItsIzerginKorepinSlavnov1993,Jie} and could even provide an
estimation of the next order correction from our numerical
data. We also checked our results by comparing with a numerical
evaluation of the correlation function using the Pfaffian
representation. Furthermore, we were able to explain even-odd effects
in the observed oscillations in the massless time-like regime by
non-linear Luttinger liquid theory.

We naturally expect, based on the success of the numerical study here, that the novel Fredholm 
determinant representation also provides an appropriate starting point for the analytic study of the XX chain.
Although we have just started the investigation in this direction, there is already evidence
that this is indeed the case \cite{GKS19I, GKS19II}.  
 
A most important issue, from our point of view, is a possible extension to a
truly interacting system, the XXZ chain. A single Fredholm determinant representation
does not seem to exist in this case. But the thermal form factor series exists
in the interacting case as well and represents the transversal correlation function
as a series of multiple integrals having certain analogies with a Fredholm determinant \cite{GKKKS2017}. Our numerical investigation suggests that the first few terms
of this series may already provide accurate estimates in appropriate limiting
cases. We hope to report concrete results in future publications.
\\[1.ex]
{\bf Acknowledgments.} F.~G\"ohmann acknowledges financial support by
the (DFG) in the framework of the research group FOR 2316 `Correlations
in Integrable Quantum Many-Body Systems'. K.~K. Kozlowski is supported
by the CNRS and by the `Projet international de coop\'eration scientifique
No.\ PICS07877': \textit{Fonctions de corr\'elations dynamiques dans la
cha{\^i}ne XXZ \`a temp\'erature finie}, Allemagne, 2018-2020. J. Sirker
acknowledges support by the DFG in the framework of the research group
FOR 2316 `Correlations in Integrable Quantum Many-Body Systems' as
well as support by the National Science and Engineering Research
Council (NSERC) of Canada. J.~Suzuki is grateful for support by a JSPS
Grant-in-Aid for Scientific Research (C) No.\ 15K05208, No.\ 18K03452
and by a JSPS Grant-in-Aid for Scientific Research (B) No.\ 18H01141.

%
\appendix
%
%
\section{A previous result}
\label{app:summary_thermal_ffseries}
In this appendix we recall a previous result, obtained in \cite{GKKKS2017},
from which we have derived the theorem in Section~\ref{sec:Fredholm}.

In \cite{GKKKS2017} we proposed a method for calculating dynamical
correlation functions at finite temperature in integrable lattice
models of Yang-Baxter type. The method is based on an expansion of
the correlation functions as a series over matrix elements of a
time-dependent quantum transfer matrix rather than the Hamiltonian.
Staying with the example of the transverse correlation functions
of the XX chain, the series is obtained, when we apply, in a slightly
sophisticated manner, respecting the integrability of the original
quantum chain \cite{Kluemper1993,SAW90,Kluemper92}, a Suzuki-Trotter
decomposition \cite{Suzuki85} to the exponential factors in
\eqref{eq:def_CmtII}. As a result, the correlation function is
represented as a (properly normalized) partition function of an
alternating vertex model acting on a fictitious space of size $2N+2$,
where $N$ is the Trotter number. This partition function can be
interpreted as a trace of a product of many staggered column-to-column
transfer matrices, the quantum transfer matrices, with two local insertions
corresponding two the two local operators $\sigma^-$, $\sigma^+$,
whose correlation function is considered. Using the `solution of the
quantum inverse problem' corresponding to the staggered monodromy
matrices appearing in the columns and expanding in terms of the
eigenstates of the QTM the thermal form factor series is obtained.

In case of the XX chains all matrix elements and correlation
lengths appearing in the series can be calculated explicitly
and the sums over classes of excitations (one hole, two holes
$+$ one particle, three holes $+$ two particles \dots) can be
turned into integrals. The integrands are composed of a number
of functions characteristic of the XX chain. 
In the first place
there are momentum and energy of the one-particle excitations
as functions of the rapidity variable,
\begin{equation} \label{eq:pepslambda}
     p(\lambda) = \frac{1}{i} \ln \frac{\tanh \lambda}{i}, \quad
     \epsilon (\la) = h - \frac{4 i J}{\sh(2 \la)}.
\end{equation}
The other functions needed in order to write the form
factor series are
\begin{align}
   &  \re (\la) = \frac{2}{\sh (2 \la)}, \label{defbaree} \\[1ex] 
     & z(\la) = \frac{1}{2 \pi i}
                \ln \biggl[\cth \biggl(\frac{\epsilon (\la)}{2T}\biggr)\biggr],
		\notag \\[1ex] 
     & \Phi (x) = \frac{\re(x)}{2} \times
                  \exp \biggl\{ 2 \int_{\cal C} \rd\mu \,\re(\mu)
		                z(\mu)  \frac{\sh(x+\mu)}{\sh(x-\mu )} \biggr\},
                   \notag \\[5ex]
     &  {\cal D} \bigl(\{x_j\}_{j=1}^{n_h}, \{y_k\}_{k=1}^{n_p}\bigr)  \notag \\
     & = \frac{\bigl[\underset {1 \le j < k \le n_h}{ \prod} \sh^2 (x_j - x_k) \bigr]
              \bigl[\underset{1 \le j < k \le n_p}{\prod} \sh^2 (y_j - y_k) \bigr]}
             {\prod_{j=1}^{n_h} \prod_{k=1}^{n_p} \sh^2 (x_j - y_k)}, \notag
\end{align}
and
\begin{align}
     & {\cal A} =
        \exp \biggl\{-\int_{{\cal C}' \subset {\cal C}} \rd \la
		     \int_{\cal C} \rd \mu\, \cth' (\la - \mu) z(\la) z(\mu) \biggr\},
		     \notag \\[1ex]
     & {\cal A} (m) = {\cal A} \times
        \exp \biggl\{ - m \int_{\cal C} \rd\mu \: z(\mu)
	                  \re(\mu)  \biggr\} . \label{defamt}
\end{align}
The contours $\cal C$ and ${\cal C}'$ in the definition of $\cal A$
are such that $\cal C$ tightly encloses ${\cal C}'$, and $\cal C$
itself simply surrounds the strip $|\operatorname{Im}(
\lambda)|<\frac{\pi}{4}$ counterclockwise with two infinitesimal
deformations at the Fermi rapidities $\lambda_F^\pm$ defined
by $\epsilon (\la_F^\pm) = 0$, $- \pi/2 < \Im (\la_F^\pm) < \pi/2$,
$\Re (\la_F^-) < 0 < \Re (\la_F^+)$.

Employing all of the above notation the thermal form factor series
for the transverse correlation function (\ref{eq:def_CmtII}) can be
written in the form
\begin{multline} \label{eq:ffseriesxxtransd}
\langle\sigma_1^- (0)\sigma_{m+1}^+ (t) \rangle 
= (-1)^{m+1} {\cal A} (m) \\ \times
        \sum_{n=1}^\infty \frac{1}{n! (n-1)!}
           \int_{\cal C} \prod_{r=1}^n \frac{\rd x_r}{\pi i}
	      \frac{\Phi_- (x_r) \re^{i (m p(x_r) - t \epsilon (x_r))}}
	           {1 - \re^\frac{\epsilon (x_r)}{T}} \\ \times
           \int_{{\cal C}_{\rm out}} \prod_{s=1}^{n-1} \frac{\rd y_s}{\pi i}
	      \frac{\re^{- i (m p(y_s) - t \epsilon (y_s))}}
	           {\Phi_- (y_s) \bigl[1 - \re^{-\frac{\epsilon (y_s)}{T}}\bigr]} \:
		   \\ \times
	      {\cal D} \bigl( \{x_r\}_{r=1}^n, \{y_s\}_{s=1}^{n-1} \bigr).
\end{multline}
%
Here we are using a contour $ {\cal C}_{\rm out}$ which tightly
encloses~${\cal C}$.  Note that as compared to the formula given
in \cite{GKKKS2017} we have introduced a few minor simplifications
by calculating two of the integrals in our original formula explicitly
(one is zero the other one a constant) and by absorbing a constant
term into the definition of the function $\Phi$.
 %
 %
\section{\boldmath The analytic properties of~$\mu$}\label{app:derivation_mu}
We first consider the massive case $h>h_c$ and assume that $\operatorname{Im} (p) >0$.
We take the integral over a closed rectangular contour ${{\cal C}_+}$ (of
infinite height and  width $\pi$), including the real axis as the bottom part,
\[
\tilde{\sigma}_{{\cal C}_+}(p) =\int_{{\cal C}_+} \frac{dq}{2\pi i} \frac{1}{\tan\frac{p-q}{2}}
   \ln \frac{1+\re^{-\epsilon(q)/T}}{1-\re^{-\epsilon(q)/T}}.
\]
Note that the integrand on the upper edge of the rectangle reduces to a constant $\frac{i}{2}$ and
the contributions from left and right sides cancel each other. We thus have
\[
\tilde{\sigma}_{{\cal C}_+} (p) = \sigma_+(p)  - \frac{\pi}{2} i.
\]
On the other hand, one can evaluate the integral directly by summing up the contributions
from the branch cuts connecting $q^u_{2j}$ and  $q^u_{2j+1}$ and the pole at $q=p$,
\begin{align*}
\tilde{\sigma}_{{\cal C}_+}(p) =& 
-2 \ln \frac{1+\re^{-\epsilon(p)/T}}{1-\re^{-\epsilon(p)/T}}
-2 \sum_{j=1}^{\infty}  \left. \log \sin \frac{q-p}{2}  \right |^{q^u_{-2j}}_{q^u_{-2j+1}}  \\
&-2 \sum_{j=0}^{\infty}  \left.  \log \sin \frac{q-p}{2}   \right |^{q^u_{2j}}_{q^u_{2j+1}}  .
\end{align*}
Thus, we conclude
\begin{multline}\label{eq:mu_positivep} 
     \mu(p) = \frac{i}{2\pi}\frac{(1-\re^{\epsilon(p)/T})}{(1+\re^{\epsilon(p)/T})^2}
              \\ \times
	      \Bigl(\prod_{j=1}^{\infty} \frac{\sin \frac{p-q^u_{-2j+1}}{2}}
	                                      {\sin \frac{p-q^u_{-2j}}{2}}
		    \prod_{j=0}^{\infty} \frac{\sin \frac{p-q^u_{2j+1}}{2}}
		                              {\sin \frac{p-q^u_{2j}}{2}} \Bigr)^2.
\end{multline}
By remembering (\ref{eq:def_qj}), one verifies that the double zeros and  
double poles at $q^u_{2k+1}$ cancel
and that the single poles at  $q^u_{2k}$ survive if $\operatorname{Im} (p)>0$.

The argument for the case $\operatorname{Im} (p)<0$ goes almost in parallel. This time we consider
a closed rectangular contour in the lower half plane.
Using a similar reasoning as above, we obtain,
\begin{multline*}
     \mu(p) = - \frac{i}{2\pi} \frac{1}{(1-\re^{\epsilon(p)/T})} 
              \\ \times
	      \Bigl(\prod_{j=1}^{\infty} \frac{\sin \frac{p-q^d_{-2j}}{2}}
	                                      {\sin \frac{p-q^d_{-2j+1}}{2}}
                    \prod_{j=0}^{\infty} \frac{\sin \frac{p-q^d_{2j}}{2}}
		                              {\sin \frac{p-q^d_{2j+1}}{2}} \Bigr)^2.
\end{multline*}
Thus, for $\operatorname{Im} (p)<0$,  $\mu(p)$ possesses single zeros at  $q^d_{2k}$ and double poles at
$q^d_{2k+1}$.
The same conclusion can be drawn from (\ref {eq:mu_positivep}), though it is derived under
the assumption that $\operatorname{Im}(p)>0$.

Next we consider the massless regime ($h<h_c$) and assume that $\operatorname{Im}(p) \ge 0$.
We again adopt a contour in the upper half plane which includes  ${\cal E}$ as a part. 
As ${\cal E}$ encircles the Fermi point $p_F$ from below,
we obtain
\begin{multline*}
\re^{\sigma_+(p)} \\
= i \Bigl ( \frac{\re^{\epsilon(p)/T}-1}{\re^{\epsilon(p)/T}+1}
\prod_{j=0}^{\infty} \frac{\sin \frac{p-q^r_{-(2j+1)}}{2}} {\sin \frac{p-q^r_{-2j}}{2}} 
\prod_{j=1}^{\infty} \frac{\sin \frac{p-q^{\ell}_{-(2j-1)}}{2}} {\sin \frac{p-q^{\ell}_{-2j}}{2}} 
\Bigr)^2,
\end{multline*}
thus,
\begin{align*}
\mu(p)=& -\frac{i}{2\pi}\frac{\re^{\epsilon(p)/T}-1}{(\re^{\epsilon(p)/T}+1)^2} \\
&\times 
\Bigl(
\prod_{j=0}^{\infty} \frac{\sin \frac{p-q^r_{-(2j+1)}}{2}} {\sin \frac{p-q^r_{-2j}}{2}} 
\prod_{j=1}^{\infty} \frac{\sin \frac{p-q^{\ell}_{-(2j-1)}}{2}} {\sin \frac{p-q^{\ell}_{-2j}}{2}} 
\Bigr)^2.
\end{align*}
From this we easily see that $\mu(p)$ has single poles at $q^{\ell}_{-2j} \,(j\ge 1)$ and 
$q^{r}_{-2j} \,(j\ge 0)$ if $\operatorname{Im} (p) \ge 0$.
The expression is valid also for $\operatorname{Im} (p) < 0$.  
Thus, we see that $\mu(p)$ has single zeros at $q^{\ell}_{2j} \,(j\ge 0)$ and 
$q^{r}_{2j} \,(j\ge 1)$, while it possesses double zeros
at  $q^{\ell}_{2j+1} \,(j\ge 0)$ and $q^{r}_{2j+1} \,(j\ge 0)$.

%
%
\section{\boldmath Static correlations for small $m$ at finite~$T$}
\label{appendix:static_finiteT}
The static correlation functions of the  $S=\frac{1}{2}$ XXZ model were studied in \cite{BDGKSW08}
in the framework of a QTM approach.
Explicit formulas for the reduced density matrix of a short chain segment in an arbitrary magnetic
field at arbitrary temperature were obtained.
They can be used to express the two-point functions in terms of two fundamental functions
$\omega(x,y)$ and $\omega'(x,y)$ (note that the prime does not mean a derivative). See eqs.~(59)
and (62) in \cite{BDGKSW08} for their definition.  
Their arguments are introduced as inhomogeneities which must be sent to zero in the end.
The homogeneous (physical) limit produces derivatives with respect to the first and the
second argument, which will be denoted by, e.g., $\omega_x= \partial_x  \omega(x,y) |_{x,y \rightarrow 0}$, 
$\omega_y= \partial_y  \omega(x,y) |_{x,y \rightarrow 0} $ and so on.
The formulas behave non-trivially when the deformation parameter $q$
that parameterizes the anisotropy of the XXZ chain as $\Delta = (q + q^{-1})/2$
approaches a root of unity, $q^n = 1$. The XX chain corresponds to $q = i$, \textit{viz}. $\Delta = 0$. It is conjectured that the expression 
for the transversal two-point function has a singularity of removable type, whenever one evaluates correlators 
points that are at distance $k m$, for some integer $k$, if $q^m=-1$.
In the present case we conjecture that l'Hospital's rule must
be applied for the computation of 
$\langle \sigma_1^x \sigma_{m+1}^x \rangle$ if $m = 2, 4, 6, \dots$.

For $m=2$, for instance,
\begin{align*}
\langle \sigma_1^x \sigma_3^x \rangle=&-\frac{\omega}{\sh (2\eta)} -\frac{\ch(2\eta)}{2\eta} \omega'_x \\
&-\frac{\ch(2\eta) \tanh (\eta) (\omega_{xx} -2\omega_{xy})}{8}+ \frac{\sh^2(\eta) \omega'_{xxy}}{8\eta},
\end{align*}
where $\eta$ is such that $\Delta=\ch(\eta)$.
The first and the third terms are obviously singular at the XX point $\eta = \frac{i \pi}{2}$.
We thus introduce a small deviation $\gamma = - i \eta = \frac{\pi}{2} -\varepsilon$
and obtain by straightforward expansion with respect to $\varepsilon$,
\begin{align}
\langle \sigma_1^x \sigma_3^x \rangle=&\frac{i(4\omega-\omega_{xx}-2\omega_{xy})}{8\varepsilon}
+\frac{4\omega_x' -\omega'_{xxy}}{4\pi i}  \nonumber \\
&+
\frac{1}{8i} (4 \partial_{\gamma} \omega +\partial_{\gamma} \omega_{xx} -2\partial_{\gamma} \omega_{xy}) +{\cal O}(\varepsilon).
 \label{exactstaticxxm2}
\end{align}
We have verified numerically that the first term vanishes, i.e., the apparent singularities cancel each other.
We, however, need to evaluate derivatives  with respect to the anisotropy parameter, which adds an extra elaboration.

For generic $\gamma$ the function $\omega$ is expressible in terms of auxiliary functions 
$\mathfrak{b}(x), \bar{\mathfrak{b}}(x), g_{\mu}^{(\pm)},  {g'}_{\mu}^{(\pm)}$ defined in
\cite{BDGKSW08},
\begin{multline*}
     \omega(\mu_1, \mu_2) =
        - \frac{ K(\mu_1 - \mu_2) }{2} \\ - \int_{- \infty}^\infty \rd k 
          \frac{\sh \bigl( (\pi - \gamma)\frac{k}{2} \bigr)
                \cos(k(\mu_1 - \mu_2))}
               {i \sh \bigl( \frac{\pi k}{2} \bigr)
                   \ch \bigl( \frac{\gamma k}{2} \bigr)}  \\[1ex]
             - \int_{- \infty}^\infty \frac{\rd x}{\gamma}
               \frac{1}{\ch \Bigl( \frac{\pi (x - \mu_2)}{\gamma} \Bigr)}
               \biggl[ \frac{g_{\mu_1}^{(+)} (x)}{1 + \mathfrak{b}^{-1} (x)}
                     + \frac{g_{\mu_1}^{(-)} (x)}{1 + \bar{\mathfrak{b}}^{-1} (x)}
               \biggl].
\end{multline*}
%
A similar expression can be written for $\omega'$.
In order to evaluate $\omega$ and $\omega'$, one needs to solve non-linear integral equations
(cf.\ eqs.\ (52)-(55) in \cite{BDGKSW08}) for the auxiliary functions.
The XX model is exceptional, however, in that one of the integration kernels, $F$, vanishes.
Then $\mathfrak{b}(x), \bar{\mathfrak{b}}(x), g_{\mu}^{(\pm)}$ can be obtained explicitly,
and  ${g'}_{\mu}^{(\pm)}$ are calculated by taking convolutions of  $g_{\mu}^{(\pm)}$ with
appropriate kernels. Even in evaluating the derivatives  with respect to  $\gamma$, one does not have
to solve the integral equations, but only needs to perform integrations over already known
auxiliary functions. For example, one needs to evaluate $\partial_{\gamma} \mathfrak{b}^{-1}
= -  \mathfrak{b}^{-1}  \partial_{\gamma} \ln \mathfrak{b}$. This is  evaluated from,
%
\begin{multline*}
\partial_{\gamma} \ln \mathfrak{b}(x)= \partial_{\gamma} D_b(x) +
\int_{-\infty}^{\infty} \frac{dy}{2\pi} \partial_{\gamma} F (x-y) \ln(1+  \mathfrak{b}(y))\\
-\int_{-\infty}^{\infty} \frac{dy}{2\pi} \partial_{\gamma} F (x-y+i (\gamma-2 \delta) )
\ln(1+  \bar{{\mathfrak{b}}}(y))
\end{multline*}
%
where $D_b$ denotes a known function and $\delta$ is a small quantity introduced for technical reasons.
The parameter $\gamma$ must be set equal to $\frac{\pi}{2}$ after taking the derivative on the right-hand side.
We thus need only known functions in order to evaluate $\partial_{\gamma} \mathfrak{b}^{-1}$. In a
similar manner, one evaluates the $\gamma$ derivatives of $g_{\mu}^{(\pm)}$,  ${g'}_{\mu}^{(\pm)}$
and then obtains $\langle \sigma_1^x \sigma_3^x \rangle$.

On the other hand, since there are no apparent singularities for $m = 1$ or $3$,
we can take the limits directly in these cases and obtain
\begin{align}
\langle \sigma_1^x \sigma_2^x \rangle &= \frac{i \omega}{2},  \label{exactstaticxxm1} \\
\langle \sigma_1^x \sigma_4^x \rangle &=
i\Bigl( -\omega - \frac{ \omega_{xxyy}}{16} + \frac{\omega_{xyyy}}{24}+ \frac{5 \omega_{xy}}{6} -\frac{\omega_{yy}}{2} 
\Bigr) \nonumber \\
&+\frac{1}{2\pi}\bigl( \omega \omega'_{xyy} +      \omega'_{y} \omega_{yy}   \bigr).
\label{exactstaticxxm3}
\end{align}
   
We compare the values of  
$\langle \sigma^-_1(0) \sigma^+_{m+1}(0) \rangle$ obtained from our Fredholm
determinant representation and those obtained from the above exact formulas\footnote{
Half of the values of (\ref{exactstaticxxm2}), (\ref{exactstaticxxm1}), (\ref{exactstaticxxm3})
for $m=2,1,3$ reflecting the symmetry between $xx$ and $yy$ correlators.
 The static autocorrelation is obtained from the magnetization 
$\mathfrak{m}(T,h) = \int_{-\pi}^{\pi}  \frac{dp}{4\pi} \tanh \frac{\e(p)}{2T}$  by
$\frac{1}{2}- \mathfrak{m}(T,h)$.}
in Table~\ref{tb:statich1} (Table~\ref{tb:statich41}) for $h=0.1J$ $(<h_c)$
($h=4.1J$ $(>h_c)$). Their agreement up to a reasonable number of digits assures
the validity of our formulation in the static limit for small $m$.
 \onecolumngrid
 \begin{center}
\begin{table}[htb]
  \begin{center}
    \begin{tabular}{|c|c|c|c|c|} \hline
    T  &                   m=0&                  m=1 &                  m=2&                   m=3 \\ \hline 
0.1J &   0.4920331867&           -0.3178821170&     0.2022244472&   -0.1710719802 \\
           &   0.4920331808&          -0.3178821255&     0.2022244549&   -0.1710719748\\ \hline
0.5 J&    0.4917867108&          -0.3094178153&     0.1915980505&   -0.1508065219\\
           &   0.4917866789&          -0.3094178186&     0.1915980506&    -0.1508065200\\ \hline
J&       0.4910914106&          -0.2793621277&     0.1561766021&    -0.1020132729 \\
       &       0.4910913226&          -0.2793621160&      0.1561766025&   -0.1020132730  \\ \hline  
5 J &       0.4953703567 &         -0.0961933456 &    0.0185079888&    -0.0036065564 \\
        &      0.4953703745&          -0.0961933456&     0.0185079891&    -0.0036065569\\
      \hline  
    \end{tabular}  \\
\caption{Results for 
$\langle \sigma^-_1(0) \sigma^+_{m+1}(0) \rangle$
at $h=0.1J$ ($<h_c$). Upper line for each $T$ shows the results from our
new formula, while lower line shows the exact values. The Fredholm determinant
is estimated by means of a $512\times 512$ matrix.}\label{tb:statich1}
\end{center}
\end{table}
\end{center}
 \begin{center}
\begin{table}[htb]
  \begin{center}
    \begin{tabular}{|c|c|c|c|c|} \hline
     T  &                   m=0&                  m=1 &                  m=2&                   m=3 \\ \hline 
0.1J &  0.0186123689&            -0.0183512823&       0.0176076636&   -0.0164562085  \\
           &  0.0186123688&            -0.0183512823&      0.0176076634&        -0.0164562090\\ \hline
0.5 J &   0.0765154951&            -0.0704244629&     0.0563287894&    -0.0402407959\\
           &  0.0765154951&            -0.0704244629&     0.0563287892&    -0.0402407964\\ \hline
 J&      0.1186618941&            -0.0983455062&     0.0624189803&    -0.0338002579 \\
       &       0.1186618940&            -0.0983455062&     0.0624189800&   -0.0338002583  \\ \hline  
5 J &      0.3181814080 &           -0.0830610673 &    0.0160102135&  -0.0029751725 \\
        &     0.3181814073&           -0.0830610672&      0.0160102133&    -0.0029751730 \\
      \hline  
    \end{tabular}  \\
\caption{Results for  
$\langle \sigma^-_1(0) \sigma^+_{m+1}(0) \rangle$  at $h=4.1J$ ($>h_c$).
Upper line for each $T$ shows the results from our new formula, while 
lower line shows the exact value.
The Fredholm determinant is estimated by means of a $512\times 512$ matrix.}
\label{tb:statich41}
\end{center}   
\end{table}
\end{center}
\twocolumngrid
%
%
\section{\boldmath Static correlations for larger $m$ at $T\ll 1$ }
\label{appendix:static_zeroT}
While exact correlations are not available for larger $m$ at finite temperatures, 
the ground state correlation function is obtained explicitly for $h=0$ \cite{KitanineMailletSlavnovTerras2002FF}.
We find that it is  neatly expressed in terms of Barnes' $G$ function,
\[
\langle \sigma^x_1(0) \sigma^x_{m+1}(0) \rangle = 
\begin{cases}
  -\frac{\Phi(\frac{m+1}{2})\Phi(\frac{m+3}{2})}{(\Phi(1))^2} &  m=\text{odd},\\[1ex]
   \Bigl( \frac{\Phi(\frac{m+2}{2}))}{\Phi(1)} \Bigr)^2& m=\text{even},  \\
\end{cases}
\]
where
\[
\Phi(x)=\frac{(G(x))^2}{G(x+\frac{1}{2}) G(x-\frac{1}{2})}.
\]

Our formulation has numerical problems in the limit $T, h\rightarrow 0$.
Nevertheless, we tried to check the consistency at larger $m$ by taking numerical limits.
For this purpose we set $h=0.01$ and extrapolated the zero temperature values
by fitting data for $T=0.03 \sim 0.06 J$ using quadratic curves (see Table \ref{tab:statich41}). 
In spite of the above difference in details, we find that the results
qualitatively agree with the exact values.

\onecolumngrid
\begin{center}
\begin{table}[htb]
  \begin{center}
    \begin{tabular}{|c|c|c|c|c|c|c|c|} \hline
                     &            m=5&                  m=6&                  m=7&           m=8 &       m=9&                   m=10 &    m=20    \\ \hline 
Extrapolated &  -0.132188&           0.119687 &         -0.111465&   0.103810&       -0.0982175&   0.0929276&  0.0659939 \\ \hline
Exact           &   -0.132195&           0.119691&          -0.111467&   0.103807&       -0.0982084 &  0.0929116 &  0.0657593  \\ 
      \hline  
    \end{tabular}  \\
\caption{Extrapolated (to $T=0$) values of 
$\langle \sigma^-_1(0) \sigma^+_{m+1}(0) \rangle$ at $h/J=0.01$.
Exact values at $h=T=0$ are given for comparison.\label{tab:statich41}}
\end{center}
\end{table}
\end{center}
\twocolumngrid
%
%
%
%
\section{The modification due to poles} \label{appendix:det_poles_formula}

The  integration contours  ${\cal E}, \bar{{\cal E}}$ are originally located near
the real axis. As discussed in the main text, we need to shift ${\cal E}$ to 
$[-\pi,\pi]+i \delta$ and $\bar{{\cal E}}$  to $[-\pi,\pi]-i \delta$, so that
they pass through the saddle points, especially when $m$ is large in the space-like regime.
In the course of this deformation, the paths cross poles and we have to take account of
these contributions and modify (\ref{eq:Fredholm_representation}). This can be easily
done by following appendix C of \cite{DugaveGohmannKozlowskiSuzuki2016}.

We denote the sets of poles of $\mu(p)$ and $\bar{\mu}(q)$, crossed by the
contours, by $\{s_j\}_{j=1}^{m}$ and $\{\bar{s}_j\}_{j=1}^{m}$, respectively. These
sets may contain, in particular, the Fermi points, namely, $p_F \in \{s_j\}$
and $-p_F \in \{\bar{s}_j\}$ in the massless case. 

The shift of ${\cal E}$ modifies the functions by analytic continuation,
\begin{align} \displaybreak[0]
\Omega(m,t)&= \int_{-\pi+i \delta}^{\pi+i\delta} \rd p\,  \mu(p) {\rm e}^{2t u_{m,t}(p)} \nonumber \\
&+2\pi i  \sum_{j} \bigl(\underset{p=s_j}{\mathrm {res} } \mu(p) \bigr) {\rm e}^{2t u_{m,t}(s_j)},  \\
\tilde{v}(q)&= \int_{-\pi+i \delta}^{\pi+i\delta} \rd p \,   \mu(p) {\rm e}^{2t u_{m,t}(p)} \varphi(p,q)  \nonumber\\
& + 2\pi i \sum_{j} \bigl(\underset{p=s_j}{\mathrm{ res}}  \mu(p) \bigr)  {\rm e}^{2t u_{m,t}(s_j)} \varphi(s_j,q).
\end{align}
Here we have used the fact that $q$ is in the lower half plane.  Note that $\tilde{V}(q,q')$
is also modified due to the expression  (\ref{eq:Vandv}).

We also have to take account of the result of the shift of $\bar{{\cal E}}$.
The  analytic continuation modifies the Fredholm determinant,
\[
\begin{vmatrix}
\delta(x-y) + {K}(x,y)&    k(x, \bar{s}_1)            &              \cdots&    k(x, \bar{s}_m) \\
{K}(\bar{s}_1,y)&       1+k(\bar{s}_1, \bar{s}_1)&                    \cdots&    k(\bar{s}_1, \bar{s}_m) \\
\vdots&                              \vdots&                          \vdots&           \vdots& \\
{K}(\bar{s}_m,y)&         k(\bar{s}_m, \bar{s}_1)            &                     \cdots&   1+ k(\bar{s}_m, \bar{s}_m) 
\end{vmatrix} ,
\]
where we introduced
\begin{align*}
K(q_i,q_j) = P(q_i,q_j)- V(q_i,q_j), \\
k(x,  \bar{s}_j)  = -2\pi i \bigl(\underset{q={\bar s}_j}{\mathrm {res}}  \bar{\mu}(q) \bigr) K(x,\bar{s}_j).
\end{align*}
In order to simplify this, we define the resolvent Kernel $R(q,q')$ by,
\[
K(q,q') = R(q,q') + \int_{-\pi-i \delta}^{\pi-i\delta}  \rd q''\, K(q,q'')R(q'',q').
\]
By the simple determinant identity
\[
\begin{vmatrix}
A&  B\\
C&  D
\end{vmatrix}
={\rm det}(A){\rm det}(D- CA^{-1} B),
\]
one concludes that
\begin{multline*} \label{eq:FredholmII}
     \underset{\bar{\cal E}} {\rm det}(1+\hat{K})
        = \underset{[-\pi-i \delta, \pi-i \delta]}{\rm det} (1+\hat{K}) \\ \times
	  \underset{1\le i,j \le m}{\rm det} 
	  \Bigl \{ \delta_{i,j} +k(\bar{s}_i,\bar{s}_j)
	           - \int_{-\pi-i \delta}^{\pi-i\delta} \rd y\,
		   R(\bar{s}_i, y)k(y, \bar{s}_j) \Bigr \},
\end{multline*}
where  $(\hat{K}f) (q) := (\hat{P}f) (q)-(\hat{V}f) (q)$.
%
%
\section{Choice of the contour: massless and time-like regime}\label{app:choice_contour}

For $t$ slightly greater than $t_c$, $p_{\pm}$ is near $\pi/2$ from (\ref{eq:def_p_pm_time}) and 
$p_F$ is not necessarily between $p_{\pm}$. In this short time region, however, the straight
contours work equally well as in the space-like regime and do not have to be deformed.
Since $p_+$ (respectively \ $p_-$) moves towards $\pi$ (respectively\ $0$),  $p_F$  soon lies between
$p_{\pm}$  (we assume this below) and we need to think about possible deformations.

\begin{figure}[!h]
\centering
\includegraphics[width=6cm]{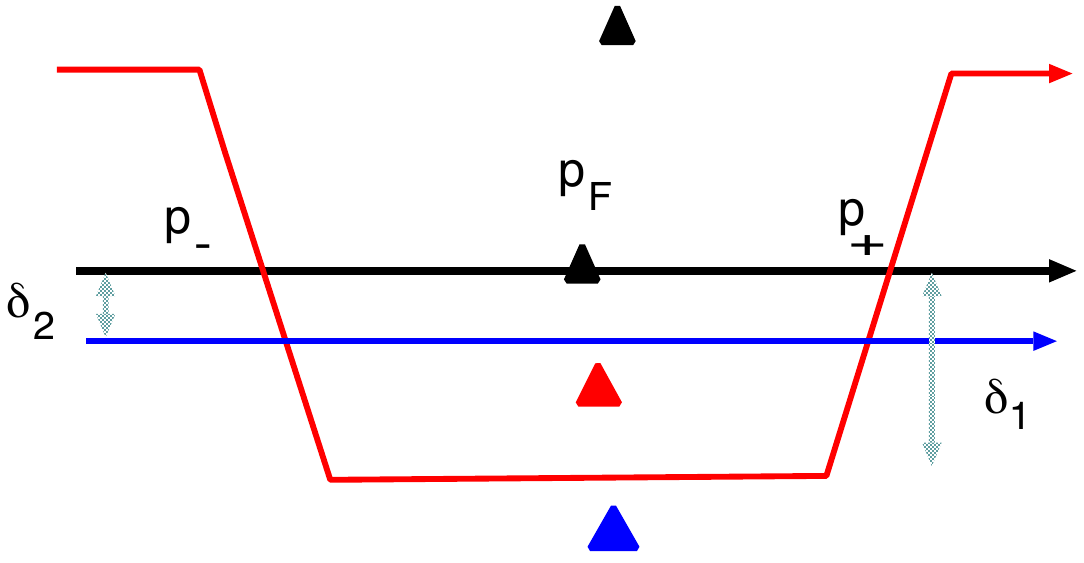} 
\caption{
Schematic picture of our choice of paths. The red path is for the $p$-variable, while
the blue one is for the $q$-variable. The black triangles represent simple poles of $\mu$ 
and the red one indicates a double pole of $\mu$. The blue triangle depicts a simple pole
of $\bar{\mu}$.
}
\label{fig:steepestdescendentII}
\end{figure}

We do not take the paths depicted in Figure \ref{fig:steepestdescendent} but use a
steepest descent path for the hole variable~$p$ and a straight line just below the real
axis for the particle variable~$q$, depicted explicitly in Figure
\ref{fig:steepestdescendentII}. We list reasons for the choice below.
\begin{enumerate}
\item
We did not deform the contour for the  $q$ variable into the upper half plane, suggested by
the steepest descent path, for two reasons. First, as already mentioned, the pole contribution
of $\varphi(p,q)$ at $p=q$  brings a divergent contribution for $t \gg 1$.
Second, $\bar{\mu}$ behaves irregularly in a region sandwiched by a double pole (in the upper half
plane, closest to the real axis) and a zero (at $q=p_F$) of $\bar{\mu}$.
\item 
On the other hand, we do deform the contour for the $p$ variable into the lower half plane,
but in a specific manner. This is due to the fact that there are two closely positioned
poles for $\mu$ (the black and the red triangles in Figure~\ref{fig:steepestdescendentII}).
In the region sandwiched by these, $\mu$ behaves quite unstable. 
We thus decided to deform he contour over the double pole of $\mu$ (the red triangle).
Note that at the blue triangle (= a pole of $\bar{\mu}$), $\mu$ is null, and around the point
it behaves smoothly.
\item
We can shift the blue contour below the red triangle, in principle.
This may be sometimes a nice choice as long as $t$ is not so large, since 
the pole contribution of $\varphi(p,q)$ at $p=q$ behaves irregularly
between the black and the red triangle. In a later stage, however, the
phase factor ${\rm e}^{-t u_{m,t}(q)}$ shows a divergent behavior if 
the contour is too far away from the real axis.
\end{enumerate}

We define the displacements  $\delta_1$ and $\delta_2$ as in Figure \ref{fig:steepestdescendentII}. 
For reasons 2 and 3, we take $\delta_1>\delta_2$, as the opposite choice $\delta_1<\delta_2$
would ruin the stability of the long time runs by reason 3.

The most subtle problem is posed by the question as to how to stabilize both factors 
$\re^{-t u_{m,t}(q)}$ and $ \mu(q_i) \mu(q_j) \bar{\mu}(q_j) \re^{ t u_{m,t}(q_i) +t u_{m,t}(q_j)}$
simultaneously. The former is the phase factor generic for the particles and diverging
for $\operatorname{Im}(q)<0$. This factor is multiplied by $\re^{-t u_{m,t}(p)}$ for which an optimal
path is already chosen and it suppresses the divergent behavior. Thus, we conclude that
the former component is already harmless.

The latter comes from the $p=q$ pole of  $\varphi(p,q)$. The naive answer is to use the
blue contour just below the real axis, like in the massive case. This does not necessarily
give a stable result: a small $\delta_2$ causes divergent behavior of $\mu(q)$.
We pay attention to the fact that the product $\bar{\mu}(q_j) \mu(q_j)$ remains finite
for $q_j\sim p_F$. We then tune $\delta_2$ so that $|\mu(q) \re^{2 t u_{m,t}(q)}| \sim 1$
for $q \sim p_F -i \delta_2$ is satisfied for each given $t$. As a consequence $\delta_2$
gets smaller for larger $t$. Therefore the calculation fails eventually at a very late stage when
$\delta_2$ turns infinitesimally small. Empirically, the correlation functions become
too small to be detected before this limitation approaches. Thus, practically, the present
method successfully yields a stable calculation for a reasonably long time range
as demonstrated in the main text. 

%
%
\section{\boldmath The choice of $n$}
\label{app:choice_on_n}

We present results of an experimental study on the choice of the dimensions $n$ of
the discretized Fredholm determinant.

We start from proposing the following:
\begin{conjecture}
Both  real  and  imaginary part of   
$(-1)^m \langle \sigma^-_1(0) \sigma^+_{m+1}(t)\rangle$ stay positive in the massless time-like regime.
\end{conjecture}

This conjecture is motivated by the empirical fact that the numerical result becomes
unstable after the real (or imaginary) part of 
$(-1)^m \langle \sigma^-_1(0) \sigma^+_{m+1}(t)\rangle$ crosses zero for  $t>0$.

Thus, we increase $n$ until the positivity of both parts of 
 $(-1)^m \langle \sigma^-_1(0) \sigma^+_{m+1}(t)\rangle$
is satisfied and until the result converges (with variation of $n$).
In the following we restrict ourselves to the parameters $m=50, T=0.05 J, h=0.5 J$ and
increase the value of $n$. 

\begin{figure}[!h]
\centering
\setlength{\tabcolsep}{20pt}
\includegraphics[height=\figheight]{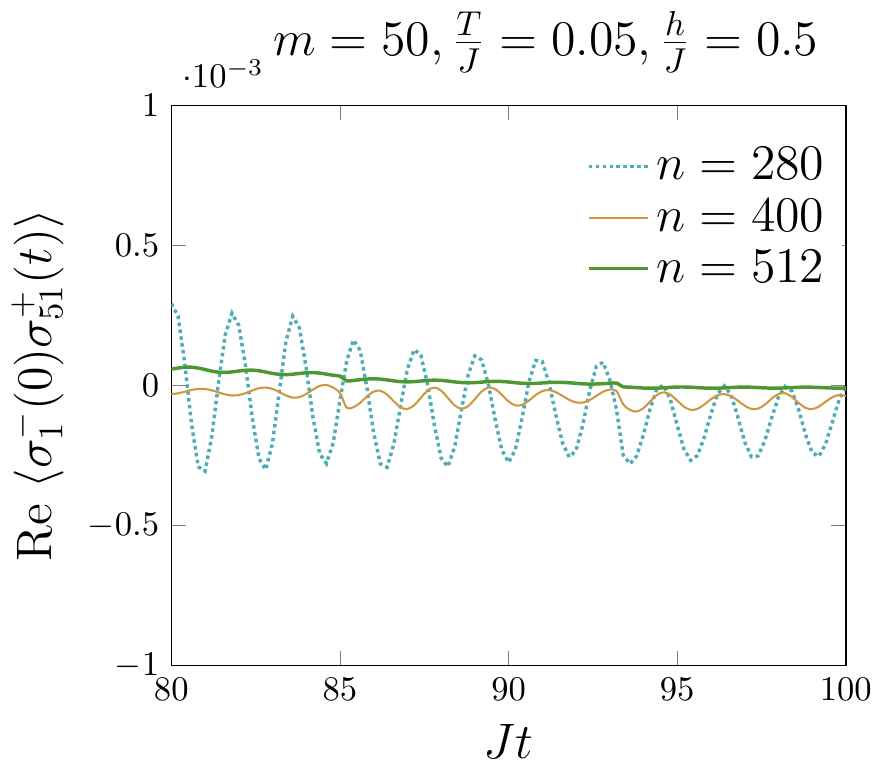}  \hspace{-0.1cm}
\includegraphics[height=\figheight]{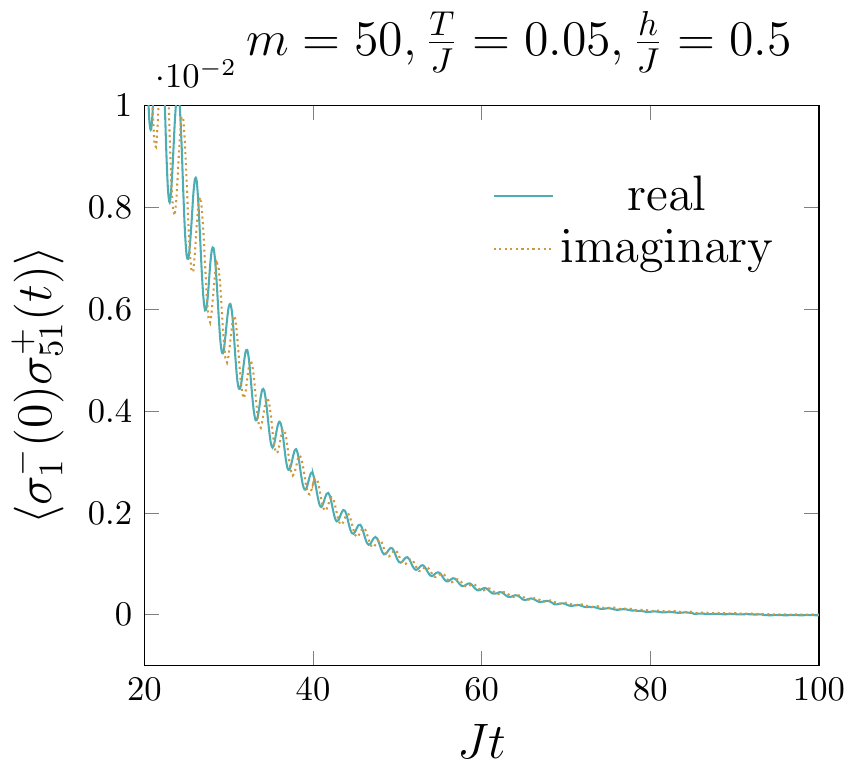}
\caption{
The real and the imaginary part of the Fredholm determinant for $m=50, T=0.05 J, h=0.5 J$.
Right panel shows the result for $n=512$, left panel a comparison of the
real parts computed with $n=280, 400$ and $512$.
}
\label{fig:m50T005h05n512}
\end{figure}

Especially, we take a closer look at the late stage and compare the real parts with 
$n=280$, $400$ and $512$. See Figure~\ref{fig:m50T005h05n512} (left). The real part
become negative for $n=280, 400$. The case with $n=280$ shows abrupt change in the
oscillation amplitude, and the amplitude is not decreasing, but seems increasing for
$n=400$. Both cases seem unreasonable. The case with $n=512$ seems well-behaved.
The real and the imaginary part of 
$\langle \sigma^-_1(0) \sigma^{+}_{51}(t) \rangle$
is plotted in Figure \ref{fig:m50T005h05n512} (right) with $n=512$.
It does not show any particular fluctuation around $20<J t<40$ any longer.

\begin{figure}[!h]
\centering
\setlength{\tabcolsep}{20pt}
\includegraphics[height=\figheight]{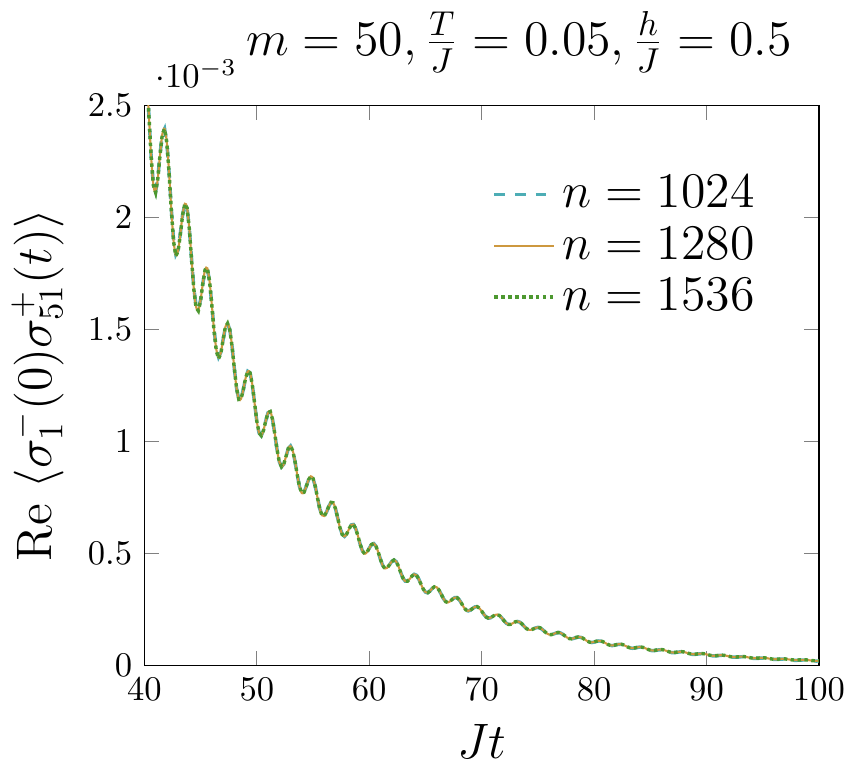} \hspace{0.1cm}
\includegraphics[height=\figheight]{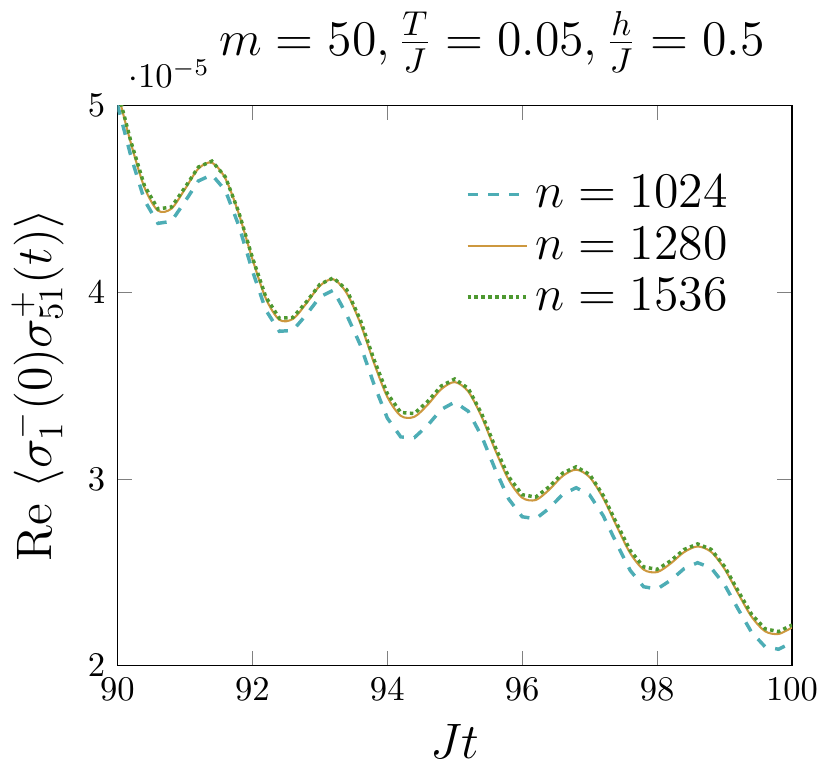}
\caption{
The real part of 
$\langle \sigma^-_1(0) \sigma^{+}_{51}(t) \rangle$ at $T=0.05 J, h=0.5 J$ with  $n=1024,
1280$ and $1536$ (left). Zoom-in is shown in the right panel.
}
\label{fig:m50T005h05n1536}
\end{figure}

We further increase the values of $n$. Figure~\ref{fig:m50T005h05n1536} supplements
the real parts of  
$\langle \sigma^-_1(0) \sigma^{+}_{51}(t) \rangle$ with $n=1024$, $1280$ and $1536$.
One can hardly see the difference. The zoom-in in the right plot convinces us that
the results with $1280$ and $1536$ have already reached numerical convergence. Thus,
we conclude that the choice $n=1536$ is a safe choice. 

We remark that when $T$ is higher, we do not have to take such large values of $n$.
The correlations will anyway decay to very small values $\sim10^{-7}$ quickly, even with
relatively small $n$ (like $n = 128 \sim 256$).

\newpage


\bibliographystyle{apsrev}
\bibliography{xx_xx_numerics_v10}

\end{document}